\documentclass[a4paper]{aa}
\usepackage{txfonts}
\usepackage{etoolbox}
\usepackage{graphicx}
\usepackage{amsmath}
\usepackage{hyperref}
\usepackage{natbib}
\newcommand{\etal}{{et al}\/.}
\newcommand\rd{{\rm d}}
\hypersetup{colorlinks=true,linkcolor=blue,citecolor=blue,filecolor=blue,urlcolor=blue}
\begin{document}

\title{Radio-loud AGN in the first LoTSS data release}
\subtitle{The lifetimes and environmental impact of jet-driven sources}
\author{M.J.~Hardcastle\inst{1}\thanks{\email{m.j.hardcastle@herts.ac.uk}}
  \and W.L.~Williams\inst{1} \and P.N.~Best\inst{2} \and J.H.~Croston\inst{3} \and
  K.J.~Duncan\inst{4} \and 
H.J.A.~R\"ottgering\inst{4} \and J.~Sabater\inst{2} \and
T.W.~Shimwell\inst{5} \and C.~Tasse\inst{6,7} \and
J.R.~Callingham\inst{5} \and R.K.~Cochrane\inst{2} \and F.~de Gasperin\inst{8} \and
G.~G\"urkan\inst{9} \and M.J.~Jarvis\inst{10,11} \and
V.~Mahatma\inst{1} \and
G.K.~Miley\inst{4} \and B.~Mingo\inst{3} \and S.~Mooney\inst{12} \and
L.K.~Morabito\inst{10}  \and S.~P.~O'Sullivan\inst{13} \and
I.~Prandoni\inst{14} \and A.~Shulevski\inst{15} \and D.J.B.~Smith\inst{1}}
\institute{Centre for Astrophysics Research, University of
  Hertfordshire, College Lane, Hatfield AL10 9AB
  \and
  SUPA, Institute for Astronomy, Royal Observatory, Blackford Hill,
  Edinburgh, EH9 3HJ, UK
  \and
  School of Physical Sciences, The Open University, Walton Hall,
  Milton Keynes, MK7 6AA, UK
  \and
  Leiden Observatory, Leiden University, PO Box 9513, NL-2300 RA
  Leiden, the Netherlands
  \and
  ASTRON, the Netherlands Institute for Radio Astronomy, Postbus 2, 7990 AA Dwingeloo, the Netherlands
  \and
  GEPI \& USN, Observatoire de Paris, Universit\'e PSL, CNRS, 5 Place
  Jules Janssen, 92190 Meudon, France
  \and
  Department of Physics \& Electronics, Rhodes University, PO Box 94,
  Grahamstown, 6140, South Africa
  \and
  Hamburger Sternwarte, Gojenbergsweg 112, D-21029 Hamburg, Germany
  \and
  CSIRO Astronomy and Space Science, PO Box 1130, Bentley WA 6102,
  Australia
  \and
  Astrophysics, University of Oxford, Denys Wilkinson Building, Keble
  Road, Oxford, OX1 3RH, UK
  \and
  Physics and Astronomy Department, University of the Western Cape,
  Bellville 7535, South Africa
  \and
  School of Physics, University College Dublin, Belfield, Dublin 4,
  Republic of Ireland
  \and
  Hamburger Sternwarte, Universit\"at Hamburg, Gojenbergsweg 112, D-21029
  Hamburg, Germany
  \and
  INAF - Istituto di Radioastronomia, Via P. Gobetti 101, 40129
  Bologna, Italy
  \and
  Anton Pannekoek Institute for Astronomy, University of Amsterdam, Postbus 94249, 1090 GE Amsterdam, the Netherlands}

\abstract{We constructed a sample of 23,344 radio-loud active
  galactic nuclei (RLAGN) from the catalogue derived from
  the LOFAR Two-Metre Sky Survey (LoTSS) survey of the HETDEX
  Spring field. Although separating AGN from star-forming
  galaxies remains challenging, the combination of spectroscopic and
  photometric techniques we used gives us one of the largest available
  samples of candidate RLAGN. We used the sample, combined with
  recently developed analytical models, to investigate the
  lifetime distribution of RLAGN. We show that large or giant powerful
  RLAGN are probably the old tail of the general RLAGN
  population, but that the low-luminosity RLAGN candidates in our
  sample, many of which have sizes $<100$ kpc, either require a
    very different lifetime distribution or have different jet physics
    from the more powerful objects. We then used analytical models
  to develop a method of estimating jet kinetic powers for our
  candidate objects and constructed a jet kinetic luminosity function
  based on these estimates. These values can be compared to
  observational quantities, such as the integrated radiative luminosity
  of groups and clusters, and to the predictions from models of RLAGN
  feedback in galaxy formation and evolution. In particular, we show
  that RLAGN in the local Universe are able to supply all the energy
  required per comoving unit volume to counterbalance X-ray radiative
  losses from groups and clusters and thus prevent the hot gas from
  cooling. Our computation of the kinetic luminosity density of
    local RLAGN is in good agreement with other recent observational
    estimates and with models of galaxy formation.}

\maketitle
\section{Introduction}
\label{sec:intro}


Radio-loud active galactic nuclei (radio galaxies and radio-loud quasars; hereafter RLAGN) are a subset of the active galaxy population in
  which accretion onto the central supermassive black hole of a galaxy
  generates a relativistic jet of charged particles (electrons,
  positrons, and/or protons) and magnetic field. These jets propagate
  into the medium permeating and surrounding the host galaxy,
  inflating ``bubbles'' of low-density, high-pressure material
  containing relativistic electrons that generate the observed radio
  emission through the synchrotron process. Basic models of the
  dynamics of these objects as they interact with the external medium
  have been available for over 40 years
  \citep{Scheuer74,Blandford+Rees74}, but have been refined and
  improved more recently both in terms of analytical models
  \citep[e.g.][]{Kaiser+Alexander97,Blundell+99,Luo+Sadler10,
    Turner+Shabala15, Hardcastle18}
  and numerical models taking account of the known environmental
  properties of these objects \citep[e.g.][]{Reynolds+02,
    Basson+Alexander03, Zanni+03, Krause05, Heinz+06, Mendygral+12,
    Hardcastle+Krause13,Hardcastle+Krause14,English+16}.

Radio galaxy physics has become important outside the active galactic
nucleus (AGN) community over the past 20 years for two closely related
reasons. The first is the role of AGN in solving the so-called cooling
flow problem. This problem was posed by observations of rich clusters
of galaxies that showed that their central hot gas, emitting in the
X-ray with temperatures $T \sim 10^7$ K, had cooling times ($\tau =
E/({\rm d}E/{\rm d}t)$) much less than the age of the Universe. This
gas should therefore cool out of the temperature regime in which it
emits X-rays and, eventually, form stars or deposit cold gas in the
central cluster galaxy at a rate, for the most rapidly cooling
clusters, of thousands of solar masses per year, while causing the gas
to flow inwards owing to the loss of central pressure (a ``cooling
flow''; \citealt{Fabian+84}). However, these large amounts of cold gas
and/or star formation were not observed, and neither, when
observational advances permitted it, was the low-temperature
X-ray-emitting gas that would have been predicted by the cooling flow
model \citep[e.g.][]{Sakelliou+02}. It was rapidly realised
(e.g.\ \citealt{Eilek+Owen06}) that essentially all cooling flow
clusters host a RLAGN with sufficient power to offset the cooling, and
so it is now widely assumed that radio galaxies provide the
``thermostat'' for rich clusters of galaxies, keeping the central gas
hot and rarefied. The precise mechanism by which the gas is coupled to
the active nucleus, and the radio lobes to the gas, is not clear. It
seems likely that at least some radio galaxies are powered by
accretion of the hot phase onto the black hole
(e.g.\ \citealt{Allen+06}, \citealt{Hardcastle+07-2}), although
increasingly the consensus is that this is mediated by a cooling
instability \citep{Pizzolato+Soker05, Gaspari+13, Voit+Donahue14}.
Hot-gas accretion thus provides the connection in one direction, while
the expansion of the radio lobes can do work on the hot gas in various
ways (see e.g.\ \cite{Fabian+00} for early imaging, and
\cite{McNamara+Nulsen12,Heckman+Best14} for recent reviews).

The second reason for the importance of RLAGN in recent times
arises in part out of the first. A major advance in our understanding
of the way all galaxies formed and evolved has come from efforts to
use numerical models to predict features of the present and past
galaxy population, such as the galaxy mass or luminosity function, the
galaxy colour-magnitude diagram or the evolution of star formation in
the Universe. Initially this work used semi-analytic models, i.e. the
properties of the baryonic matter in the Universe were inferred from a
hydrodynamical simulation of the dark matter
(e.g.\ \citealt{Bower+06,Croton+06}). With increasing computing
power, it is now possible to model the baryons and dark matter
together and in a self-consistent way
(e.g.\ \citealt{Vogelsberger+14,Schaye+15}) and semi-analytic
modelling has also become more sophisticated \citep{Croton+16}. However, all these models
agree in predicting a very different galaxy luminosity function from
what is observed, if only the physics of dark matter, gas, and stars is
taken into account; far too many luminous galaxies are produced, and
the most luminous galaxies in the simulations are an order of
magnitude more luminous than anything we observe today. Motivated in
part by the observational evidence that RLAGN indeed solve
the cooling flow problem and prevent the formation of massive
cluster-centre galaxies in the local Universe, modellers can reproduce
the observed galaxy luminosity function by introducing AGN feedback
into their models. In modern models, this takes the form of an
injection of energy into the baryonic matter driven by accretion onto
the galactic-centre black hole. This AGN feedback takes place not just
in the local Universe, but over all cosmic time, and, as it is a
crucial ingredient of all modern models of galaxy formation,
it is vital that the nature and energetics of the feedback predicted
be tested against observations.

Cosmological models that deal with a scale large enough to reproduce
the galaxy luminosity function do not simultaneously deal with the
scales at which detailed AGN physics can be modelled. Even if they
did, we still lack a basic understanding of what causes some AGN to
have powerful radio jets. Therefore models do not predict the relative
importance of radio-loud and radio-quiet AGN in heating the baryons
and inhibiting star formation: the oft-quoted division by
\cite{Croton+06} into ``jet-mode'' and ``quasar-mode'' AGN does not
imply that radiative feedback is known observationally to terminate
star formation in major mergers. There are several reasons to think
that RLAGN may be important, however. Firstly, we know (as discussed
above) that RLAGN, not radio-quiet ones, are responsible for the
maintenance of hot cluster haloes in the local Universe: few if any of
these host a luminous quasar but effectively all host a powerful radio
galaxy. Secondly, RLAGN have a clear mechanism, the interaction
between the jets and the external medium, for efficiently coupling the
AGN output (in the form of the kinetic power of the jets) to the
baryonic matter, and this is directly observed to drive hot and cold
gas out of galaxies (see
e.g.\ \citealt{Morganti+05,Nesvadba+08,Hardcastle+12,Russell+17}). On
the other hand, radio-quiet AGN, which produce all of their energetic
output as photons, can only drive outflows in dusty galaxies where the
radiation from the accretion disc is efficiently absorbed before it
can escape from the galaxy, meaning that, for example, almost all
optically selected quasars cannot be efficiently optically coupled to
their host galaxies. The true answer to the question of which AGN are
implicated in feedback processes can only be provided by observation.
In order to understand the contribution of RLAGN to these processes,
we need the ability to measure the kinetic power, and thus the kinetic
luminosity function, of large, well-constrained samples of RLAGN.

At present, although significant advances have been made in recent
years, this is still a difficult undertaking even in the local
Universe. Two approaches to measuring the jet power from the radio
luminosity are commonly used. Firstly, analytic models of the source
can be used to predict the radio luminosity for a given jet power
(e.g.\ \citealt{Willott+99}). Secondly, estimates of the jet kinetic
power can be derived from X-ray observations that show cavities in the
hot gas inflated by the radio lobes are used to infer the $p\Delta V$
work done to inflate the cavity, which can be combined with some
estimate of the source age to infer the jet power; these cavity powers
can then be empirically related to the radio luminosity
\citep[e.g.]{Birzan+04,Cavagnolo+10}. Both methods have significant
problems. The cavity power method relies on a poorly known source age
and can only work when cavities are observed, which rules out the use
of this approach in the case of the most powerful classical double
AGN, in which typically the lobes are brighter in inverse Compton than
their surroundings (see \citealt{Hardcastle+Croston10} for a
discussion of why this is so). This method is, moreover, biased
towards small sources in rich cluster environments \citep{Birzan+12}
and relies on expensive X-ray observations that are not available for
large samples of sources. Therefore there is at least some possibility
that the relationships that are derived from cavity estimates are
biased for the population in general. It is not even clear whether the
correlations between radio luminosity and cavity power that are
observed in these samples are driven by physics rather than a common
correlation with distance \citep{Godfrey+Shabala16}. On the other
hand, a single conversion based on a theoretical model giving the
radio luminosity is also unrealistic, since it is clear on simple
physical grounds that the radio luminosity must depend strongly
  on the source age (since the luminosity depends on the energy density in the
  lobes, the lobe volume and the magnetic field strength in the lobes,
  all of which evolve with time), on environment, and on redshift (due
  to inverse-Compton losses). Both numerical and analytical models of
this evolution exist, at least for certain types of RLAGN
\citep[e.g.][]{Kaiser+Alexander97,Blundell+99,Mocz+11,Turner+Shabala15,Hardcastle18},
but they have generally not been applied to large numbers of sources
in a consistent way to infer jet powers.

To measure kinetic powers in the local Universe, large sky areas are
needed, but large-area statistical studies of RLAGN have
been hindered in the past by the capabilities of
previous-generation radio instruments. Existing very wide-area
  radio surveys that have had a resolution high enough to allow
adequate identification of RLAGN with their host galaxy or quasar have
not simultaneously had the range of short baselines necessary for
high-fidelity imaging of extended structures. To date the
highest resolution wide-area radio survey is the Very Large
  Array (VLA) survey Faint Images of the Radio Sky at Twenty-Centimetres (FIRST), with a resolution of 5 arcsec
\citep{Becker+95}. As this is insensitive to structures on scales
larger than around 1 arcmin, however, it is not possible to generate a complete
sample from FIRST alone, and in the past it has been necessary to
combine catalogues from the NRAO VLA Sky Survey (NVSS)
\citep{Condon+98} and FIRST to achieve this
\citep[e.g.][]{Best+05,Hardcastle+12,Best+Heckman12}. With this approach,
though it is possible to obtain flux densities and optical
identifications for radio sources, it is not possible (without a
  great deal of work on the archival FIRST and NVSS $uv$ data) to make fully
spatially sampled high-resolution images of them; this means
that insufficient information about, for example, source size, a proxy of
age, is available for jet power inference.

 The LOw Frequency ARray (LOFAR; \citealt{vanHaarlem+13} is in the
  process of solving this problem. The LOFAR survey of the northern
sky, the LOFAR Two-metre Sky Survey (LoTSS; \citealt{Shimwell+17}, when
complete, will provide an unrivalled resource for wide-area
low-frequency (144-MHz) selection of extragalactic samples, both of
star-forming galaxies (hereafter SFG) and of RLAGN\footnote{See \url{http://lofar-surveys.org/}.}. At optimal
declinations for LOFAR LoTSS is approximately ten times deeper than
FIRST for typical observed spectral indices ($\alpha \sim 0.7$), while having a similar resolution (6 arcsec) and, crucially,
possessing the short baselines necessary to image all but the
largest scale structures in the radio sky\footnote{In the imaging
    that supports this paper we use a short-baseline cut of 100 m,
    allowing good imaging of structures on scales up to $\sim
    1^\circ$. In practice, we are limited in imaging such structures
    by surface brightness sensitivity rather than short baselines.}.
Low-frequency selection for RLAGN is extremely valuable because it
minimizes the effect on the total flux density of flat-spectrum beamed
structures such as the core, jets, and hotspots: at low frequencies emission from a
RLAGN is dominated by the much more isotropic large-scale
lobes. Thus, although the forthcoming Evolutionary Map of the Universe
(EMU) survey
\citep{Norris+11} with the Australian Square Kilometre Array
Precursor (ASKAP) will cover a larger sky area at comparable (slightly
lower) resolution to LoTSS and very similar sensitivity to typical
sources at its operating frequency of 1.3 GHz, LoTSS as a
  low-frequency survey will remain competitive until the (currently
hypothetical) long-baseline extension of the low-frequency Square
Kilometer Array (SKA) itself.

The present paper is concerned with the properties of RLAGN selected
from the LoTSS survey of the Hobby-Eberly Telescope Dark Energy
eXperiment (HETDEX; \citealt{Hill+08}) Spring field (hereafter the
HETDEX survey; \citealt{Shimwell+18}), the first
  full-quality data release of LoTSS (DR1). We investigate what can be learned
about RLAGN physical properties, and in particular their effect on
their environments, from the LOFAR-detected RLAGN population without
spectroscopic information other than what is provided by the Sloan Digital Sky Survey (SDSS: \citealt{Eisenstein+11}). We begin by
constructing an RLAGN sample based on the spectroscopic data where
available and on photometric redshifts and {\it WISE} colours
otherwise. This allows us to construct a very large sample of
  objects with radio luminosity and (projected) physical size
  information. We then show that a simple model of the RLAGN lifetime
  function, essential input into an inference of jet power from
  radio observations, adequately explains the observed distribution of
  source sizes for luminous sources. Furthermore, there is no evidence
  for any difference in host galaxy properties as a function of
  physical size, which is consistent with a simple model in which the powerful
  radio galaxies are a single physical population observed at
  different times in their life cycle. This conclusion allows us to
  carry out bulk jet power inference using a dynamical model of radio
  source evolution and to construct a jet kinetic luminosity function
  in the local Universe whose integral can be compared to the current
  radiative output of groups and clusters.
Throughout this paper we use a cosmology in which $H_0 = 70$ km
s$^{-1}$, $\Omega_{\rm m} = 0.3$ and $\Omega_\Lambda = 0.7$. The
spectral index $\alpha$ is defined in the sense $S \propto \nu^{-\alpha}$.

\section{The data}

\subsection{Radio data used in this paper}
\label{sec:radiodata}

This paper is based on DR1 of the LoTSS survey, which covers 424
deg$^2$, i.e. about 2\% of the total planned northern sky coverage. As
described by \cite{Shimwell+18}, we have devised an observation
and imaging strategy for this area that permits high-fidelity imaging
over wide areas down to a typical rms noise level of 70 $\mu$Jy
beam$^{-1}$ at the full 6-arcsec resolution of the Dutch LOFAR
baselines\footnote{The component of the International LOFAR
    Telescope (ILT) located in the Netherlands has a maximum baseline
    of 120 km. The observations of the HETDEX field did not include
    the longer baselines of LOFAR, using telescopes in international
    partner countries, although they are generally present in other data for the
    LoTSS survey.}. \cite{Williams+18b} describe the processing of
the raw catalogues derived from the Python Blob Detector and Source
Finder ({\sc PyBDSF}) software
\citep{Mohan+Rafferty15} to give a sample of 318,520 radio
sources that are believed to be real (i.e. not artefacts from the
limited dynamic range of the survey) and physical (i.e. lobes of radio
galaxies are associated and unassociated sources are de-blended).
These authors also describe the combination of the radio images and catalogues
with the available optical and near- to mid-infrared data from PanSTARRS DR1
\citep{Chambers+16} and {\it AllWISE} \citep{Wright+10,Mainzer+11}, a
process that gives plausible optical/IR counterparts for 72\% of these
objects (231,716). The vast majority of these sources are derived from likelihood-ratio
cross-matching with a combined optical/IR catalogue (for
  simplicity we refer to these as optical counterparts in what
  follows). Finally \cite{Duncan+18} describe the algorithms used to
estimate photometric redshifts for these optical counterparts; 162,249
sources (51\% of the input catalogue and 70\% of those objects with
optical identifications; IDs) have some kind of redshift estimate, using spectroscopic
redshifts where available (principally from the SDSS; \citealt{Eisenstein+11}) and photometric redshifts otherwise.

Our starting point in this paper is the 318,520 sources in the
``value-added'' radio and optical catalogue of \cite{Williams+18b}. We describe this as the value-added catalogue because it
  contains optical, infrared, and redshift information that is not
  present in the raw radio catalogues. \cite{Shimwell+18} and
\citeauthor{Williams+18b} describe the measurement of the radio
properties of these objects, but it is worth briefly summarising these properties
here. In essence, objects in the sample fall into two categories:
objects for which we adopt the PyBDSF properties of an original radio
detection, and objects in which a number of original PyBDSF sources have
been amalgamated (or, in very rare cases, where one PyBDSF source has
been split into components) after human visual inspection. In the
former case (simple sources), the flux density is the result of a
Gaussian fit or fits to the image data by PyBDSF, and we adopt a
largest angular size for the source that is twice the full width at
half maximum (FWHM) of the deconvolved fitted Gaussian (this is
roughly correct for uniform-brightness projected spherical or
ellipsoidal sources). In the latter case (composite sources) the
total flux density of the resulting source is taken to be the sum of
the total flux densities of all the components used, and the largest
angular size is taken to be the maximum distance across the convex
hull enclosing the elliptical regions with semi-major and semi-minor
axes corresponding to the deconvolved major and minor axes (FWHM) of
the fitted Gaussians. This definition has the property that it would
be consistent with the simple-source definition if there were only one
Gaussian in the composite source. In order to permit the convex hull
to be calculated, unresolved sources that are part of a composite
object are given a very small size (0.1 arcsec).

Our definition of composite source size differs from that of
\cite{Hardcastle+16} (hereafter H16), which was the previous
largest area AGN survey with LOFAR. In their work, H16 used the maximum pairwise
distance between the centres of all components of a composite source.
However, visual inspection of sources from H16 established that, while
summing the flux densities of composite components gives results that
are consistent with flux-density measurements from hand-drawn regions,
the H16 size definition tends to systematically underestimate true
source sizes. Our present definition is likely to be closer to
the truth than that of H16 in many cases and is good enough for the
  purposes of the present paper. More computationally complex size
  definitions will be discussed in other papers.

\cite{Shimwell+18} give as the criterion for deciding whether a
simple source is genuinely resolved a relationship between peak and
integrated flux density: a source is unresolved if
\begin{equation}
  \frac{S_{\rm int}}{S_{\rm peak}} > 1.25 + 3.1 \left(\frac{S_{\rm
        peak}}{RMS}\right)^{-0.53}
,\end{equation}
where RMS is the local RMS noise level and the coefficients of
  the relationship are best-fitting parameters of an envelope that
  encompasses 95\% of the apparently compact LoTSS-DR1 sources,
  checked by comparison with the properties of bright FIRST sources.
We adopt this definition and apply it to both simple and composite
sources, with two additional criteria: we say that sources are always
resolved if they are composite sources with two or more components,
and that they are never resolved if they are less than 1 arcsec in
size (this catches composite sources with one bright unresolved
component). By these criteria, there are 38,230 resolved sources and
280,290 unresolved sources in the catalogue.
\label{sec:resolved}

Throughout the rest of the paper we refer to the LOFAR observing
  frequency as 150 MHz, for ease of comparison with the many other
  surveys that have used this observing frequency. The small
  difference between 150 MHz and the true central frequency of around
  144 MHz at the pointing centre has no effect on the scientific
  interpretation of the data. The effective frequency varies
    slightly across the field because in a given LOFAR observation
    the lower frequencies, corresponding to a larger station primary
    beam, contribute more to the image at large off-axis distances.
    Given the mosaicing strategy described by \cite{Shimwell+18}, this
    gives rise to only a small effect on the data.

\subsection{Catalogues}

\begin{table*}
  \caption{Samples considered in this paper}
  \label{tab:samples}
  \begin{tabular}{llr}
    \hline
    Name&Description&Number of objects\\
    \hline
    Full&Complete sample of \cite{Williams+18b}&318,520\\
    FC&Flux-complete, flux cut at 0.5 mJy&239,845\\
    O&Optical ID exists&231,716\\
    FCO&Intersection of FC and O&172,898\\
    Z&Some redshift estimate exists&162,249\\
    ZG&``Good'' photometric redshift exists&89,671\\
    FCOZG&Intersection of FCO and ZG&71,955\\
    FCOZGM&Cross-match of FCOZG with the MPA-JHU sample&12,803\\
    FCOZGM RLAGN&RLAGN selected from FCOZGM&3,706\\
    FCOZGM SFG&SFG selected from FCOZGM&9,097\\
    RLAGN&RLAGN selected from FCOZG&23,344\\
    SFG&SFG selected from FCOZG&41,998\\
    \hline
  \end{tabular}
\end{table*}

\begin{figure}
  \includegraphics[width=1.0\linewidth]{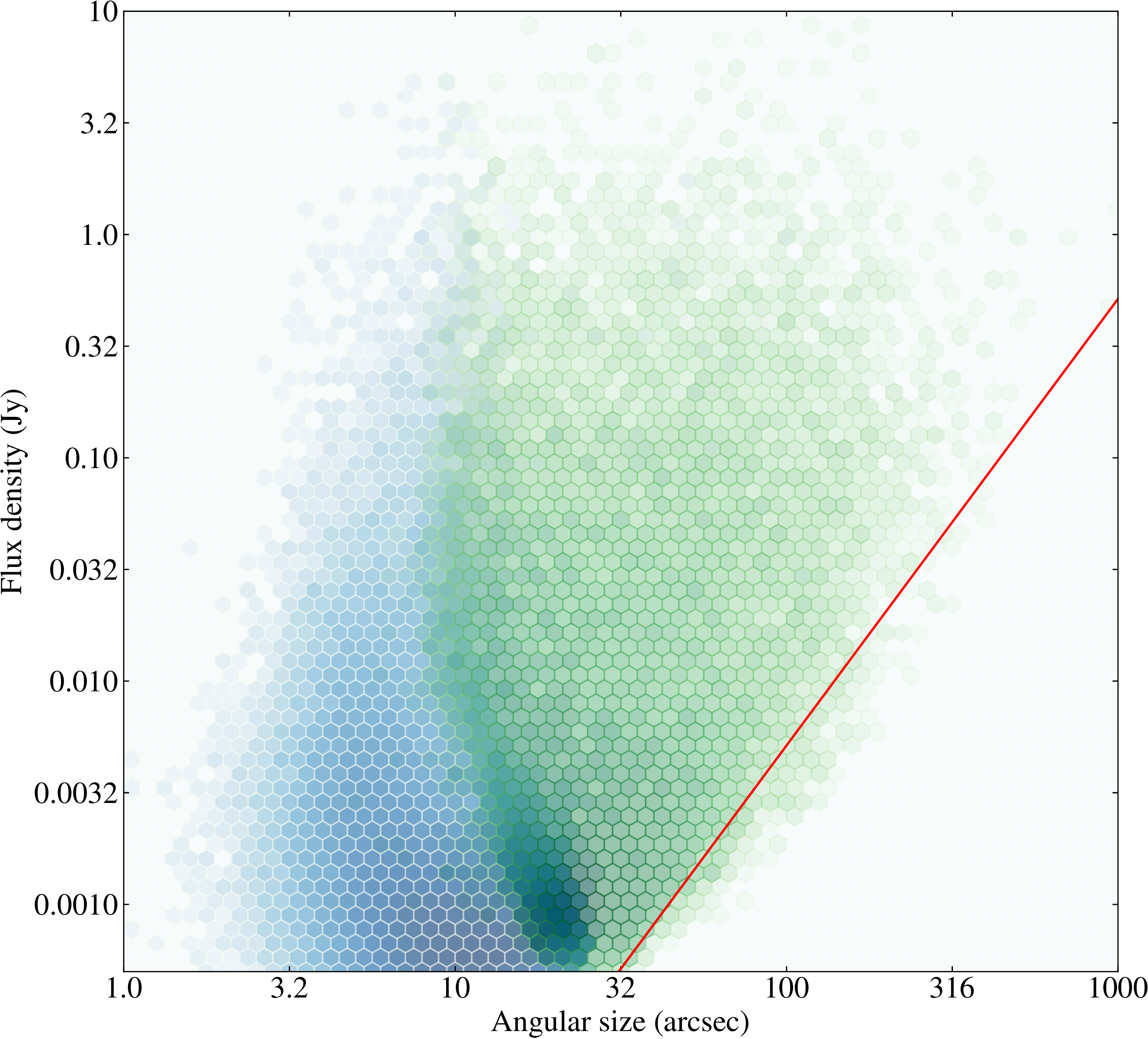}
\caption{Total flux density of
sources above the point-source completeness cut as a function of their
total angular size. The density plot shows the distribution of the FCO sample
sources with measured angular sizes; blue sources are
unresolved and green sources resolved. The red line shows an empirically
normalised line of $S \propto \theta^2$ as expected for a
surface-brightness limited sample. There are 280,290 unresolved and
38,230 resolved sources; the two colour scales are adjusted to make
both populations visible.}
\label{fig:sblimit}
\end{figure}

We generated catalogues for further study by imposing cuts on the
complete catalogue of 318,520 sources, which generates new samples.
For reference, a list of all the samples considered in this work is tabulated
in Table \ref{tab:samples}.

The first point to consider is the flux completeness of the survey.
\cite{Shimwell+18} show that the survey is better than 99\% complete for
point sources having flux densitities greater than 0.5 mJy at 150 MHz.  As PyBDSF selects sources above a $5\sigma$ detection
  threshold and the worst rms noise levels in the mosaiced images are around 100 $\mu$Jy
  beam$^{-1}$, this number seems reasonable and adopting it is
  equivalent to adopting a uniform noise floor across the survey. If
we apply a flux density cut at 0.5 mJy the total number of sources
is reduced to 239,845 (hereafter the FC sample), with a very similar
optical identification and redshift fraction; basing our analysis on
this sample ensures that we can make unbiased statements about the
fractions of sources as a function of measured flux density or some
associated quantity. It is important to remember that the survey is
not complete for resolved sources at this level, however, as it
is surface-brightness limited rather than flux limited. The fact that
surface-brightness limitations affect us is clearly seen in
Fig.\ \ref{fig:sblimit}, where we plot the total flux density of
sources above the point-source completeness cut as a function of their
total angular size. A boundary to the right of this plot imposed by
surface-brightness limitations is visible and shows approximately the
expected slope. However, we can also see that we are sensitive to at
least some comparatively large sources even at the lowest flux
densities. This is a consequence of the fact that many sources are
not uniform in surface brightness. The dependence of this
observational limit not just on the average surface brightness but on
its distribution is an insuperable problem for this sort of
survey (in absence of a much more sensitive survey from which we
can estimate the incompleteness) and its effects must be borne in mind
in what follows.

The next set of criteria to be applied is on optical identifications
and redshifts. For any study of the physical nature of these sources
we need an optical identification, so at this point we have to
restrict ourselves to the 231,716 sources with an optical counterpart
in the {\it WISE} or PanSTARRS data (sample O). The nature of the
remaining objects cannot be determined at this point; a large fraction
of these sources are expected to be high-redshift galaxies, but they will also
include low-redshift objects where the optical identification is
ambiguous or the radio structure is not clear enough to permit an ID.
Further investigation of this population is important but is beyond
the scope of this paper. If we restrict ourselves to sources that are
both in FC and in O, we obtain 172,898 sources (sample FCO).

\begin{figure*}
  \includegraphics[width=0.46\linewidth]{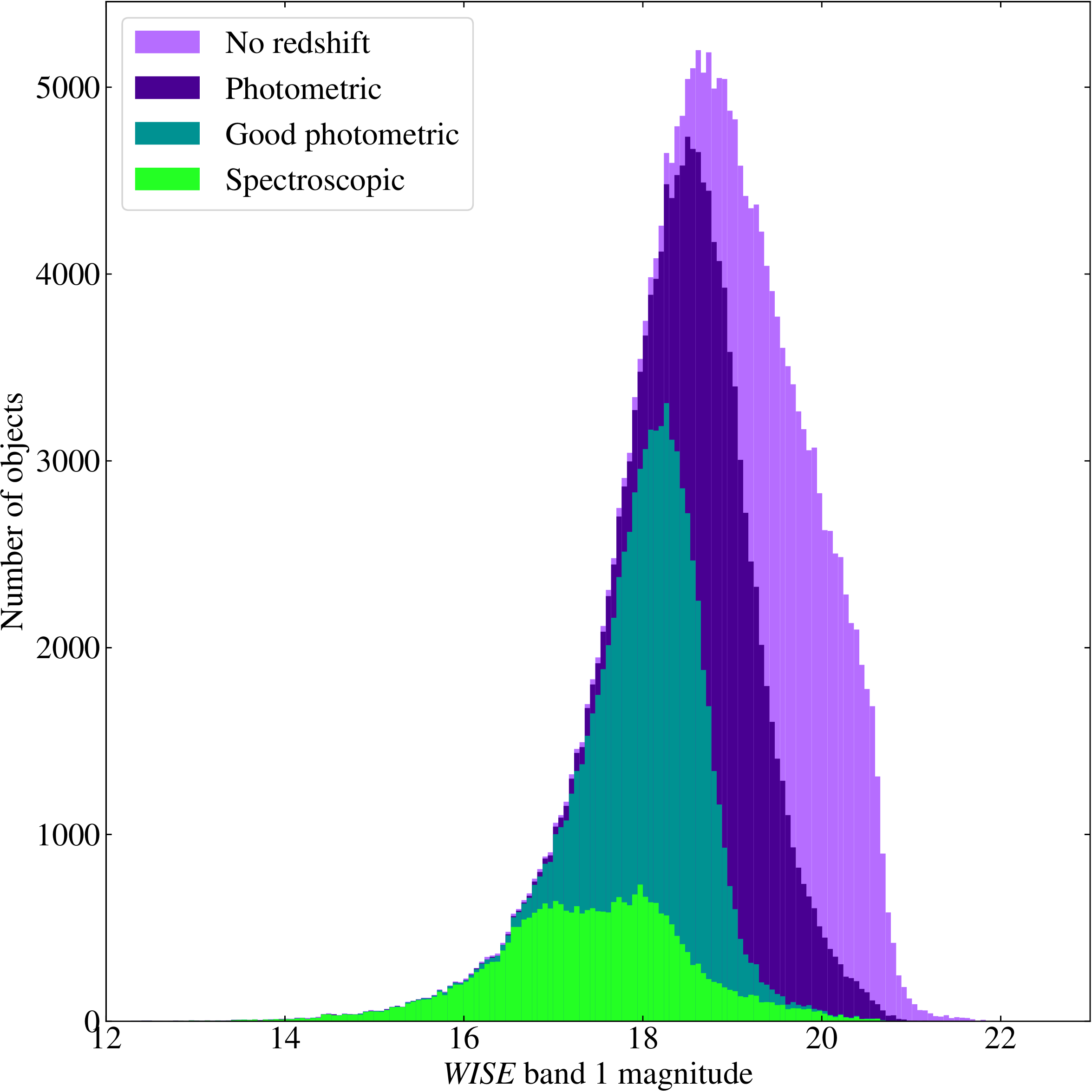}
  \hskip 25pt
  \includegraphics[width=0.46\linewidth]{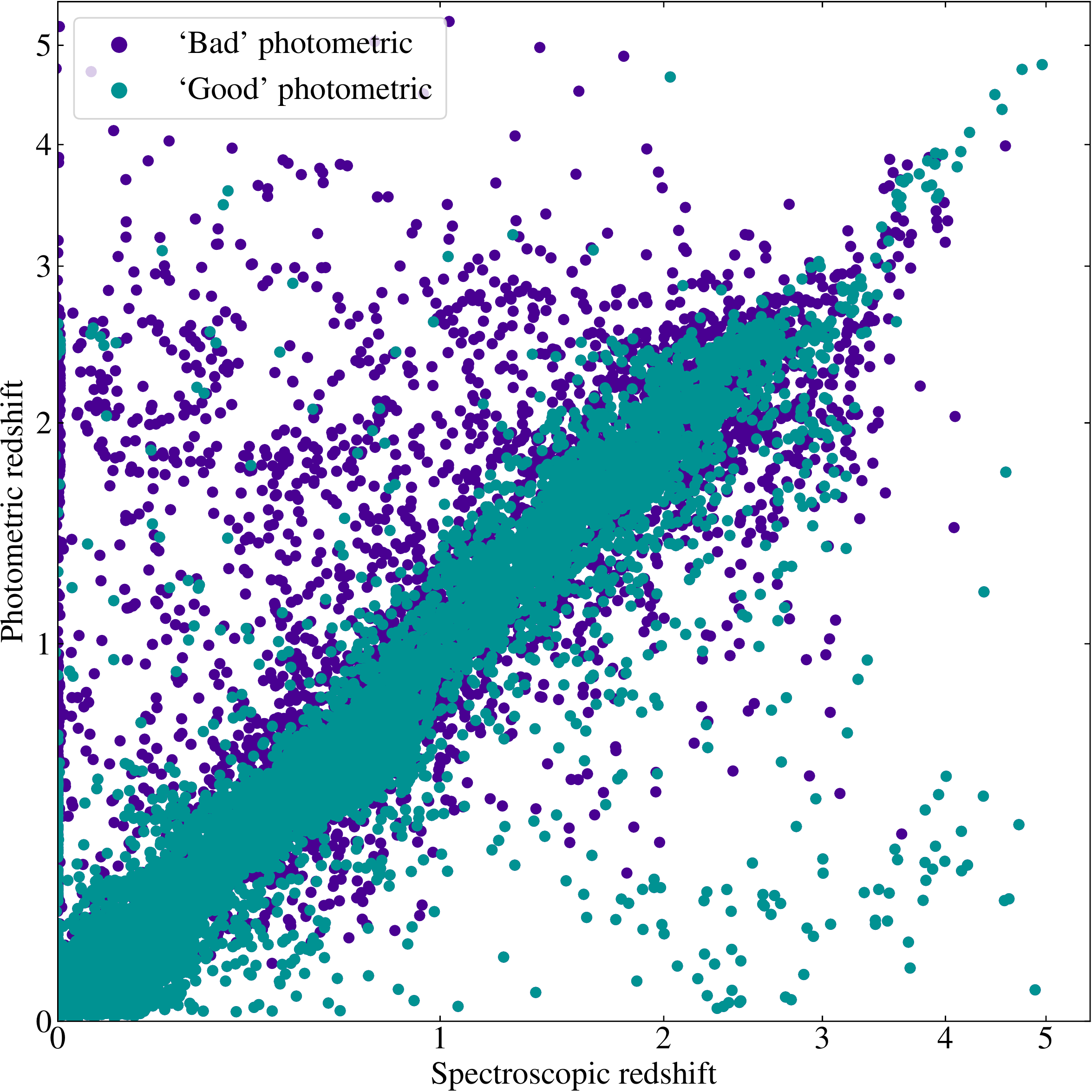}
\caption{Left: Histogram of the distribution of {\it WISE} band-1 AB
  magnitudes for optically identified objects in the sample (sample
  O), colour-coded by the quality of available redshifts
  (spectroscopic, good photometric with $\Delta z/(1+z) < 0.1$, any
  photometric, or none). The plot shows 218,600 sources with WISE
  detections. Right: Photometric vs. spectroscopic redshift for
  sources where both are available, showing the distribution of all
  photometric redshifts and of the good sample.}
\label{fig:zqualhist}
\end{figure*}

We also require a redshift, and so we needed to make a decision on the
quality of photometric redshifts that we were prepared to accept. In
total 162,249 sources have either a spectroscopic redshift or some
photometric redshift estimate (sample Z). The smaller size of Z
compared to O is essentially because an optical detection, which gives
us matched photometry across all of the optical bands, is required to derive a photometric redshift and a large number of the
detections in O are in {\it WISE} only. The errors on some of the
redshift estimates are large. We chose to use $\Delta z/(1+z)$ as our
figure of merit for photometric redshifts, where $\Delta z =
(z_{1,{\rm max}} - z_{1,{\rm min}})/2$ is the half-width of the 80 per
cent credible interval defined by \cite{Duncan+18}, and is therefore
slightly larger than the $1\sigma$ error of standard error analysis.
``Good'' photometric redshifts then have $\Delta z/(1+z)$ less than some
threshold value\footnote{We note that the error estimates do not take
  into account some systematic effects. For example, contamination of
  the photometry by emission lines has a complex,
  redshift-dependent effect that it is difficult to model and
  remove.}. For example, 89,671 sources (sample ZG) have either a
spectroscopic redshift or a photometric redshift with $\Delta z/(1+z)
< 0.1$. The relative numbers of sources with different redshift
quality as a function of optical brightness are shown in
Fig.\ \ref{fig:zqualhist}, which also shows the effect on the outliers
of applying this cut on $\Delta z$. Generally the effect is to reduce
the number of sources with grossly discrepant redshifts, although a
small number of sources remain (in the bottom right of
  Fig.\ \ref{fig:zqualhist}) with photometric redshifts much less
than their spectroscopic redshifts. These objects are all
  high-redshift quasars and are discussed by \cite{Duncan+18}; other
  quasars are well fitted by the photometric redshift code and the
  issues that affect these particular objects include very bright
  broad emission lines that affect the optical spectral energy distribution (SED), or lines of sight
  with particularly low absorption due to intervening inter-galactic
  medium (and hence weak Lyman break features). Because bright quasars
  are very likely to be selected as such by SDSS spectroscopy, it
  seems unlikely that they represent a significant contaminating
  population at low redshift.
There are a total of 71,955 sources in the flux-complete catalogue
that also have an optical ID and a good redshift. From this
population (FCOZG) we can start to select samples of RLAGN.

\section{AGN selection}
\label{sec:agnsel}

The separation of RLAGN from SFG is one of the
biggest problems faced by this and all other current-generation
extragalactic radio surveys in which SFG are present in
significant numbers (i.e. any survey, like LoTSS, with the equivalent
of a sub-mJy flux limit at 150 MHz). Emission due to the stellar
population is always going to be present at least in cases in which we do not
have the ability to separate this emission spatially from RLAGN activity,
which requires resolution substantially better than the spatial
  scales of the galactic disc. Therefore a perfect RLAGN selection would
involve selecting as AGN all those, and only those, galaxies whose
radio emission significantly exceeds the level expected from star
formation or other stellar processes \citep{Hardcastle+16,Calistro-Rivera+17,Smolcic+17}.
It should be noted that this is significantly different from other AGN
selection methods and produces a different population. Many
radiatively efficient AGN, selected as such using X-ray emission, SED
fitting \citep[e.g.][]{Calistro-Rivera+16} or
traditional emission-line classifications, appear to lie on the
star-forming main sequence, perhaps with no significant radio emission
that is not due to star formation \citep{Mingo+16,Gurkan+18,Gurkan+18b}, while many RLAGN have
little radiative nuclear output and would not easily be selected as
AGN in any band other than the radio. Similarly, AGN selections
  using mid-infrared colour/colour criteria
  \citep[e.g.][]{Assef+10,Jarrett+11,Stern+12,Mateos+12,Secrest+15} cleanly
  select sources dominated by very luminous (quasar-like) AGN; these
  selections, however,
  have been shown to under-represent the radio-loud population, as
  they are biased against lower luminosity and
  higher redshift AGN, both of which are preferred hosts for radio
  sources \citep[e.g.][]{Gurkan+14,Rovilos+14,Mingo+16}.

There are two problems in practice with selection based on the
expected level of emission from stellar processes. Firstly, the
relationship between radio emission and star formation is still poorly
understood. It may depend not just on star formation but on a number
of galaxy parameters \citep{Gurkan+18} and, because of the complex
chain of physical processes and timescales connecting low-frequency
radio emission to star formation, it certainly has a
good deal of irreducible, intrinsic scatter that will always act to
blur the distinction between strong star formation and weak AGN
activity. Thus there are physical reasons why there will never
be a unique right answer for objects on the SFG/AGN boundary,
irrespective of the accuracy of the available star formation rate estimates.
Secondly, in our particular case, we do not have good information
about the star formation rates of most of the HETDEX host galaxies.
Working in the H-ATLAS NGP field, H16 were able to make use of the
{\it Herschel} data to select RLAGN using the radio/far-infrared relation;
\cite{Gurkan+18} in the same field expanded this to select
radio-excess AGN candidates based on star formation rates inferred
from spectral fitting to the broad-band far-infrared through to optical
photometry for the (low-$z$) galaxies in their parent sample, using
the {\sc magphys} code \citep{daCunha+08} in a manner similar to that
described by \cite{Smith+12}. However, we do not have {\it Herschel}
data for HETDEX and inference of star formation rates from SED fitting
is much less robust without it.

One approach is simply to apply a luminosity cut. However, starburst
galaxies with star formation rates of $\sim 10^3\,M_\odot$
yr$^{-1}$ would have LOFAR luminosities of $\sim 10^{25}$ W
Hz$^{-1}$ at 150 MHz according to the radio to star formation rate
relation of \cite{Gurkan+18}, although it should be noted that this is
an extrapolation as such extreme objects do not exist in their sample.
A simple cut in luminosity thus needs to be placed at relatively high
luminosities to avoid contamination. For example, cutting FCOZG at $10^{25}$
W Hz$^{-1}$, which should remove most star-forming objects, leaves
6,660 sources -- still a large sample but less than a tenth of the
parent population. Many low-luminosity RLAGN would be excluded by
  such a cut.

\begin{figure}
\includegraphics[width=\linewidth]{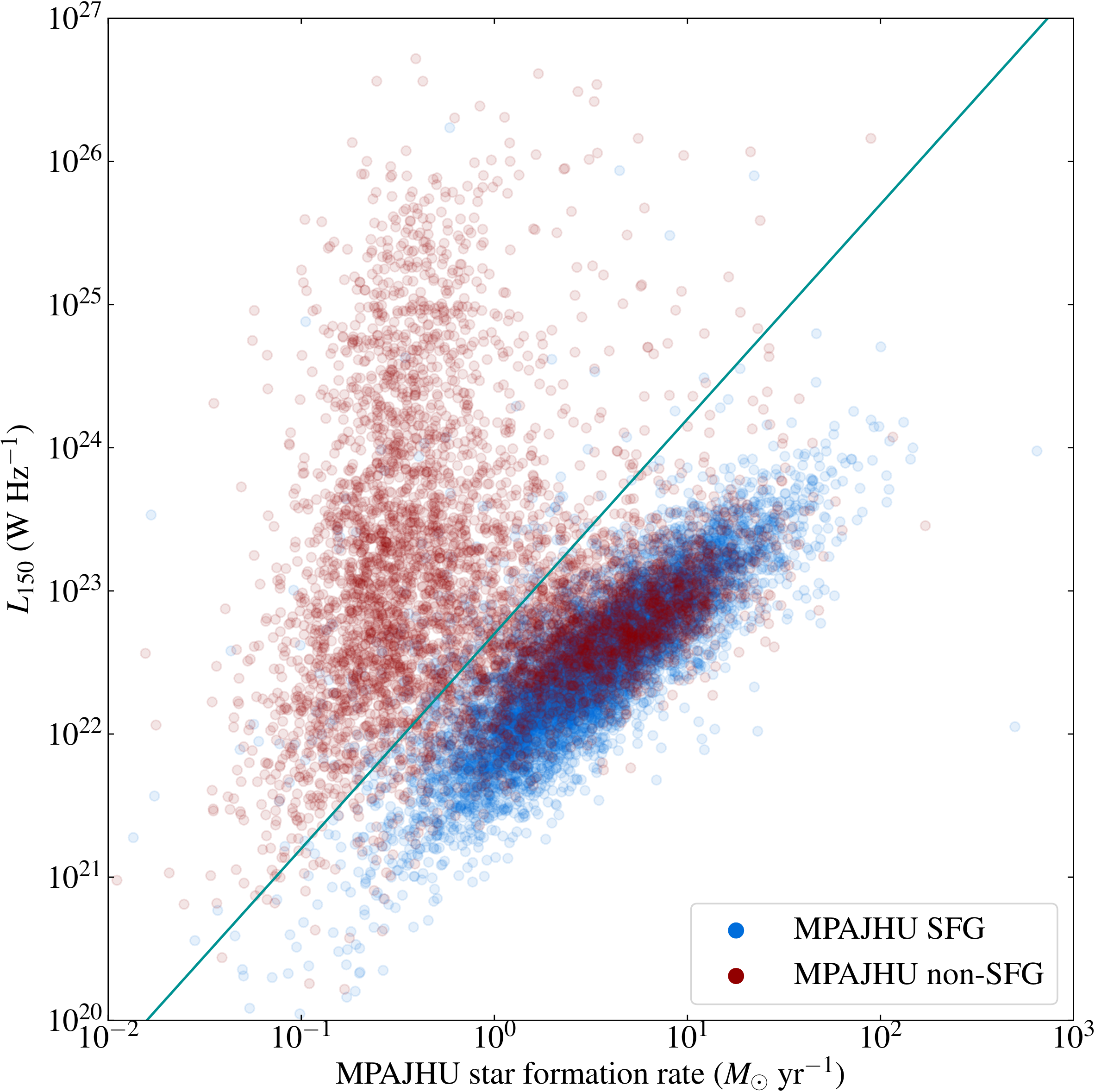}  
\caption{Radio luminosity as a function of star formation rate for the
  sources in the FCOZGM sample. Objects are colour-coded according to
  whether they are classed as star forming in the MPA-JHU catalogue
  (classifications ``STARBURST'' or ``STARFORMING''); some objects not
  classified as SFG in this way are clearly nevertheless on the
  radio to star formation rate relation for star-forming objects. All objects not so
    classified (including unclassified objects) are placed in the
    ``non-SFG'' sample. The line shows a
  plausible by-eye selection of a division between the two classes.}
\label{fig:butterfly}
\end{figure}

A subset of the objects in the FCOZG sample (12,803 objects: sample
FCOZGM) have emission-line measurements and estimates of host galaxy
properties from SDSS, provided by the MPA-JHU
catalogue\footnote{\url{https://wwwmpa.mpa-garching.mpg.de/SDSS/DR7/}}.
Data available for the FCOZGM objects include spectroscopic source
classifications and estimates of star formation rates using the
methods of \cite{Brinchmann+04}, which combine emission-line and
continuum (4000-\AA\ break) information. For these objects, which are
typically at low redshift given the requirement for SDSS spectroscopy,
it would in principle be possible to follow \cite{Gurkan+18} and
select as RLAGN sources that lie significantly above the locus for SFG
in a plot of star formation rate versus radio luminosity. Such a plot
(Fig.\ \ref{fig:butterfly}) indeed appears to show a good division
between two distinct populations. However, a problem with this is that
star formation rates for quiescent galaxies may be underestimated
relative to, for example, the SED-fitting results of \cite{Gurkan+18},
as we have verified by considering the same plot using the H-ATLAS NGP
data. Use of the MPA-JHU star formation rates could artificially
accentuate the differences between sources at low star formation
rates. We therefore do not use this method directly. Instead, we use
the classification scheme developed by \citet{Sabater+18}, which
builds upon the work of \citet{Best+05} and \citet{Best+Heckman12}. In
brief, \citeauthor{Sabater+18} consider four different diagnostic
diagrams to separate radio AGN from galaxies whose radio emission is
primarily powered by star formation. These are (1) the comparison
between the 4000\AA\ break strength and the ratio of radio power per
unit stellar mass, developed by \citet{Best+05}; (2) the widely used
BPT emission line ratio diagnostic diagram
\citep{Baldwin+81,Kauffmann+03,Kewley+06}; (3) the radio luminosity
versus H$\alpha$ line luminosity; and (4) the W2-W3 {\it WISE} colour
\citep[as used by e.g.][]{Wright+10,Mateos+12,Gurkan+14,Herpich+16}.
The first and third of these diagnostics are based on the same
principle as the use of the radio/far-IR relation: the two parameters
are expected to be related for SFGs as they both broadly trace
specific star formation rate (diagnostic 1) or star formation rate
(diagnostic 3); the RLAGN are identified as those sources offset
from this relation due to an additional (jet-related) contribution to
the radio luminosity. Diagnostic 2 is well established to separate AGN
from SFGs in galaxies with measured emission lines, but fails to
distinguish radio-quiet from RLAGN. Diagnostic 4 is less precise, but
provides a valuable discriminant where the other diagnostics give
contradictory results. \citet{Sabater+18} then combine the results
from these four diagnostics to produce an overall AGN/SFG
classification, using a comparison with the classifications determined
by \citet{Gurkan+18} for the H-ATLAS NGP sample to optimise this
combination. Using this classification scheme, 3706 of the FCOZGM
sources are classified as being radio-loud AGN, and 9097 are
classified as SFGs (where the latter category may include radio-quiet
AGN).

\begin{figure}
\includegraphics[width=\linewidth]{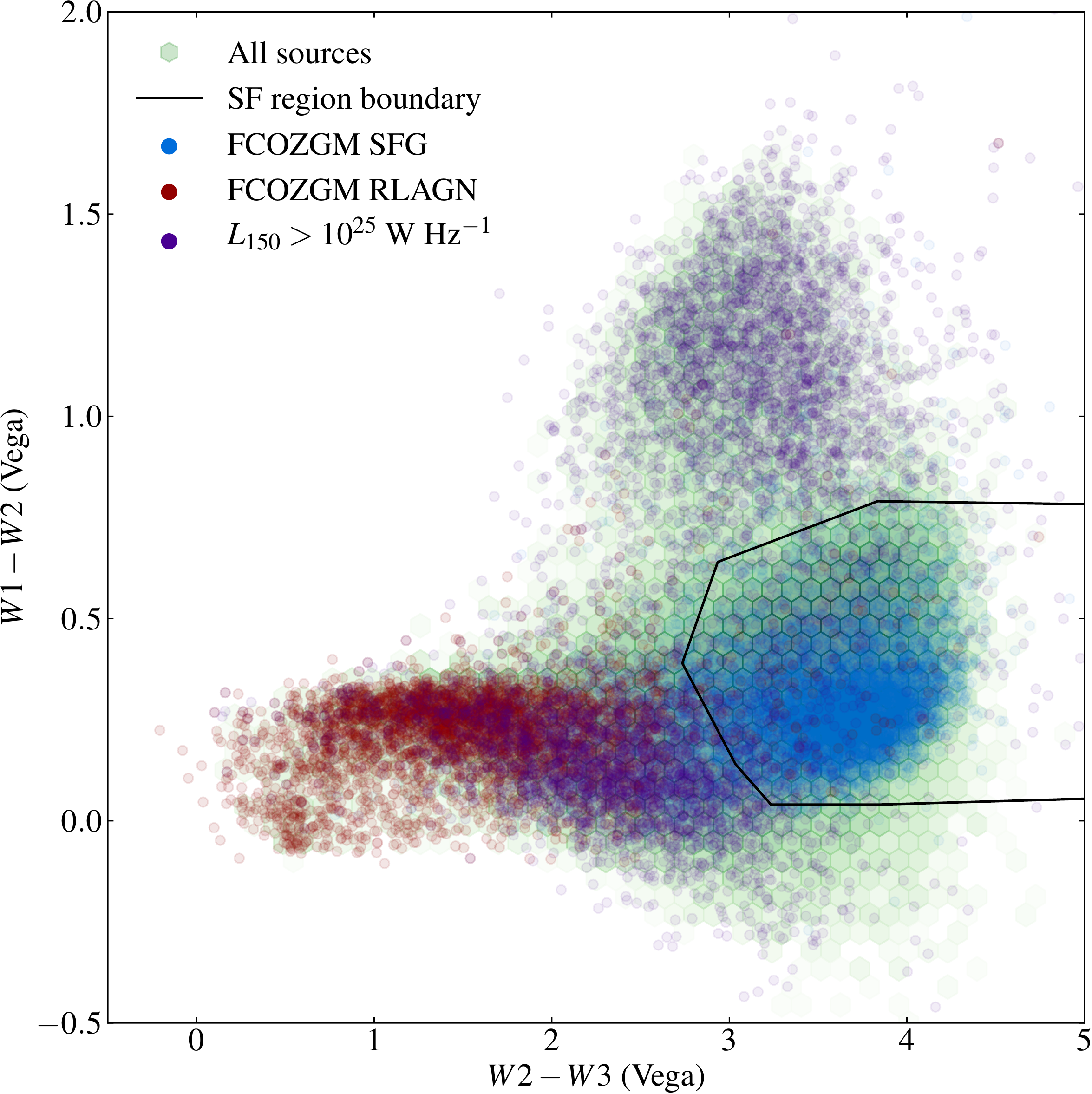}  
\caption{Observational {\it WISE} colour-colour diagram for the
  FCOZG sample. Overlaid on the green density plot showing the full sample
  are the locations of FCOZGM RLAGN, FCOZGM SFG, and luminous radio
  sources. Lines indicate the locus populated by SFG and avoided by
  RLAGN discussed in the text.}
\label{fig:wisecc}
\end{figure}

This leaves us with the problem of classifying the remaining sources
that do not have this spectroscopic information. For this purpose we
considered only the {\it WISE} data, as {\it
  WISE} data are available for almost all the FCOZG sample (only 2,600
sources do not have {\it WISE} photometry); the three bands we used,
$W1$, $W2,$ and $W3$, correspond to 
3.4, 4.6, and 12 $\mu$m and so sample the rest-frame near- and mid-IR
wavelengths for the redshifts of our sample. To plot this diagram in
the traditional way we converted the catalogued AB {\it WISE} magnitudes
for our sources into Vega
magnitudes. 
Fig.\ \ref{fig:wisecc} shows a density plot for the whole FCOZG
sample with the classified FCOZGM sources overlaid. As expected, the
hosts of FCOZGM objects lie in very different locations depending on
their classification as RLAGN or SFG. Moreover, when we add in luminous
($L_{150} > 10^{25}$ W Hz$^{-1}$) sources, we see that these also tend
to avoid a well-defined location in the colour-colour diagram around
the location of the FCOZGM SFG. We therefore exclude objects that
  have {\it WISE} colours consistent with the SFG locus, defined as
  lying in a polygonal region in colour-colour space chosen to give
  the best separation between SFG and other objects, as shown in
Fig.\ \ref{fig:wisecc}.

At high redshifts quasars present a particular problem. Although these
are AGN by construction, they need not show any excess radio emission
over the expectation from star formation \citep{Mingo+16}. Indeed, \cite{Gurkan+18b} argue that the majority of
LOFAR-selected quasars have radio emission consistent with star
formation if we assume that star formation scales with AGN power as
observed at low redshift. If this is the case we should exclude these
objects from the RLAGN sample, which we do by making an empirical cut in
radio luminosity/absolute magnitude space. We can very easily select
quasars by their bright rest-frame magnitudes; anything with
$Ks$-band absolute magnitude $<-25$ is likely to be a quasar.

The full selection method is as follows. Starting from the FCOZG sample,

\begin{enumerate}
  \item Sources with $Ks$-band rest-frame magnitudes outside the range
    $-33 < Ks < -17$ are removed. This disposes of sources with
    outlier absolute magnitudes that presumably indicate aberrant
    redshifts.
  \item Sources classed as SFG from FCOZGM are removed.
  \item Sources with WISE colours in the SFG locus of
    Fig.\ \ref{fig:wisecc}, or with no available {\it WISE} data, are
    removed unless {\it either}:
    \begin{itemize}
    \item They are classified in FCOZGM as RLAGN
    \item Their luminosity is $ > 10^{25}$ W Hz$^{-1}$ and their
      $Ks$-band magnitude is $>-25$ (non-quasars), or
      \item Their $Ks$-band rest-frame magnitude is $<-25$ (quasars), and their
        radio luminosity is such that $\log_{10}(L_{150}) > 25.3 - 0.06 (25+Ks).$
    \end{itemize}
\end{enumerate}

\begin{figure}
\includegraphics[width=\linewidth]{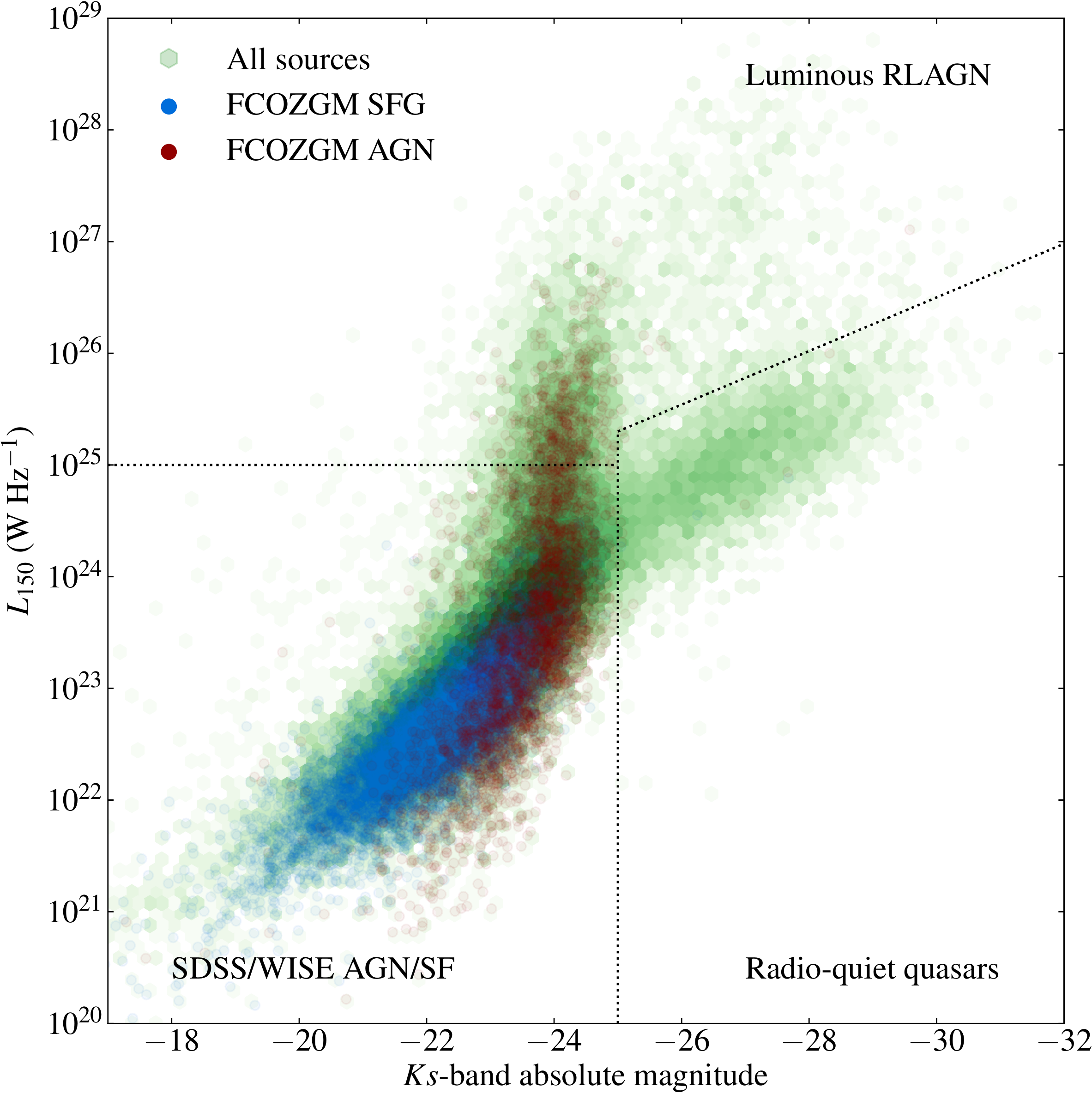}  
\caption{Radio luminosity vs. $Ks$-band absolute magnitude for the
  FCOZG sample. Overlaid on the density plot showing the full sample
  are the locations of FCOZGM RLAGN and FCOZGM SFG. Lines show the
  divisions on the plot used to select optically classified or radio-luminous AGN.}
\label{fig:ksl150}
\end{figure}

The motivation for the cuts used in radio and optical luminosity is
illustrated in Fig.\ \ref{fig:ksl150}, which shows a plot of the
sample in radio/optical luminosity space indicating the positions of
the FCOZGM-classified objects. After these cuts are applied we are
left with 23,344 sources, which form our RLAGN sample.

Clearly there are a number of ways in which this selection is not
ideal. The {\it WISE} colours have errors and so classifications cannot be
exact at the boundary. The FCOZGM RLAGN and SF overlap to some extent in
the {\it WISE} colour space and so we know that it does not provide an
exact separation between the populations. We are using apparent
colours and therefore the precise boundary between populations should in
principle be redshift dependent, but we have not attempted to take
this into account in any way. Some high-excitation RLAGN
\footnote{These are objects with radiatively efficient nuclei and
 thus strong optical emission lines, including
quasars and broad- and narrow-line radio galaxies (see \citealt{Hardcastle+09}
and references therein) and are contrasted with low-excitation radio
galaxies (LERG), which have colours and emission-line properties
more typical of ordinary ellipticals.} with
intermediate nuclear absorption are expected to lie in the SFG colour
location, and these are excluded from the RLAGN sample. And, most
obviously, we are essentially selecting based on the colour of the host
galaxy and not on the radio properties of the source, such that, for example,
we cannot select as AGN strongly SFG that also host
RLAGN unless their radio luminosity is very
high. For all these reasons our RLAGN sample is likely to be neither
clean nor complete, but it represents the best sample we are able to
construct with the available data given that we lack the data to
  select radio-excess sources directly. It should be noted that the RLAGN
  luminosity functions of \cite{Sabater+18} and Williams
  \etal\ (in prep), which use respectively the FCOZGM and the full
  RLAGN sample, agree well with those of H16, which  used a
  radio-excess selection method. Thus we can be confident that the
  necessarily more complex selection used in this work is not significantly
  biasing the RLAGN selection.

\begin{figure*}
\includegraphics[width=\linewidth]{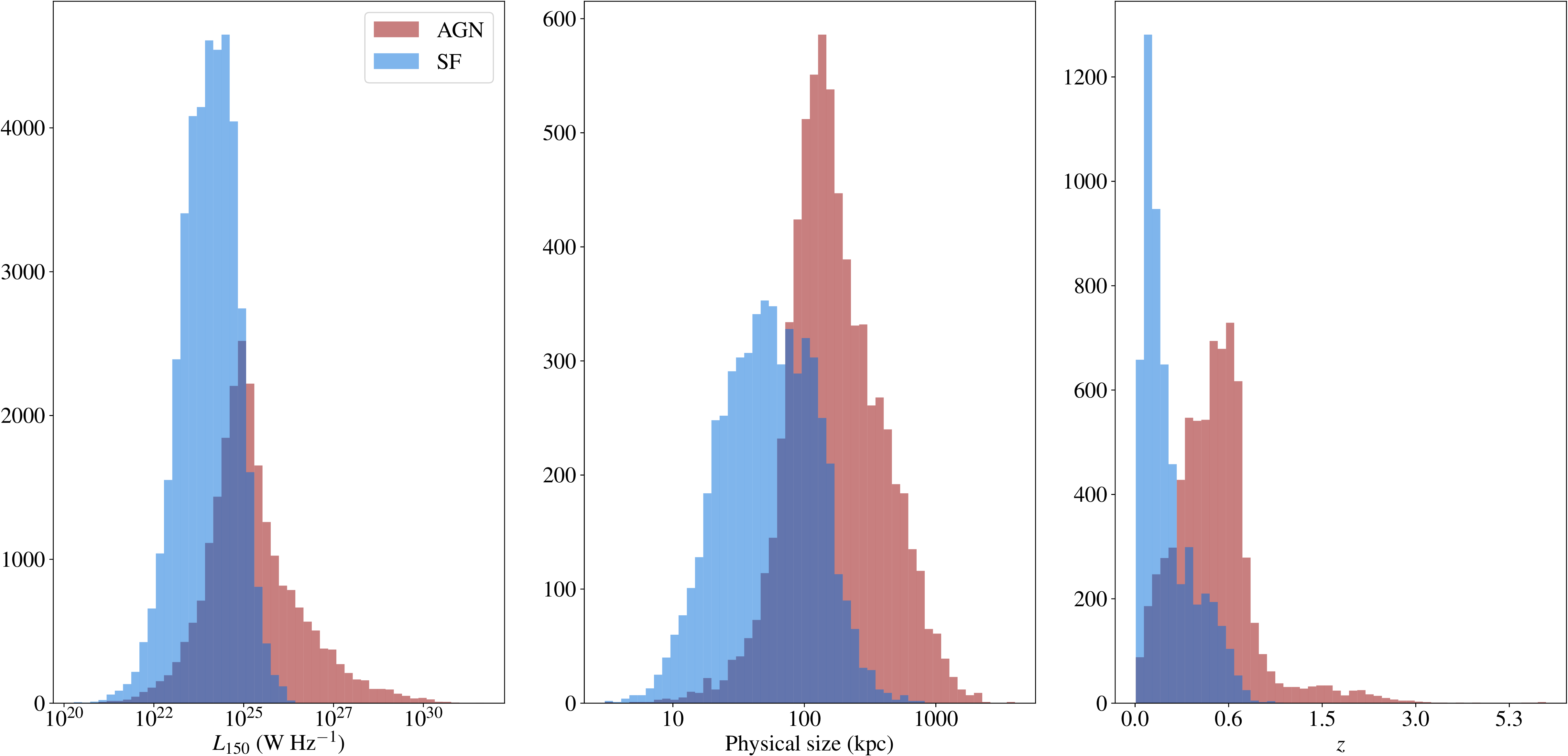}  
\caption{Distributions of (left) radio luminosity, (middle) physical
size, and (right) redshift for objects selected and not selected as RLAGN from the FCOZG
sample using the criteria described in the text. All sources are shown
in the left- and right-hand histograms, whereas in the middle only resolved sources
(as defined in Section \protect\ref{sec:resolved}) are plotted.}
\label{fig:agn_sf_hists}
\end{figure*}

Fig.\ \ref{fig:agn_sf_hists} illustrates the differences between the
23,344 objects selected as RLAGN from FCOZG (hereafter the ``RLAGN sample'')
and the 41,998 objects selected as SFG on the basis of FCOZGM
classifications or {\it WISE} colours; the 3,460 candidate
SF-dominated quasars are excluded from both plots as they are
  neither RLAGN nor typical SFG. We see that, as
expected, RLAGN are generally more luminous and at higher redshift and
that resolved SFG have a characteristic size of tens of kpc. A small
tail of very large ($>100$ kpc) SFG must either indicate
misclassification, misidentification, incorrect size measurements, or
incorrect redshifts and visual inspection of some of these sources
shows that all of these factors are involved, and at least some of
the SFG show RLAGN-like structures on scales larger than
  those of the host galaxy. On the whole, however, these
plots show that the separation gives the expected behaviour in terms
of physical properties of the radio sources.

\section{Results and modelling}

\subsection{Powers and linear sizes of RLAGN}
\label{sec:pdd}

\begin{figure}
\includegraphics[width=\linewidth]{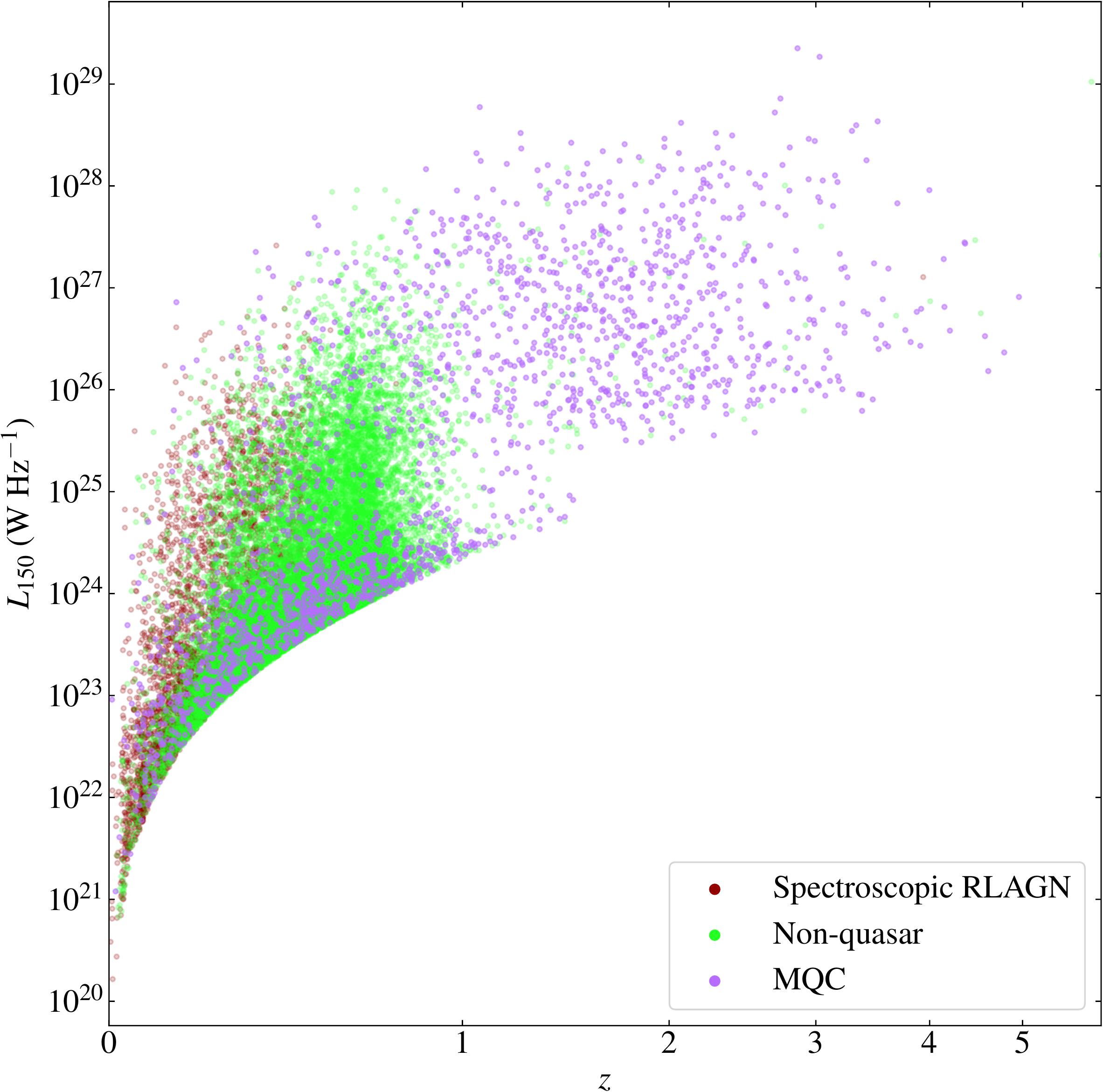}  
\caption{Sampling of the redshift/luminosity plot by the RLAGN
  sample. The figure distinguishes between FCOZGM RLAGN
  (``spectroscopic AGN''), objects classified as quasars in the Million
  Quasar Catalogue (MQC; \protect\url{http://quasars.org/}), which
  are flagged as such in the value-added catalogue \citep{Duncan+18} and non-quasar AGN selected using the other criteria
  discussed in Section \protect\ref{sec:agnsel}. We note that the
    $x$-axis shows $\log(1+z)$, labelled linearly.}
\label{fig:l_z}
\end{figure}

Fig.\ \ref{fig:l_z} shows the sampling of the luminosity-redshift
plane by objects in the RLAGN sample. The sample luminosity spans over nine
orders of magnitude due to the wide range in redshift present in the
data. However, the high-luminosity objects are dominated by quasars
due to the requirement for an optical or {\it WISE} detection. Only
below a luminosity of $10^{27}$ W Hz$^{-1}$ do we have large numbers
of galaxies, which occupy the space below $z \approx 0.8$. Below a
luminosity of around $10^{24}$ W Hz$^{-1}$ the sample is limited by
the radio flux density limit rather than optical detectability in
  the sense that objects below this radio luminosity cannot be seen
  at all $z<0.8$.

\begin{figure*}
\includegraphics[width=\linewidth]{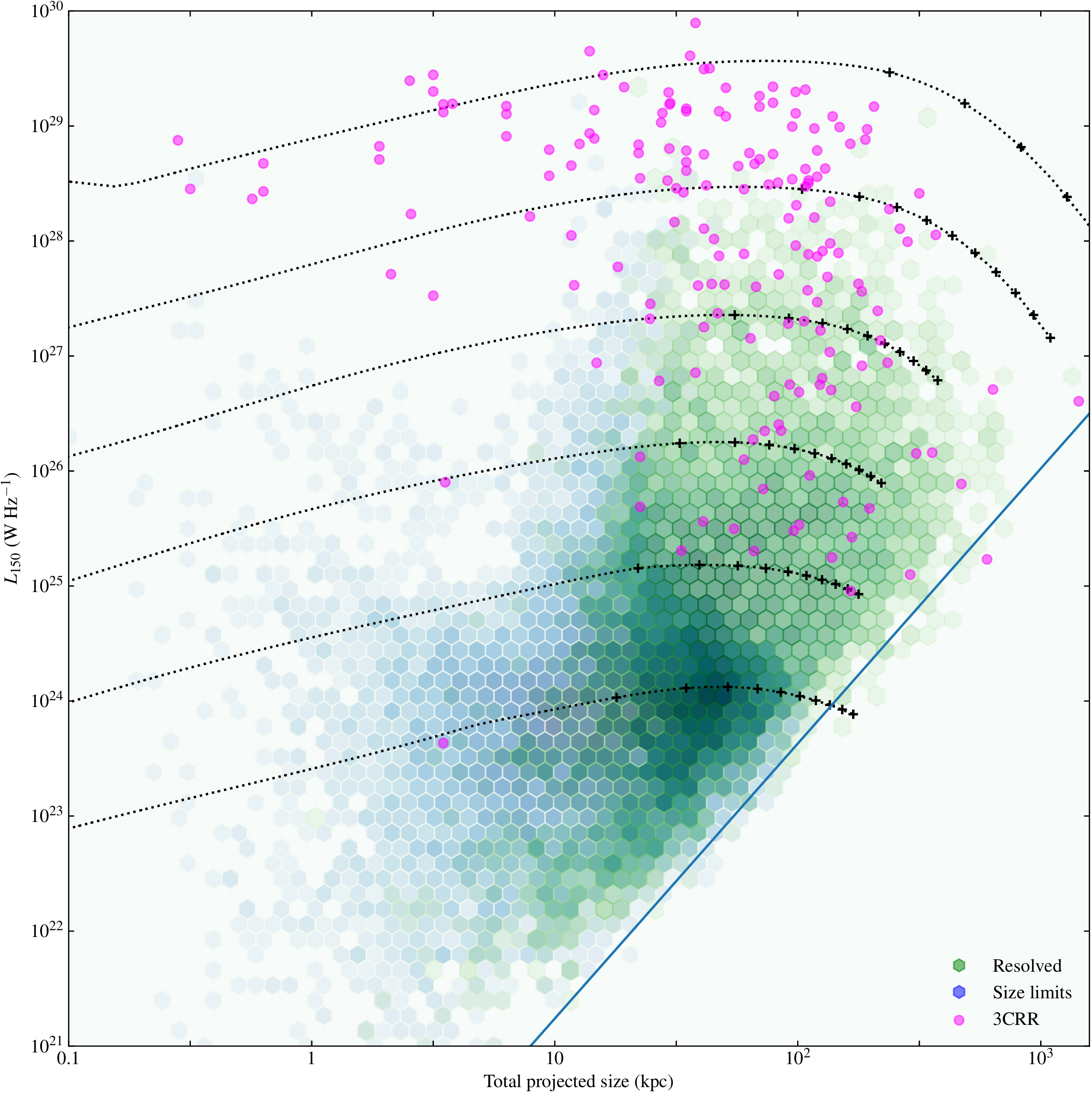}  
\caption{Power/linear size plane ($P$-$D$ diagram) for the RLAGN
  sample. Sources that are resolved as defined in Section
  \ref{sec:resolved} are shown in the green density plot; unresolved
  sources, where the sizes are upper limits, are in blue; and the 3CRR
  sample \citep{Laing+83} is overplotted for comparison. There
    are a total of 6,850 resolved and 16,494 unresolved sources on the
    plot; the colour scales are adjusted so that both groups can be
    seen. The diagonal blue line shows
  (qualitatively) the area of the plot in which surface brightness
  limitations become important, following the analysis of H16.
  Overplotted are theoretical tracks for $z=0$ sources lying in the
  plane of the sky in a group environment ($M_{500} = 2.5 \times
  10^{13} M_\odot$, $kT = 1$ keV) for two-sided jet powers (from
  bottom to top) $Q=10^{35}, 10^{36}, \dots, 10^{40}$ W; see the text
  for details. Crosses on the tracks are plotted at intervals of 50
  Myr, where linear size increases monotonically with time; each track lasts for 500 Myr in total.}
\label{fig:pdd}
\end{figure*}

The power-linear size plane or $P$-$D$ diagram for RLAGN
\citep{Baldwin82,Kaiser+97,Blundell+99,Turner+18} is 
analogous to the Hertzsprung-Russell diagram for stars or the stellar
mass/star formation rate plot for galaxies, in the sense that the
location of a source is indicative of both its initial conditions and its evolutionary state and tracks
in the diagram can be associated with particular phases of evolution.
In the $P$-$D$ diagram, objects with
particular properties describe tracks on
the plane that are defined purely by source physics, while remnant
sources in which the jets have switched off follow a distinct set of
evolutionary tracks
\citep{Godfrey+17,Hardcastle18}. However, the interpretation of the position of a 
particular source on the $P$-$D$ diagram is complicated for
several reasons. Firstly, the RLAGN environments have a non-negligible
effect on their tracks in the $P$-$D$ diagram (see
\citealt{Hardcastle+Krause13,Hardcastle+Krause14}). Secondly, radio galaxies
  are not spherical, with the effect that the position of real sources on the
$P$-$D$ diagram is dependent on unknown observing factors such as
Doppler boosting and the source angle to the line of sight. Thirdly, the
theoretical tracks used to interpret the diagram tend to assume that
there is a single phase of constant-jet-power evolution followed by a
phase of zero-jet-power evolution, whereas we know observationally both
that activity of sources can stop abruptly and restart and
  that optical AGN activity can vary on very short timescales, so that
there is no reason to suppose that the jet power $Q$ cannot vary with
time on a wide range of timescales. Nevertheless, the $P$-$D$
diagram remains one of the key tools for interpreting the evolution of
populations of RLAGN.

Fig.\ \ref{fig:pdd} shows this plot for the 23,344 sources of the
RLAGN sample, which represent by far the largest sample to have been
interpreted in this way at the time of writing, along with the 3CRR
sources of \cite{Laing+83} for comparison\footnote{Data from
  \url{https://3crr.extragalactic.info/}.}. For LOFAR sources, resolved and unresolved sources are plotted; for the unresolved
sources we take as an upper limit on size the measured deconvolved
major axis plus three times the formal error on the major axis; this value is plotted on the density plot for these sources rather
than the best estimate of the size (which is zero in many cases).
Another feature of the $P$-$D$ diagram is that it is strongly affected
by surface-brightness limitations, as noted by H16. Physically large,
low-luminosity (and therefore low-redshift) sources cannot be detected
and catalogued even by LOFAR because their surface brightness falls
below the detection threshold for our full-resolution imaging. Only
for luminosities
around $10^{26}$ W Hz$^{-1}$ and above does this 
limitation have a negligible effect on the observed size distribution.
It can be seen, in spite of this bias, that the LOFAR data span a
  far wider range in luminosity than the 3CRR sources, while covering
  much the same range in linear size.

Also overplotted on Fig.\ \ref{fig:pdd} are theoretical evolutionary
tracks from the models of \cite{Hardcastle18} (hereafter H18). These,
in common with a number of other models in the literature
discussed in Section \ref{sec:intro}
are derived from a model that predicts the time evolution of both
luminosity and physical size in a given environment and for a given
jet power $Q$ (defined as the two-sided power, i.e. the total
kinetic power of both jets). To simplify the plot we use a single
environment, a group with $M_{500} = 2.5 \times 10^{13} M_\odot$
(corresponding to an X-ray gas temperature of $\sim 1.0$ keV),
and evolve sources with jet powers $Q=10^{35}, 10^{36}, \dots,
10^{40}$ W for a lifetime of 500 Myr assuming $z=0$; the choice of
redshift affects the radiative losses due to inverse-Compton
emission. Looking just at the normalisation of the tracks, we
  can see that the powerful 3CRR sources in
  these models correspond to jet powers $\ga 10^{39}$ W, while the LOFAR
  survey is dominated by sources with jet powers $\la 10^{38}$ W.
  The positions of the time evolution markers on the tracks show that, if all RLAGN have long lifetimes, we
expect them to spend most of their lifetime with (unprojected) sizes
between a few tens and a few hundreds of kpc and that these
predictions seem to be qualitatively consistent with the size
distribution of LOFAR sources with luminosities $\ga 10^{25}$ W Hz$^{-1}$;
at lower radio luminosities this is much less clearly the case, with
many smaller sources being present. The H18 work showed that the expected size
distribution is sensitive to the lifetime function, i.e. the fraction
of sources in the population that have active jet lifetimes less than some
limiting value. To investigate the lifetime distribution in the
current sample we have to carry out more detailed modelling.

\subsection{Modelling the linear size distribution}
\label{sec:sizedist}

There are two possible approaches to trying to infer population
properties of RLAGN by combining models and data. In the first, we
would try to estimate the physically interesting parameters of each
source (such as jet power $Q$ and source age $t$) from the available
data for that source. Since the easily available observables (radio
luminosity and linear size) depend not just on $Q$ and $t$ but also on
the unknown source environment, the angle of the source to the line
of sight $\theta$, and redshift, the inference of $Q$ is a
poorly constrained inverse problem and necessarily will not produce
particularly accurate answers for any given source. Better results
would be achieved with per-source environmental measures if they were
available. We begin by trying the second approach, which is to
forward-model subsets of the whole population using known constraints
on the distribution of environments, redshifts, and angles to the
line of sight. This approach has the advantage that observational
limitations like the surface brightness limit can easily be taken into
account, but the disadvantage that it is computationally expensive and
cannot provide a full exploration of all the underlying distributions.
However, it is well suited to the current goal of understanding
whether the observed projected linear size distributions are
consistent with models.

In order to investigate the implications of the size distribution we
first restrict ourselves to sources with $z<0.8$, motivated by
Fig.\ \ref{fig:l_z}. Above this redshift the sample becomes
increasingly dominated by quasars, which are biased in their angle to
the line of sight: excluding high redshifts also makes us insensitive
to the treatment of radio-quiet quasars discussed above. We then
consider three slices in the $P$-$D$ diagram in the luminosity ranges
$10^{24}$ -- $10^{25}$ W Hz$^{-1}$, $10^{25}$ -- $10^{26}$ W
Hz$^{-1}$, and $10^{26}$ -- $10^{27}$ W Hz$^{-1}$. As
Fig.\ \ref{fig:pdd} shows, the last of these should be basically
unaffected by surface brightness limitations and thus allows us to
constrain the upper end of the lifetime function. These three
luminosity ranges sample similar redshift ranges, limited by the
optical data (see Fig.\ \ref{fig:l_z}) and therefore results can be compared
without worrying excessively about the cosmological evolution of the
population.

Simulated samples were created as described by H18, but we drew
the distribution of redshifts from the observed redshift distribution
in each luminosity bin, smoothed using a Gaussian kernel density
  estimator (KDE) with bandwidth 0.05. In general cluster masses
  can be described by a mass function, which conventionally gives the
  number of clusters above a given mass as a function of mass
  \citep[e.g.][]{Reiprich+Bohringer02}. We took cluster masses from the
mass function of \cite{Girardi+Giuricin00}, who show that at
  $z=0$ a single Schechter function can describe the local mass
function of both groups and clusters. Of course the mass function of
RLAGN-hosting clusters and groups may be different from that of
clusters and groups in general, but the approach we used should give us
a reasonable approximation; we drew environments in the mass range
$10^{13}$ to $10^{15}M_\odot$ from their distribution, which of course
implies a strong bias towards the sort of group-mass environments that
RLAGN are known to tend to favour based on optical clustering and
  X-ray studies
\citep[e.g.][]{Lilly+84,Prestage+Peacock88,Hill+Lilly91,Hardcastle+Worrall99,Harvanek+01,Best04,Ineson+15}.
We took the probability of a source having a given jet power
$p(Q) \propto Q^{-1}$, motivated by the slope of the steep end of the
RLAGN luminosity function (see also below, Section \ref{sec:jetlf}).
For the trial lifetime functions, we followed H18 and adopted two
possibilities: (i) lifetimes are distributed uniformly in linear space
between 0 and 1000 Myr and (ii) lifetimes are distributed uniformly in
log space between 1 and 1000 Myr. Starting times were distributed
uniformly between 0 and 1200 Myr before the time of observation, and remnant sources were included
in the models, as they are presumably present in the data;
   as noted by H18, however, they are expected to constitute only a
  small fraction of the total for powerful objects. We simulated
10,000 sources for each luminosity range, tuning the range of input
jet powers simulated to be appropriate for the luminosity range, and
then simulated observations that matched the completeness flux cut of our
RLAGN sample and the surface brightness limits that applied to the real
data (Fig.\ \ref{fig:sblimit}) and the appropriate luminosity
cuts.

\begin{figure*}
\includegraphics[width=0.46\linewidth]{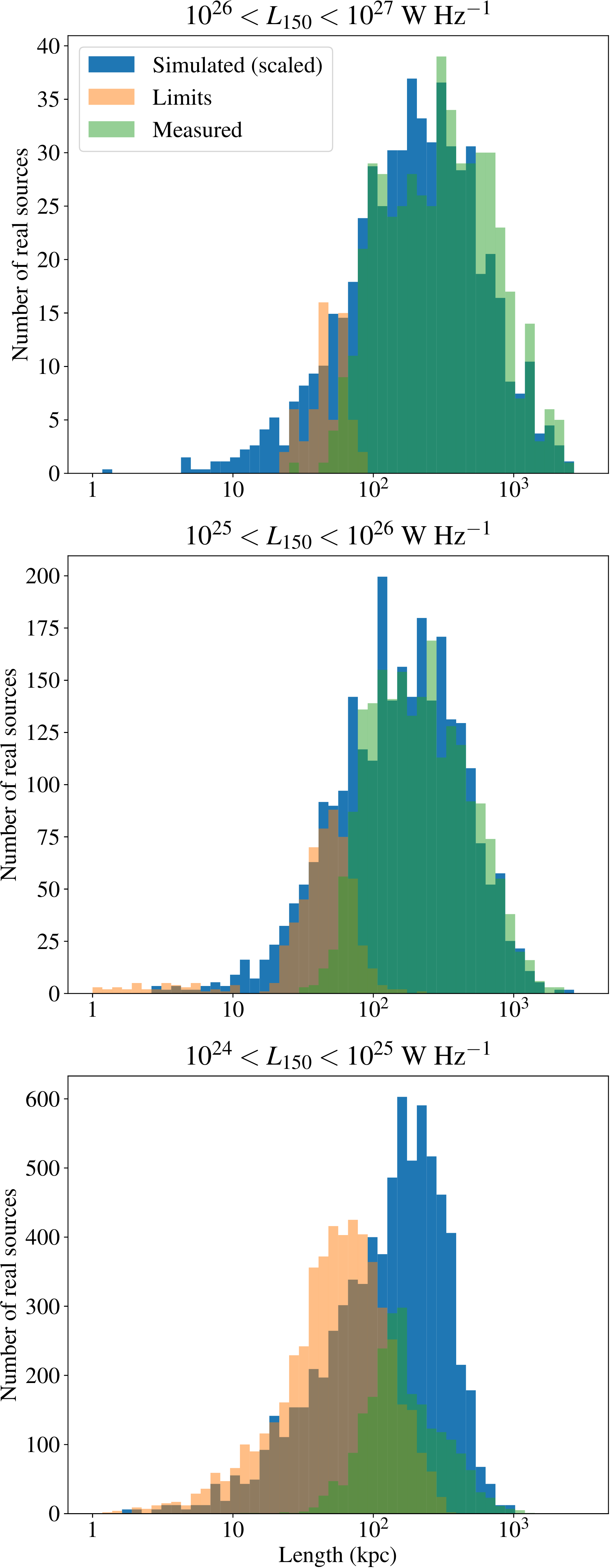}  
\includegraphics[width=0.46\linewidth]{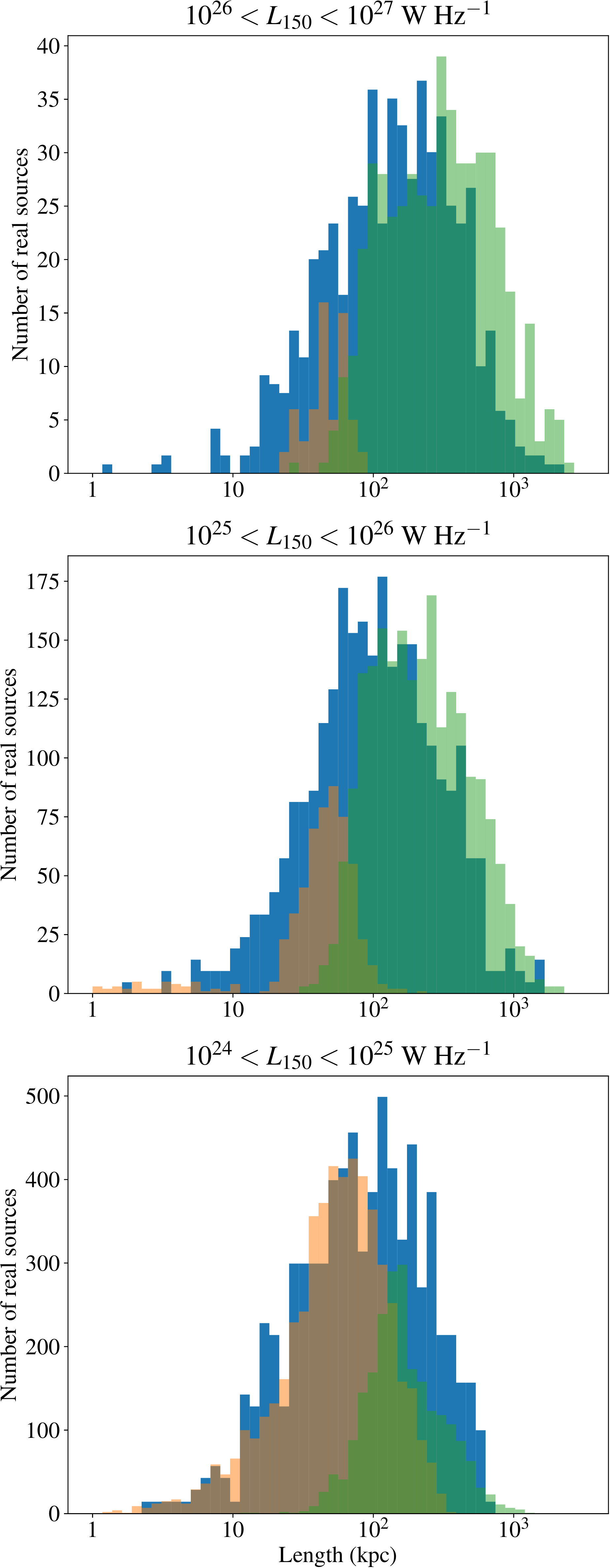}  
\caption{Distributions of real and simulated projected sizes for three
luminosity ranges (from top to bottom, three slices of the $P$-$D$ diagram of
Fig.\ \ref{fig:pdd}) and two lifetime distributions (left: model (i), uniform,
right: model (ii), log-uniform.}
\label{fig:simcomp}
\end{figure*}

\begin{table*}
  \caption{Numbers and fractions of real and simulated sources with
    size $<100$ kpc as a function of radio luminosity.}
\label{tab:sizecomp}
  \begin{tabular}{lrrrrr}
    \hline
    Luminosity range&\multicolumn{3}{c}{Real
      sources}&Simulation (model
      i)&Simulation (model ii)\\
    W Hz$^{-1}$&Total&Small&Small fraction&Small fraction&Small
    fraction\\
    \hline
    $10^{26}$--$10^{27}$&566&124&$0.22 \pm 0.02$&$0.26 \pm 0.01$&$0.38
    \pm 0.02$\\
    $10^{25}$--$10^{26}$&944&2683&$0.35 \pm 0.01$&$0.33 \pm 0.01$&$0.49
    \pm 0.02$\\
    $10^{24}$--$10^{25}$&4443&7457&$0.59 \pm 0.01$&$0.38 \pm 0.01$&$0.54
    \pm 0.02$\\
    
    \hline
    \end{tabular}
\end{table*}

Results are shown in Fig.\ \ref{fig:simcomp}. As was already implied
by the $P$-$D$ diagram presented in Fig.\ \ref{fig:pdd}, we see that
model (i), the uniform-lifetime model, reproduces extremely well the
linear size distribution of the most powerful sources ($L_{150} >
10^{25}$ W Hz$^{-1}$). It perhaps slightly underpredicts the number of
very large sources but we have not attempted to adjust the maximum
lifetime to fit the observations. The differences in models are
  clearest when we compare the numbers of small sources (where we
  define ``small'' as $<100$ kpc  to include all the upper
  limits on size in this bin) and so in Table \ref{tab:sizecomp} we
  compare real fractions of small sources as a function of radio
  luminosity with simulated sources. We see that model (i) agrees very
  well (to within a few per cent) with the fraction of small sources
  observed above $10^{25}$ W Hz$^{-1}$, but is not at all consistent
  with the fraction of small sources in the low-luminosity bin. By
  contrast we see that model (ii) substantially overpredicts the
  number of small sources in the more luminous subsamples, while doing
  a better job with the numbers in the lowest luminosity bin. Model
  (ii) also substantially underpredicts the number of very large
  sources observed in the two higher luminosity bins while
  overpredicting the numbers of large sources in the lowest luminosity
  bin.

These results have several interesting implications. Firstly, the fact
that we can reproduce the size distribution of the most powerful
sources with such a simple model as model (i) is striking. Equally, it
is clear that the data for the most luminous sources are not
consistent with a model, like model (ii), where there are many more
short-lived objects than there are long-lived objects. While the very
youngest sources are expected to be affected by absorption effects that are not included in the analytical model, this is only relevant
for a small fraction of the lifetime of a source (consistent with the
  small fraction of sources with a low-frequency spectral turnover
  detected by \citealt{Callingham+17}) and cannot explain the low
numbers of small, luminous sources seen in the LOFAR samples. If the
models are anywhere near correct, we must assume that the typical
lifetime of a powerful radio galaxy is long, of the order of several
hundred Myr at least, such that most of these systems spend most of
their lifetimes extended on $\ga 100$ kpc scales.

We can then ask why the results are so different at lower
luminosities, particularly for the $10^{24} < L_{150} < 10^{25}$ W
Hz$^{-1}$ sample. This difference cannot be a redshift-dependent
effect, partly because the redshift distributions for the three
samples are not very different (Fig.\ \ref{fig:l_z}) and partly because
the modelling takes account of the different redshift distribution of
each sample. Several possible explanations may be considered:
\begin{itemize}
\item SFG contaminate the samples at low luminosities. This is
  likely to be the case at some level given the limitations
    on the colour selection that we discuss above and the lowest
  luminosity range we consider is such that moderately powerful SFGs
  might well be present, although we cannot say in what numbers.
  However, if this is the case then the {\it WISE} colour selection
  must be failing badly for a large population of SFG. Alternatively,
  some other less obvious contaminating population that generates
  low-luminosity, compact sources may be present.
\item Identifications are worse at low luminosities. This seems
  unlikely to be the case since the contaminating population are
  mostly compact sources that usually have a good identification
  with a nearby galaxy.
  \item There is a genuine luminosity (or rather jet-power) dependent
    difference in the lifetime function of low-power and high-power
    sources, such that low-power sources are genuinely more
    short-lived and have a lifetime function more like that of model
    (ii). One possibility is that this difference is related to the
    different fuel sources available to RLAGN; perhaps sources powered
    by accretion from the hot phase of the inter-galactic medium have a significantly
    different lifetime function. Testing this model requires more
    environmental and AGN accretion mode information than we currently
    have for this sample; \cite{Croston+18b} show that most
      objects in the sample are not members of the
      available optical group and cluster catalogues.
\item The models get the source physics wrong at low luminosities. To some extent we
  expect this to be the case; the model overpredicts the radio
  luminosity of FRI-type sources, which should dominate the
  lowest luminosity bin, where a significant amount of the energy
  input of the jet appears to go into non-radiating particles
  \citep{Croston+18}. But it is difficult to see how this solves the
  problem; if we are overpredicting radio luminosities in this
  r\'egime then the jet powers in this luminosity band should actually
  be higher than in the models and the sources in the simulated
  sample, if corrected for this, correspondingly larger.
\item The models get the environment wrong in a way that induces a
  luminosity dependence. There are several ways in which this might be
  possible. For example, the models do not contain the dense, cold
  central gas that is invoked in ``frustration'' models of compact
  steep-spectrum sources, and such a component would have a larger
  effect on sources of lower jet power. Other,
    more subtle luminosity-dependent effects include a tendency for
    lower luminosity sources to lie away from their host group or
 cluster centre and a dependence of radio luminosity on host environment \citep{Ineson+15,Ching+17,Croston+18b}.
\item The measured sizes are wrong. This is very likely to be the case
  in faint sources in the low-power, FRI regime, since the surface
  brightness of lobes or plumes drops off rapidly with distance from
  the nucleus. We may simply lack the surface brightness sensitivity
  to map extended structures in many of these sources
  \citep[cf.][]{Shabala+17}. The H18 model is based on the dimensions
  of the shocked shell driven out by the momentum flux of the jet,
  which may well extend beyond the limits of any observable jet for
  FRI sources, while it is almost always going to be close to the
  hotspots of resolved FRIIs. If this is the sole explanation for the
  large number of apparently compact RLAGN then we would expect deeper
  LOFAR observations still to start to reveal extended structures
  around many RLAGN that are compact at our current observational
  sensitivity. Existing surveys at higher frequencies, even with
    high sensitivity, are likely to be less sensitive to extended
    structure than LOFAR and would also miss this extended emission.
  Such an explanation will be testable with ``Tier 2'' LOFAR surveys
  data with sensitivities of tens of $\mu$Jy, or with deep surveys
  with MeerKAT \citep{Jarvis+17} or the SKA.
\end{itemize}

\subsection{RLAGN host properties with size}

\begin{figure*}
  \includegraphics[width=0.46\linewidth]{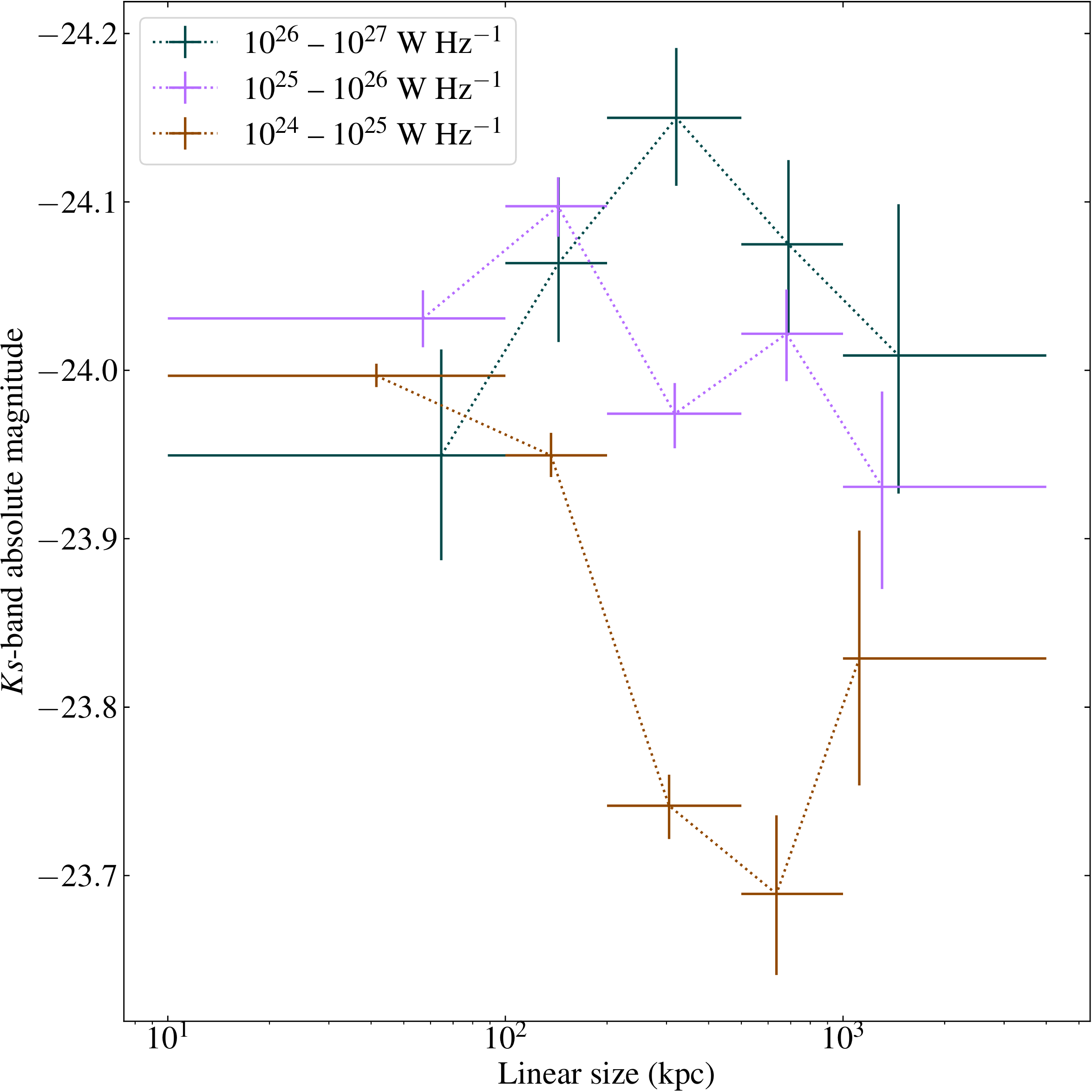}
  \hskip 25pt
  \includegraphics[width=0.46\linewidth]{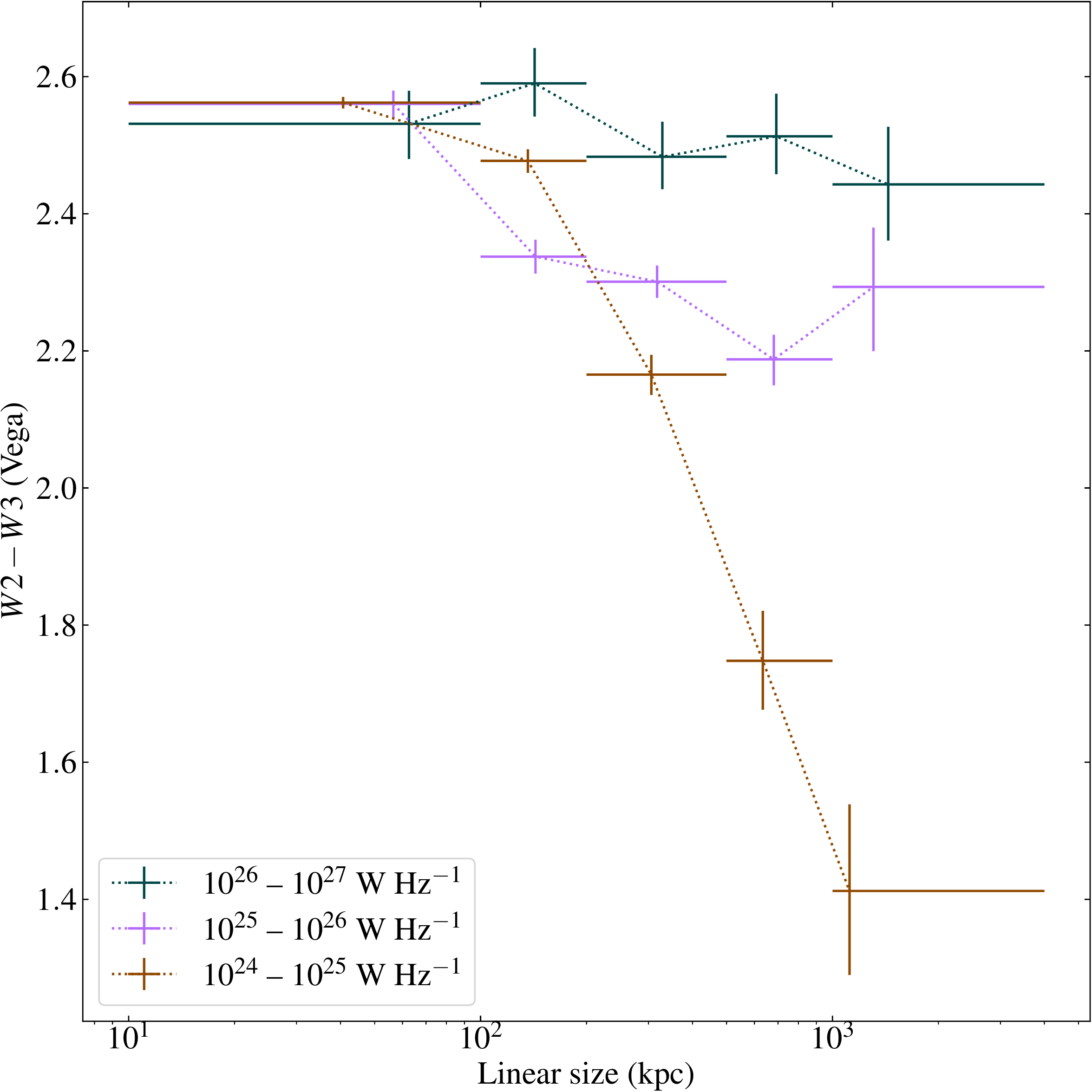}
  \caption{Mean properties of the host galaxies of RLAGN as a function
    of projected linear size. Left: $Ks$-band rest-frame magnitude.
    Right: {\it WISE} band 2 minus band 3 colours (Vega magnitudes).
    In both plots the vertical error bars represent $1\sigma$ errors
    derived from bootstrap, the horizontal bars represent the bin
    range, and the position of the central point represents the mean
    projected linear size within the bin. Dotted lines joining points
    in a particular sample are to guide the eye only.}
\label{fig:hostsize}
\end{figure*}

In the models discussed in the previous section, which successfully
describe powerful radio sources in the RLAGN sample, large physical
size is just a marker of a long-lived source rather than indicating
something special about the host galaxy or its environment. The H18
models produce a very few extreme giants (high-power sources in
low-density environments) but generally giant radio galaxies are
expected to be a natural consequence of observing normal powerful
sources towards the end of their lives. The RLAGN sample contains 126
objects with projected physical size $>1$ Mpc in our adopted
cosmology, satisfying the classical definition of a giant radio galaxy
(GRG); as noted by H16, LOFAR's combination of low-frequency selection
(GRGs are likely to have steep radio spectra) and excellent
surface-brightness sensitivity makes it a very productive instrument
for studies of such large sources. The sky density of candidate GRGs
in the HETDEX survey (about 1 per 4 square degrees) exceeds even that
reported by H16 by a factor $\sim 5$ thanks to the improved image
fidelity, uniform sensitivity, and better optical data of the
HETDEX survey. We emphasise that these are giant candidates only, as
their sizes have been measured automatically and many of the redshifts
are photometric; \cite{OSullivan+18} report a case in which the use of
a newly obtained spectroscopic redshift instead of the photometric
redshift used in this work reduces the projected size of one of these objects
from 4 Mpc to 3.4 Mpc. However, a substantial fraction of the GRG
redshifts are spectroscopic and there is no reason to suppose that a
large fraction of them will be reclassified below the 1 Mpc threshold
either because of their redshifts or because of their automatically
measured angular sizes.

The RLAGN sample therefore provides an excellent opportunity to test
the hypothesis that the hosts of these objects are not special
and that they merely represent the late-time evolution of normal
powerful radio galaxies. In this hypothesis properties of the host
galaxies, such as their colours and absolute magnitudes, should be
close to independent of source projected physical size\footnote{For
  powerful radio sources there is evidence \citep{Best+97-2} that the
  early stages of radio galaxy evolution are associated with an
  aligned, blue component in the host galaxy, which may be connected to, for example jet-induced star formation, and which disappears later in the
  lifetime of a source. However, this effect is much less obvious in the
  infrared bands that we use for this test, and as this effect is also seen in
  sources much more powerful than those in our sample, we neglect it
  here.} Fig.\ \ref{fig:hostsize} shows such a test. We divided the RLAGN sample into the three luminosity bins of the
previous section and then binned in projected linear size, taking the
average of rest-frame $Ks$-band magnitude and {\it WISE} band 2/band 3
colour (see Figs \ref{fig:ksl150} and \ref{fig:wisecc} for
distributions of the whole sample in these parameters). The upper limits
on physical size are treated as measurements for purposes of binning
in these plots; as almost all of these limits are less than 100 kpc
(Fig.\ \ref{fig:simcomp}) there is very little ambiguity in the
binning. A tiny minority of sources without {\it WISE} photometry are
ignored.

What we see in the first panel of Fig.\ \ref{fig:hostsize} is
that the absolute magnitudes of all three samples show very little
variation with physical size, barring a slight deviation from the mean
in the 200--500 kpc bin for the lowest luminosity sources for which we
have no explanation. Broadly this plot is consistent with the idea
that all powerful RLAGN hosts have an absolute magnitude around $-24.0$, and
scatter of a few tenths of a magnitude at most irrespective of their
radio luminosity or size. This is consistent with what is seen for the
whole population in Fig.\ \ref{fig:ksl150} and this standard
  infrared magnitude is of course the basis of the well-known K-$z$
  relation for radio galaxies \citep{Lilly+Longair84}. \cite{Sabater+18} discuss in more detail the distributions of the host
  galaxy masses of RLAGN.

The second panel of Fig.\ \ref{fig:hostsize} shows that the mean {\it
  WISE} colour of the highest luminosity sample is constant with
length, that of the intermediate-luminosity sample deviates from a
constant value in the lowest size bin, and for the lowest luminosity
sample the colour is very strongly dependent on projected linear size
over the whole range of sizes studied. It is very striking that the
population that shows such a deviation from the hypothesis that all
RLAGN hosts are the same is precisely the population that we previously suggested may be contaminated by some other type of source,
such as SFG. The colour deviations seen in this figure are in the sense that
sources move closer to SFG colours as their sizes get smaller. We emphasise that the average colours never become as extreme as
colours that we expect from SFG, which would be impossible given the
{\it WISE} colour selection we used for the RLAGN sample, and that
type 1 and type 2 quasars and Seyfert galaxies also have higher $W2-W3$ colours due to the
torus. We conclude that it is plausible that the low-luminosity RLAGN
sample contains more than one population. However, the constancy of
host galaxy colours and masses as a function of size for the
highest luminosity bins provides strong evidence that powerful RLAGN are
homogeneous: there is no evidence that the largest, oldest
RLAGN have different hosts from their smaller counterparts.
Investigation of the related question about environment -- some
relationship between size and environment is a prediction of the
models -- will require a data set with more environmental information
than is currently available.

\subsection{Bulk inference of jet power}

Noting that tracks of constant jet power $Q$ describe characteristic
curves in the $P$-$D$ diagram for a given environment and redshift
(Fig.\ \ref{fig:pdd}), we can now investigate a simple model-dependent
method for inferring jet power $Q$ from the observed redshift, $L_{150}$ and
projected linear size $D$ for the RLAGN sample. We do not have direct
measurements of environmental richness for most of these objects (see
\citealt{Croston+18b} for a discussion of the available constraints) and similarly almost no
information about the angle to the line of sight for a given source; other potentially
useful parameters such as the axial ratio of the lobes or their
integrated spectral index (H18) have not yet been measured. Thus we
focus on what can be inferred from $z$, $L_{150}$ and $D$.

Our approach, as in Section \ref{sec:sizedist}, is to generate
populations of simulated sources that match the LOFAR observations in
terms of observational selection criteria and populate the observable
regions of the $P$-$D$ diagram using the models of H18. In the absence
of any environmental information we assume the same distribution of
source environments as earlier and the same distribution of angles to
the line of sight. We can then estimate the jet power corresponding to
any particular position in the $P$-$D$ diagram by looking at the mean
jet power of simulated sources that lie close to that location: the
uncertainty in the inference comes from the distribution of the local
simulated sources. This method automatically takes into account the
unknown angle to the line of sight and the unknown environment, as
long as the distributions we use are approximately correct. To take
into account the strong redshift dependence of radio luminosity as a
result of inverse-Compton losses, we generate populations for a number
of redshifts in the range $0<z<0.8$ where we have a uniform population
of RLAGN, and interpolate between the nearest one or two for any given
source.

\begin{figure*}
  \includegraphics[width=0.46\linewidth]{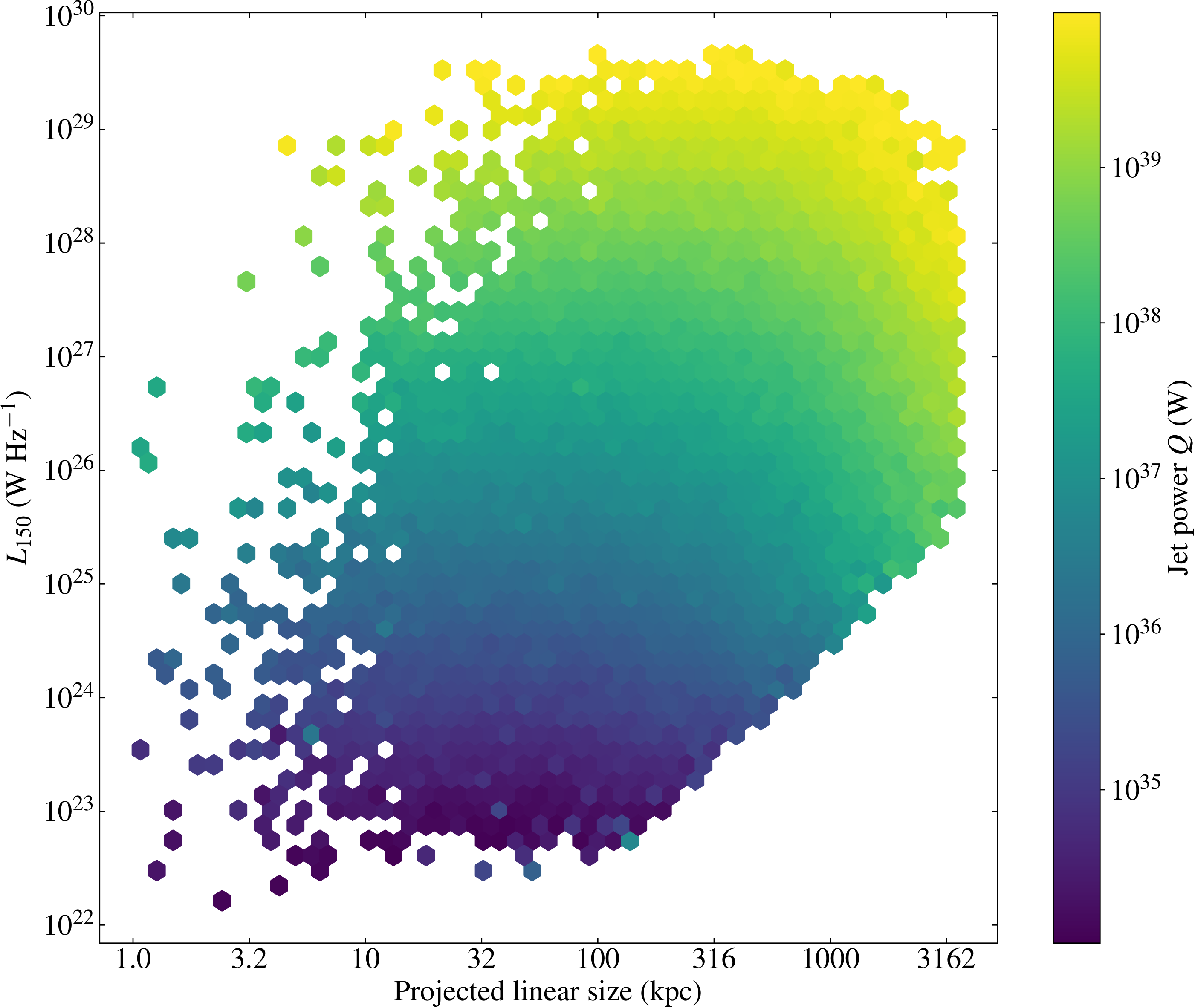}
  \hskip 25pt
\includegraphics[width=0.46\linewidth]{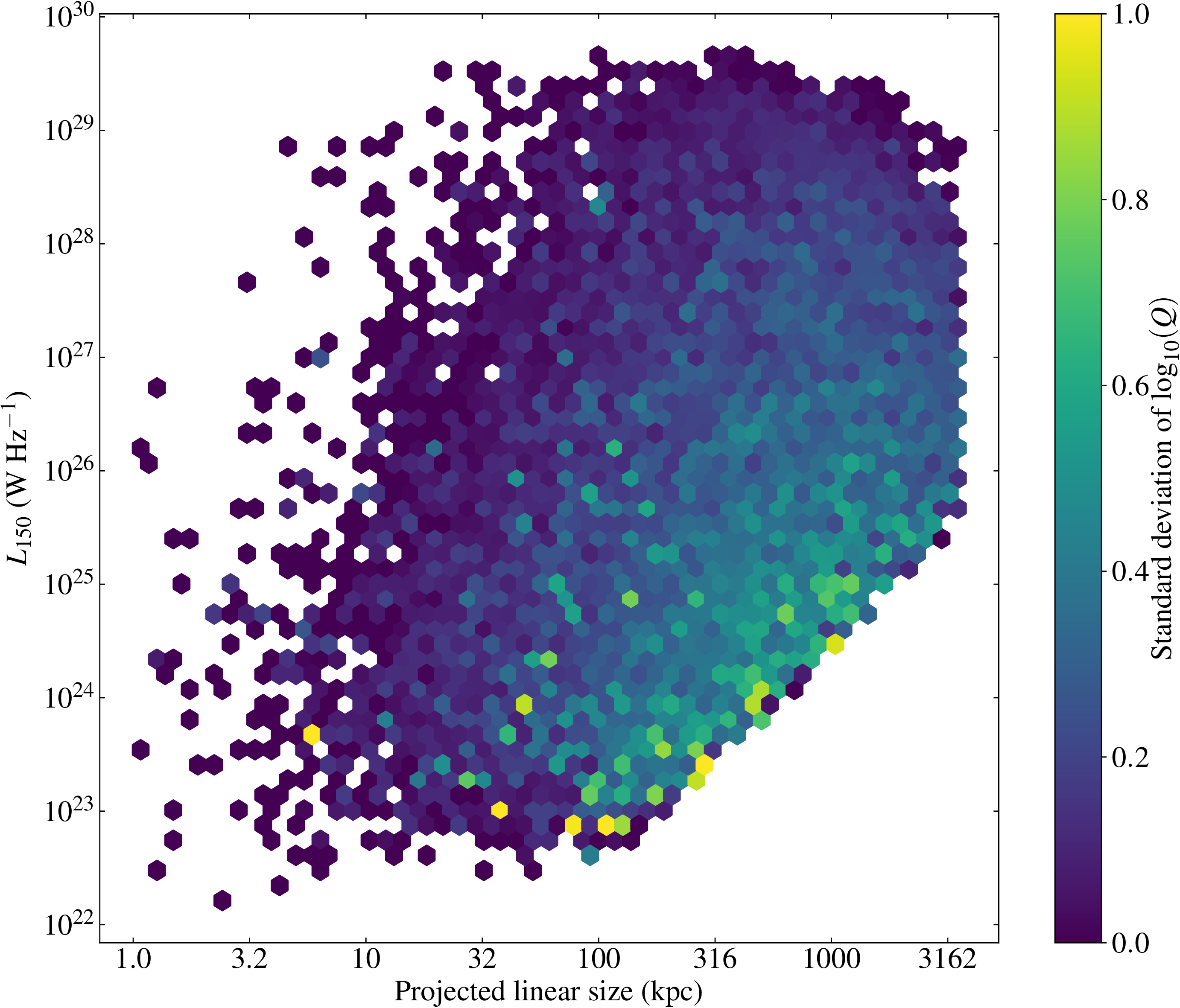}  
\caption{Left: Mean jet power $Q$ as a function of radio luminosity and
  projected linear size for the full simulated sample over all
  redshifts. Right: Standard deviation of $\log_{10}(Q)$ for each bin
  in the left-hand panel, showing the scatter in the inference
  introduced by different environments, projection angles, ages, and
  redshifts. See the text for details of the modelling.}
\label{fig:inference_pd}
\end{figure*}

In detail, we take a set of redshifts ($0.05, 0.15, \dots, 0.75$) and,
for each redshift, populate a $P$-$D$ diagram using jet powers in the
range $10^{34} < Q < 10^{40}$, where we assume a uniform distribution
of $Q$ in log space to make sure that all of the luminosity
range is populated. We take the lifetime function to be a uniform
distribution of lifetimes in linear space, as in model (i) of
Section \ref{sec:sizedist}. We apply the LOFAR observational selection
criteria to the simulated sources, giving us of order 5,000 sources
per redshift slice. A plot showing the binned mean $Q$ as a function
of position in the $P$-$D$ diagram for the stacked simulated sample,
and the dispersion in inferred $Q$ introduced by different
environments, projection angles, ages, and redshifts, is shown in
Fig.\ \ref{fig:inference_pd}.

We then restrict the RLAGN sample to $z<0.8$ and $L_{\rm 150} >
10^{23}$ W Hz$^{-1}$ giving us a total of 18,948 objects; below that luminosity we regard the jet models
as uncertain and linear sizes above 100 kpc are not expected to be
present. Then, for each resolved
object in the restricted sample, we take the Gaussian-weighted mean
$Q$ in log space of all of the simulated points within $3\sigma$ of
the position of the real object in $P$, $D$ space, where we define the
width of the weighting Gaussian $\sigma = 0.04$ dex, corresponding to
a fractional error of 10\%. This is reasonable at least for the
luminosities, where the absolute flux calibration uncertainty is
probably of this order: we have no real constraints on the
uncertainties on projected physical size but a 10\% uncertainty seems
plausible. For unresolved objects we instead use the upper limit on
size from earlier in this section and consider all simulated sources
consistent with that limit and within $3\sigma$ of the position
defined by the radio luminosity. In both cases an error on $Q$ can be
estimated by bootstrapping from the sample of simulated sources: this
automatically accounts for the uncertainties on inference in parts
of the $P$-$D$ plane that can be populated by a large range of jet
powers. Typically the errors estimated in this way are of the order of
10\% in $Q$, which is reasonable given the assumed input uncertainties
on $L_{150}$. In a few cases the errors are much larger ($>0.5Q$) or
there are not enough points in simulated $P$-$D$ space for the
estimation or bootstrap process to work: in this case we flag the
measured values of $Q$ as bad. In total 19,356 objects have a good
estimated jet power.

\begin{figure}
\includegraphics[width=\linewidth]{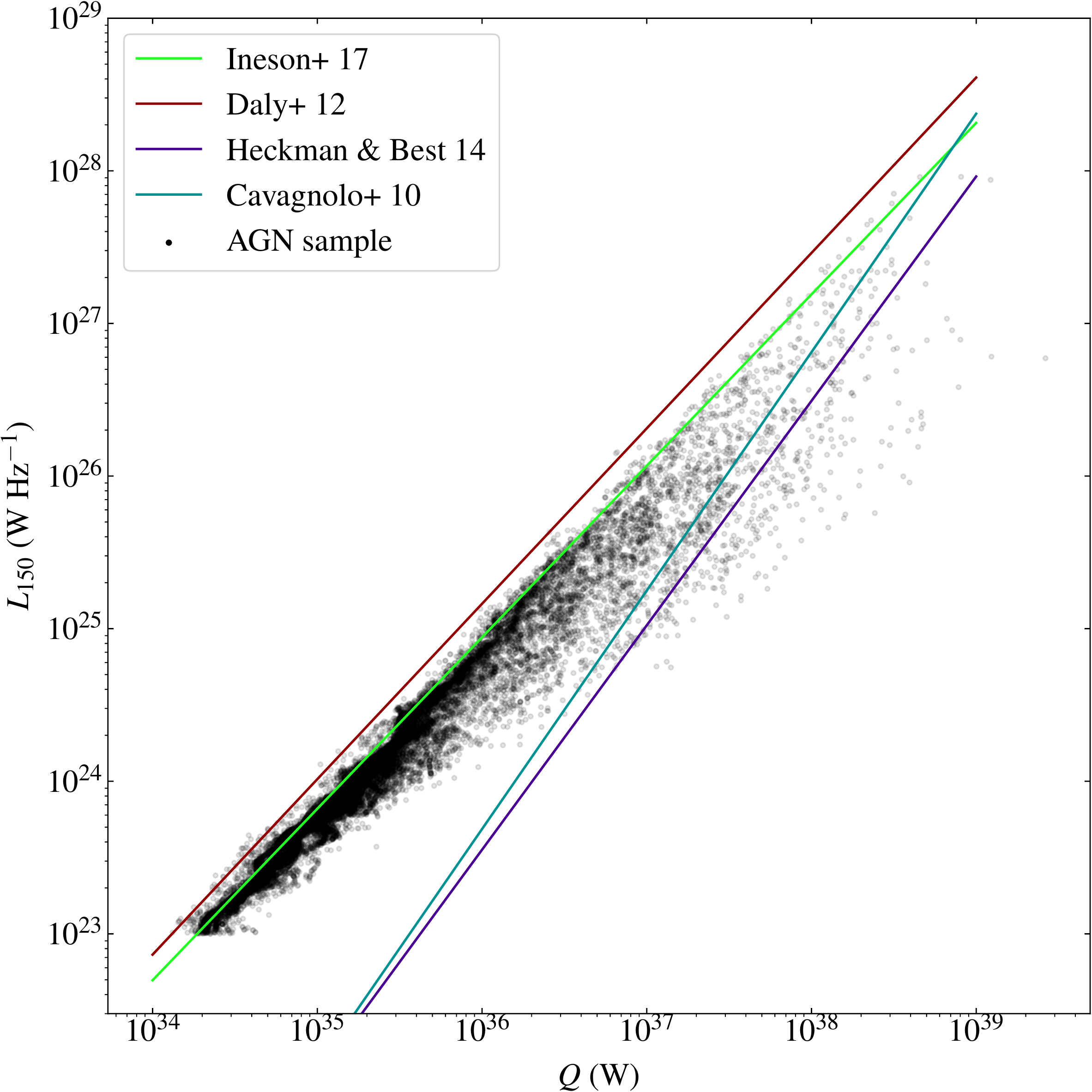}  
\caption{Radio luminosity plotted against the inferred jet power. The
  relations obtained by \protect\cite{Cavagnolo+10},
  \protect\cite{Daly+12}, \protect\cite{Heckman+Best14}, and \protect\cite{Ineson+17} are also shown.}
\label{fig:l150q}
\end{figure}

Fig.\ \ref{fig:l150q} shows the relationship between radio luminosity
and jet power that we infer for the RLAGN sample. We overplot for
  comparison the relations derived by \cite{Cavagnolo+10},
  \cite{Daly+12}, \cite{Heckman+Best14}, and \cite{Ineson+17}. We see
  that the \cite{Ineson+17} relation agrees well with our inference,
  which is not surprising since its methods are closest to the
  assumptions of the H18 models. The inference in this work is for slightly
  higher jet powers for a given radio luminosity when compared to the
  relations of \cite{Ineson+17} or H18. But this is probably a
  result of the assumptions that we make regarding environment -- the
  typical environment of our simulated sources is poor and so less
  radio emission is produced for a given jet power -- and partly due
  to the higher average redshift, $z\sim0.5$ of the RLAGN sample,
  which gives rise to lower radio luminosity for a given jet
    power compared to $z=0$ because of stronger inverse-Compton
    losses. Our results are similar to, but generally predict
  slightly higher powers than, the results of \cite{Daly+12}, which are
  based on powerful FRIIs; their method \citep{O'Dea+09} uses spectral
  ages involving minimum-energy magnetic fields and so would be
  expected to underestimate both the age and energetic content
  given the observed sub-equipartition field strengths \citep{Hardcastle+02,Kataoka+Stawarz05,Croston+05,Ineson+17}, but clearly these
  effects cancel to some extent in practice. There is much less good
  agreement with the cavity-based relations of \cite{Cavagnolo+10} or
  \cite{Heckman+Best14} at low luminosities. For the most powerful
  sources in our sample, with luminosities $L_{150} \approx 10^{26}$ W
  Hz$^{-1}$, however, all jet power estimates are of the same order of magnitude
  (see discussion by \citealt{Heckman+Best14}) and the cavity
  relations are actually reasonably consistent with our inference, although it
  should be noted that the cavity relations are not generally
  supported by much data at these radio luminosity values. For
  low-luminosity sources, as already noted, the H18 model is likely to
  overestimate the radio luminosity for a given jet power and the
  truth is likely to lie somewhere in between our inferred values and
  the cavity models.

Better environmental information for our sample would improve our
inference process and decrease the uncertainties on the inferred jet
powers. Without this information -- or other information that
  we might be able to make use of, such as constraints on angle to the
  line of sight for individual sources -- these jet powers are still
  only estimates that are not expected to be particularly accurate
  for any given source. Nevertheless this work demonstrates the
feasibility of bulk estimation of $Q$ without resorting to simple
scaling relationships based on radio luminosity.

\subsection{Jet kinetic luminosity function}
\label{sec:jetlf}

\begin{figure*}
  \includegraphics[width=0.46\linewidth]{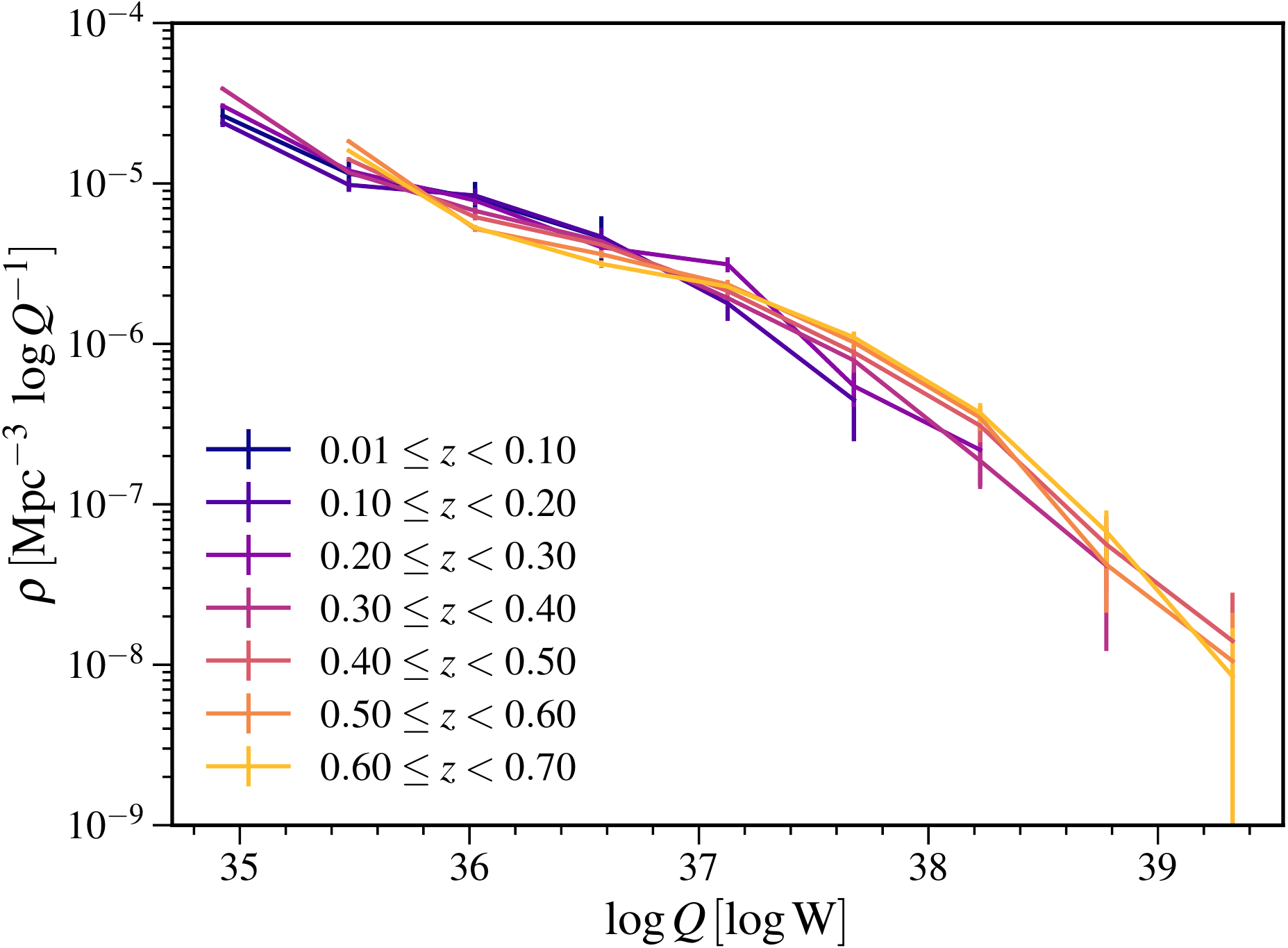}
  \hskip 25pt
  \includegraphics[width=0.46\linewidth]{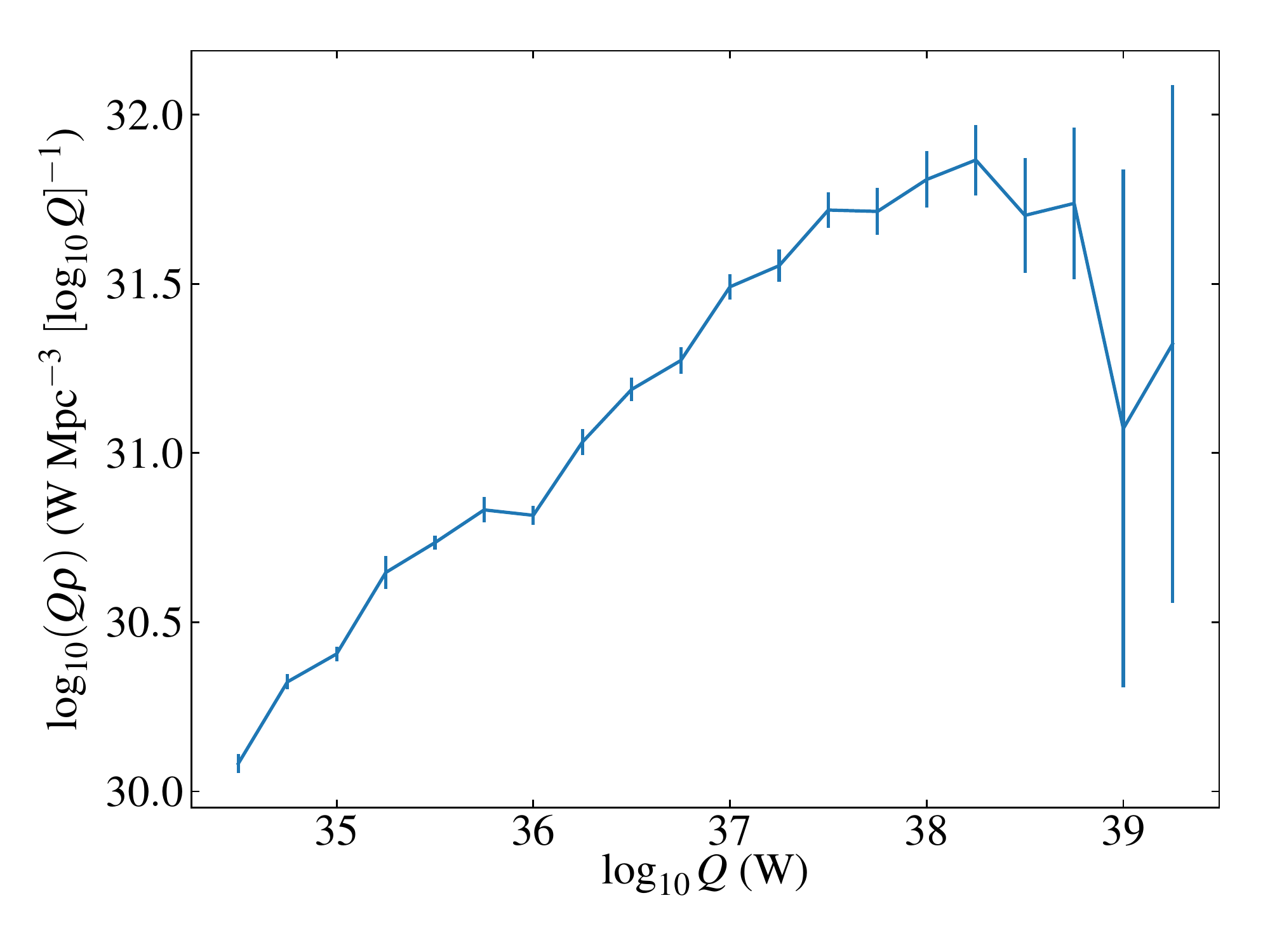}
\caption{Jet kinetic luminosity function for the $z<0.7$ LOFAR RLAGN
  sample. Left: The kinetic luminosity function
  divided in bins of redshift. Right: The full $z<0.7$ luminosity
  function multiplied by $Q$ to show the peak at $\sim 10^{38}$ W.}
\label{fig:klf}
\end{figure*}

We can use the inferred jet powers from the previous subsection to
construct a jet kinetic luminosity function that represents our
current best estimate of this quantity for the LOFAR RLAGN sample,
bearing in mind that we excluded very low-luminosity radio
  sources and those with poor jet model fits and that the inclusion
  of these sources would probably slightly increase the normalisation
  of the luminosity function. To do this we simply apply the
standard $V_{\mathrm{max}}$ method \citep{Schmidt68,Condon89} to the jet
power $Q$, calculating the volume based on the combined radio and optical
constraints. Only sources within $0.01 < z < 0.7$ are considered and an
$i$-band limit of 21.5 mag is imposed to ensure photometric redshift
completeness. The radio $V_{\mathrm{max}}$ is calculated as  $\int
d_{\mathrm{max}} dA$, where the completeness function, $dA$, is determined
from the LoTSS rms map. For the optical, $V_{\mathrm{max}}$ is calculated from
our imposed $i$-band limit of 21.5 mag after the optical magnitudes are
corrected for the Galactic reddening calculated by \cite{Schlafly+Finkbeiner11}
and $K$-corrected based on the rest-frame magnitudes calculated by Duncan et
al. (2018). Our approach automatically takes into account the unknown RLAGN
  duty cycle since only the density of LOFAR-detected, luminous
  sources is calculated. The results are shown in
Fig.\ \ref{fig:klf}. We see, as expected, that the kinetic luminosity
function appears very like the radio luminosity function, in that it
is flatter at low jet powers and steepens at higher powers; there is
little evolution in the kinetic luminosity function with redshift,
as expected since the dominant LERG population is known not to
  evolve strongly \citep{Best+14,Williams+18}.

The kinetic luminosity function $\rho(Q)$ is a physically important
quantity in that the integral $\int Q \rho(q) \rd \log(Q)$ tells us
the energy per comoving volume injected by all RLAGN jets into their
host environments, and this work represents the first attempt to construct
$\rho(Q)$ from bulk inference of jet powers for a large sample. Integration
of our kinetic luminosity function over the observed range gives a
total RLAGN kinetic luminosity density, including the effects of
  all the sources in our luminosity range not excluded by
  surface-brightness selection, of $7 \times 10^{31}$ W Mpc$^{-3}$.
The integral is dominated (Fig.\ \ref{fig:klf}, right-hand panel) by
powerful sources, peaking at jet powers around $10^{38}$ W, and so is not sensitive to the uncertain jet powers
of low-power objects; indeed, the peak lies in the region in which
inference, FRII power estimates, and cavity power estimates all give
similar results (Fig. \ref{fig:l150q}). The integral of the kinetic
luminosity function may be compared to the total
radiative (cooling) luminosity density of groups and clusters.
Integrating the Schechter function fitted by \cite{Bohringer+14} to
the local cluster luminosity function between cluster luminosities of
$10^{42}$ and $10^{46}$ erg s$^{-1}$, we obtain a cooling luminosity
of $2 \times 10^{31}$ W Mpc$^{-3}$, a result that is insensitive
  to the limits of integration because of the form of the Schechter
  function. Thus the RLAGN population found in this work can in principle
completely offset, in statistical terms, all the local radiative
cooling of the environments that they are expected to occupy, even
allowing for the fact that some of the kinetic luminosity goes into
cosmic rays that may not play much of a role in heating the thermal plasma
in groups and clusters\footnote{In numerical models, the fraction
    of the jet kinetic power that heats the external environment is a
    little over 0.5 during the active source lifetime
    \citep{Hardcastle+Krause14,English+16}, but much of the energy
    stored in the lobes is then lost to the large-scale environment in
    the remnant phase (English \etal\ in prep); the fraction of energy
    that remains in cosmic rays at late times is not well known,
    although it is an important ingredient in models of cluster
    evolution.}. This is strong support for feedback models in
  which RLAGN provide the ``maintenance mode'' required to prevent the
  hot phase of the environment of their host galaxies cooling back
  onto the central galaxy and reinvigorating star formation.

We emphasise that this is only the first step towards the construction
of a truly reliable kinetic luminosity function. A key problem is the
effect on the kinetic luminosity function of the large number of low-luminosity sources
that probably in reality do not lie on the relation between $Q$ and
$L_{150}$ that is implied by the H18 models (see Section
  \ref{sec:sizedist}). Their radio luminosities for a given $Q$ are
expected to fall significantly below the model expectations, since
some of their internal pressure is provided by a non-radiating
particle population, and so their inferred jet powers should actually
be higher than the values we used. In addition, if larger
versions of these sources exist, it is entirely possible that we are
systematically missing numbers of them due to the surface-brightness
limitations on our survey, which feed through into limitations on the
sources that may be observed in the power/linear-size diagram
(Fig.\ \ref{fig:pdd}). Modelling of these two effects will be
important in order to make progress, though as we noted above, these
sources will have only a small effect on the integral of the
luminosity function unless their jet powers are $\sim 2$ orders of
magnitude higher than we infer them to be. On the other hand, powerful
sources are likely to live in richer environments than we have
modelled \citep{Ineson+15} and to have lower jet powers than we infer,
which would have a stronger effect, because of the shape of the
luminosity function, on the integrated kinetic power input we find. It
is entirely possible that some of the curvature in the observed radio
luminosity function is due to these combined effects of radio galaxy
physics, observational selection, and environment.

Earlier calculations of the kinetic luminosity density were carried
out by for example \cite{Best+06} and \cite{Smolcic+17}, and these compare
very well to the results we derive from integrating the kinetic
luminosity function. \cite{Best+06} computed jet powers for the
$z\approx 0$ SDSS/FIRST-based sample of \cite{Best+05}, i.e. a very
comparable, although smaller, sample, using a version of the cavity
power estimates discussed in Section \ref{sec:intro}, which gives a
very flat jet power/radio luminosity dependence compared to ours. They
nevertheless obtained a luminosity density of $4 \times 10^{31}$ W
Mpc$^{-3}$, which agrees with ours to within a factor 2.
\cite{Smolcic+17} used deep VLA data from the COSMOS field with
excellent multiwavelength counterparts and so were able to probe out to much
higher redshifts than we can achieve, allowing them to investigate the
cosmic evolution of the kinetic luminosity function. On the other hand
their sampling of the local Universe was necessarily limited by the
small volume available to them. They made use of the \cite{Willott+99}
radio luminosity/kinetic luminosity relation, which, as discussed in
Section \ref{sec:intro}, can only ever be an approximation; however,
as shown by H18, suitable choices of normalisation of the Willott
relation can bring it into agreement with more sophisticated models
for large, mature sources at a particular redshift, and the values
adopted for the ``uncertainty parameter'' of the Willott relation by
\cite{Smolcic+17} span the range that would be appropriate for
powerful, mature sources at $z=0$ in the H18 models. Given these
differences in the model and the data, there is excellent agreement
between our kinetic luminosity density of $7 \times 10^{31}$ W
Mpc$^{-3}$ at $z<0.7$ and their estimates ranging between $\sim 2$ and
$\sim 5 \times 10^{31}$ W Mpc$^{-3}$ for their preferred uncertainty
parameter over the same redshift range. We caution, however, that in
the H18 models the uncertainty factor is a function of environment
and redshift, and therefore it is not safe to assume that it is constant
over the lifetime of the Universe. \cite{Smolcic+17} further estimated
the luminosity density required by the SAGE model of \cite{Croton+16}
to be $\sim 7 \times 10^{31}$ W Mpc$^{-3}$ (roughly constant or
slightly declining over the range $0<z<1$), which is again in excellent
agreement with our calculation, although it somewhat exceeds the
observed group/cluster X-ray cooling luminosity density. Combining our
work with that of \cite{Best+06} and \cite{Smolcic+17}, we can
conclude that estimates of the effects of RLAGN on their local
environment are in remarkably good agreement with both X-ray
observations and models. \cite{Sabater+18}, using cavity-based jet power estimates, come to a similar
conclusion in their study of the nearby AGN population, showing that
the jet power of RLAGN is more than sufficient to offset the cooling
of gas in their host ellipticals.

\section{Summary and future work}

In this paper we have constructed a sample of RLAGN from
the value-added catalogue drawn from the LoTSS survey of the HETDEX
Spring field, based on a combination of radio properties, spectroscopic information where
available, and {\it WISE} colour information or radio luminosity
otherwise; this is not a true radio-excess sample of the type selected
  by H16 or \cite{Smolcic+17}, but is expected to be very comparable to
  such a sample. Although only a small fraction of the total radio catalogue
can be classed robustly as RLAGN using our methods because many objects are
SFGs or do not yet have good enough optical
identifications or redshifts to be classified, this process still
yields one of the largest homogeneous RLAGN data sets in existence,
and one from which many interesting individual objects can be drawn.

In the current paper we have focussed on new conclusions that can be
drawn about the properties of the RLAGN population using this large
sample. To do this we require a model of radio galaxy evolution, which
can give us observable quantities such as radio luminosity and total
linear size from model inputs such as jet power, redshift, environment,
and time. We chose to work with the models of H18 but it is
important to note that this is not the only radio source model
available; different analytical models make different approximations
and it will be important in future to cross-calibrate these models and
to see what differences the use of a different model makes to the
inference of population properties.

Comparison with the H18 model tracks in the power/linear-size
($P$-$D$) plot showed that the distribution of source sizes in the
luminosity range best sampled by the LOFAR data was perhaps
surprisingly consistent with a model in which most sources in the
luminosity range $10^{25} < L_{150} < 10^{27}$ W Hz$^{-1}$ are
long-lived objects in relatively poor (group-like) environments. The
critical unknown distribution in this case is the lifetime function, the
distribution of total lifetimes of RLAGN. We showed that a uniform
distribution in the range 0--1000 Myr reproduced well the
distributions of projected linear sizes of the powerful sources; there
are relatively few physically small sources in this radio luminosity
range. On the other hand, at the lower end of the LOFAR luminosity
range there are many more small sources, even when surface-brightness
selection effects are taken into account as we are able to do with our
modelling. This low-luminosity, compact population has been noted
previously \citep[e.g.][]{Sadler+14,Baldi+15,Whittam+17} but
we find large numbers of these objects, requiring either a very
different lifetime distribution at low luminosities, a breakdown of
the underlying models, or some contamination by a separate population
of objects. Based on analysis of the colour and magnitude of host
galaxies binned by linear size, we show that it is entirely plausible
that the luminous LOFAR RLAGN are a homogeneous population in which large
sizes (including those of the many $>1$ Mpc giants in our sample) are
simply an effect of old age; but at low luminosities the strong
dependence of colour on physical size suggests that more than one
population is present. This may be the result of contamination by
star-forming objects or it may indicate that more than one RLAGN
population, perhaps with different large-scale fuelling mechanisms, is
present.

Finally, we used the H18 models to attempt to infer the jet powers $Q$
for LOFAR sources based only on their positions on the $P$-$D$
diagram, marginalising over the unknown environments and angles to the
line of sight of the LOFAR sources. This is a proof of principle for
bulk inference of $Q$ and maybe other source parameters from large
volumes of data. Again, the results are model-dependent and also
dependent on our assumptions, particularly relating to environment. It seems likely, as noted by H18, that observations, for example those of
  \cite{Ineson+15}, require some intrinsic
relationship between jet power and environmental richness that is not
present in the models used in this work. Nevertheless we are able to derive
jet powers that agree reasonably well with results already present in
the literature and allow us to construct the first large-scale jet
kinetic luminosity function based on inference of jet powers rather
than simple scaling relations with radio luminosity. The distribution
of jet powers, and the integral of the jet kinetic luminosity
function, are key parameters in models of galaxy formation and
evolution and, as shown by \cite{Smolcic+17}, it is now possible
  to compare these quantities to the assumptions
made in such models. Integration of the existing luminosity
function, which will be substantially refined in future, suggests that
the energy input from RLAGN is more than adequate to offset all of the
observed X-ray radiative cooling of the group and cluster population
in which we assume the RLAGN to lie; the value we obtain is consistent both with
independent observational estimates of the kinetic luminosity density
by \cite{Best+06} and \cite{Smolcic+17} but also with galaxy evolution models.

Forthcoming developments in LoTSS observations and ancillary data will
allow substantial improvements to be made in all of these areas in the
near future. Star formation/RLAGN separation, as well as the quality
of redshifts and thus luminosities and physical sizes, should be
greatly improved by the WEAVE-LOFAR project \citep{Smith+16}, which will provide both spectroscopic redshifts and emission-line
diagnostics for large numbers of LOFAR sources, including those at
high redshift for which we currently have little information.
High-resolution images using the LOFAR international baselines will
help with source size measurements, optical identifications, and
RLAGN/star formation separation; the Very Large Array Sky Survey
  (VLASS\footnote{\url{https://science.nrao.edu/science/surveys/vlass}}),
  when complete, will also be very useful for the identification of
  flat-spectrum cores in LOFAR objects and for resolving bright,
  compact sources. A key missing ingredient in our bulk inference in
this paper is information on the environments of the RLAGN.
Environmental information can be obtained, for example from SDSS, at low
redshifts \citep{Croston+18b} but the HETDEX sky area is too
small to obtain a representative sample of powerful AGN. The much
larger sky areas provided by the full LoTSS survey, which will reach
10,000 deg$^2$ of coverage in the next two years, will allow us to
probe a larger range of radio luminosities and environments at low
redshift and to take full account of environmental information both
from SDSS and from the forthcoming {\it e-ROSITA} X-ray survey. The
lessons learned from this and the planned subsequent LoTSS work will
inform the even larger surveys that will be carried out with the
SKA, but, as the present paper demonstrates, the
era of big data for RLAGN surveys is already here.

\begin{acknowledgements}
We would like to thank Sarah Needleman (University of Cambridge: summer
student in 2016) and Sinan Hassan (University of Hertfordshire: BSc
investigation in 2018) for their contributions to this work.

MJH and WLW acknowledge support from the UK Science and Technology
Facilities Council (STFC) [ST/M001008/1]. JHC acknowledges support
from the STFC under grants ST/R00109X/1 and ST/R000794/1. KJD
acknowledges support from the ERC Advanced Investigator programme
NewClusters 321271. PNB and JS are grateful for support from STFC via
grant ST/M001229/1. MJJ acknowledges support from the Oxford Hintze
Centre for Astrophysical Surveys which is funded through generous
support from the Hintze Family Charitable Foundation. GG acknowledges
a CSIRO OCE Postdoctoral Fellowship. SM acknowledges funding through
the Irish Research Council New Foundations scheme and the Irish
Research Council Postgraduate Scholarship scheme. FdG is supported by
the VENI research programme with project number 1808, which is
financed by the Netherlands Organisation for Scientific Research
(NWO). IP acknowledges support from INAF under PRIN SKA/CTA
``FORECaST''. RKC is grateful for support from STFC. SPO acknowledges
financial support from the Deutsche Forschungsgemeinschaft (DFG) under
grant BR2026/23.

LOFAR, the LOw Frequency ARray designed and constructed by ASTRON, has
facilities in several countries, which are owned by various parties
(each with their own funding sources), and are collectively operated
by the International LOFAR Telescope (ILT) foundation under a joint
scientific policy. The ILT resources have benefited from the
following recent major funding sources: CNRS-INSU, Observatoire de
Paris and Universit\'e d'Orl\'eans, France; BMBF, MIWF-NRW, MPG, Germany;
Science Foundation Ireland (SFI), Department of Business, Enterprise
and Innovation (DBEI), Ireland; NWO, The Netherlands; the Science and
Technology Facilities Council, UK; Ministry of Science and Higher
Education, Poland.

Part of this work was carried out on the Dutch national
e-infrastructure with the support of the SURF Cooperative through
grant e-infra 160022 \& 160152. The LOFAR software and dedicated
reduction packages on https://github.com/apmechev/GRID\_LRT were
deployed on the e-infrastructure by the LOFAR e-infragroup, consisting
of J.\ B.\ R.\ Oonk (ASTRON \& Leiden Observatory), A.\ P.\ Mechev (Leiden
Observatory) and T. Shimwell (ASTRON) with support from N.\ Danezi
(SURFsara) and C.\ Schrijvers (SURFsara). This research has made use of
the University of Hertfordshire high-performance computing facility
(\url{http://uhhpc.herts.ac.uk/}) and the LOFAR-UK computing facility
located at the University of Hertfordshire and supported by STFC
[ST/P000096/1]. This research made use of {\sc Astropy}, a
community-developed core Python package for astronomy
\citep{AstropyCollaboration13} hosted at
\url{http://www.astropy.org/}, of {\sc Matplotlib} \citep{Hunter07},
of {\sc APLpy}, an open-source astronomical plotting package for
Python hosted at \url{http://aplpy.github.com/}, and of {\sc topcat}
and {\sc stilts} \citep{Taylor05}.

The Pan-STARRS1 Surveys (PS1) have been made possible through
contributions by the Institute for Astronomy, the University of
Hawaii, the Pan-STARRS Project Office, the Max-Planck Society and its
participating institutes, the Max Planck Institute for Astronomy,
Heidelberg, and the Max Planck Institute for Extraterrestrial Physics,
Garching, The Johns Hopkins University, Durham University, the
University of Edinburgh, the Queen's University Belfast, the
Harvard-Smithsonian Center for Astrophysics, the Las Cumbres
Observatory Global Telescope Network Incorporated, the National
Central University of Taiwan, the Space Telescope Science Institute,
and the National Aeronautics and Space Administration under Grant No.
NNX08AR22G issued through the Planetary Science Division of the NASA
Science Mission Directorate, the National Science Foundation Grant No.
AST-1238877, the University of Maryland, Eotvos Lorand University
(ELTE), and the Los Alamos National Laboratory.

Funding for SDSS-III has been provided by the Alfred P. Sloan
Foundation, the Participating Institutions, the National Science
Foundation, and the U.S. Department of Energy Office of Science. The
SDSS-III web site is \url{http://www.sdss3.org/}.

SDSS-III is managed by the Astrophysical Research Consortium for the
Participating Institutions of the SDSS-III Collaboration including the
University of Arizona, the Brazilian Participation Group, Brookhaven
National Laboratory, Carnegie Mellon University, University of
Florida, the French Participation Group, the German Participation
Group, Harvard University, the Instituto de Astrofisica de Canarias,
the Michigan State/Notre Dame/JINA Participation Group, Johns Hopkins
University, Lawrence Berkeley National Laboratory, Max Planck
Institute for Astrophysics, Max Planck Institute for Extraterrestrial
Physics, New Mexico State University, New York University, Ohio State
University, Pennsylvania State University, University of Portsmouth,
Princeton University, the Spanish Participation Group, University of
Tokyo, University of Utah, Vanderbilt University, University of
Virginia, University of Washington, and Yale University.

The National Radio Astronomy Observatory (NRAO)
is a facility of the National Science Foundation operated under
cooperative agreement by Associated Universities, Inc.

This publication makes use of data products from the {\it Wide-field
Infrared Survey Explorer}, which is a joint project of the University
of California, Los Angeles, and the Jet Propulsion
Laboratory/California Institute of Technology, and NEOWISE, which is a
project of the Jet Propulsion Laboratory/California Institute of
Technology. {\it WISE} and NEOWISE are funded by the National Aeronautics
and Space Administration.

\end{acknowledgements}

\bibliographystyle{aa}
\renewcommand{\refname}{REFERENCES}
\bibliography{../bib/mjh,../bib/cards}

\begin{thebibliography}{125}
\expandafter\ifx\csname natexlab\endcsname\relax\def\natexlab#1{#1}\fi

\bibitem[{{Allen} {et~al.}(2006){Allen}, {Dunn}, {Fabian}, {Taylor}, \&
  {Reynolds}}]{Allen+06}
{Allen}, S.~W., {Dunn}, R.~J.~H., {Fabian}, A.~C., {Taylor}, G.~B., \&
  {Reynolds}, C.~S. 2006, \mnras, 372, 21

\bibitem[{{Assef} {et~al.}(2010){Assef}, {Kochanek}, {Brodwin}, {Cool},
  {Forman}, {Gonzalez}, {Hickox}, {Jones}, {Le Floc'h}, {Moustakas}, {Murray},
  \& {Stern}}]{Assef+10}
{Assef}, R.~J., {Kochanek}, C.~S., {Brodwin}, M., {et~al.} 2010, \apj, 713, 970

\bibitem[{{Astropy Collaboration} {et~al.}(2013){Astropy Collaboration},
  {Robitaille}, {Tollerud}, {Greenfield}, {Droettboom}, {Bray}, {Aldcroft},
  {Davis}, {Ginsburg}, {Price-Whelan}, {Kerzendorf}, {Conley}, {Crighton},
  {Barbary}, {Muna}, {Ferguson}, {Grollier}, {Parikh}, {Nair}, {Unther},
  {Deil}, {Woillez}, {Conseil}, {Kramer}, {Turner}, {Singer}, {Fox}, {Weaver},
  {Zabalza}, {Edwards}, {Azalee Bostroem}, {Burke}, {Casey}, {Crawford},
  {Dencheva}, {Ely}, {Jenness}, {Labrie}, {Lim}, {Pierfederici}, {Pontzen},
  {Ptak}, {Refsdal}, {Servillat}, \& {Streicher}}]{AstropyCollaboration13}
{Astropy Collaboration}, {Robitaille}, T.~P., {Tollerud}, E.~J., {et~al.} 2013,
  \aap, 558, A33

\bibitem[{{Baldi} {et~al.}(2015){Baldi}, {Capetti}, \& {Giovannini}}]{Baldi+15}
{Baldi}, R.~D., {Capetti}, A., \& {Giovannini}, G. 2015, \aap, 576, A38

\bibitem[{{Baldwin} {et~al.}(1981){Baldwin}, {Phillips}, \&
  {Terlevich}}]{Baldwin+81}
{Baldwin}, J.~A., {Phillips}, M.~M., \& {Terlevich}, R. 1981, \pasp, 93, 5

\bibitem[{{Baldwin}(1982)}]{Baldwin82}
{Baldwin}, J.~E. 1982, in IAU Symposium, Vol.~97, Extragalactic Radio Sources,
  ed. D.~S. {Heeschen} \& C.~M. {Wade}, 21--24

\bibitem[{{Basson} \& {Alexander}(2003)}]{Basson+Alexander03}
{Basson}, J.~F. \& {Alexander}, P. 2003, \mnras, 339, 353

\bibitem[{{Becker} {et~al.}(1995){Becker}, {White}, \& {Helfand}}]{Becker+95}
{Becker}, R.~H., {White}, R.~L., \& {Helfand}, D.~J. 1995, \apj, 450, 559

\bibitem[{{Best}(2004)}]{Best04}
{Best}, P.~N. 2004, \mnras, 351, 70

\bibitem[{{Best} \& {Heckman}(2012)}]{Best+Heckman12}
{Best}, P.~N. \& {Heckman}, T.~M. 2012, \mnras, 421, 1569

\bibitem[{{Best} {et~al.}(2006){Best}, {Kaiser}, {Heckman}, \&
  {Kauffmann}}]{Best+06}
{Best}, P.~N., {Kaiser}, C.~R., {Heckman}, T.~M., \& {Kauffmann}, G. 2006,
  \mnras, 368, L67

\bibitem[{{Best} {et~al.}(2005){Best}, {Kauffmann}, {Heckman}, {Brinchmann},
  {Charlot}, {Ivezi\'c}, \& {White}}]{Best+05}
{Best}, P.~N., {Kauffmann}, G., {Heckman}, T.~M., {et~al.} 2005, \mnras, 362,
  25

\bibitem[{{Best} {et~al.}(2014){Best}, {Ker}, {Simpson}, {Rigby}, \&
  {Sabater}}]{Best+14}
{Best}, P.~N., {Ker}, L.~M., {Simpson}, C., {Rigby}, E.~E., \& {Sabater}, J.
  2014, \mnras, 445, 955

\bibitem[{{Best} {et~al.}(1997){Best}, {Longair}, \&
  {R\"ottgering}}]{Best+97-2}
{Best}, P.~N., {Longair}, M.~S., \& {R\"ottgering}, H.~J.~A. 1997, \mnras, 292,
  758

\bibitem[{{B{\^i}rzan} {et~al.}(2004){B{\^i}rzan}, {Rafferty}, {McNamara},
  {Wise}, \& {Nulsen}}]{Birzan+04}
{B{\^i}rzan}, L., {Rafferty}, D.~A., {McNamara}, B.~R., {Wise}, M.~W., \&
  {Nulsen}, P.~E.~J. 2004, \apj, 607, 800

\bibitem[{{B{\^i}rzan} {et~al.}(2012){B{\^i}rzan}, {Rafferty}, {Nulsen},
  {McNamara}, {R{\"o}ttgering}, {Wise}, \& {Mittal}}]{Birzan+12}
{B{\^i}rzan}, L., {Rafferty}, D.~A., {Nulsen}, P.~E.~J., {et~al.} 2012, \mnras,
  427, 3468

\bibitem[{{Blandford} \& {Rees}(1974)}]{Blandford+Rees74}
{Blandford}, R.~D. \& {Rees}, M.~J. 1974, \mnras, 169, 395

\bibitem[{{Blundell} {et~al.}(1999){Blundell}, {Rawlings}, \&
  {Willott}}]{Blundell+99}
{Blundell}, K.~M., {Rawlings}, S., \& {Willott}, C.~J. 1999, \aj, 117, 677

\bibitem[{{B{\"o}hringer} {et~al.}(2014){B{\"o}hringer}, {Chon}, \&
  {Collins}}]{Bohringer+14}
{B{\"o}hringer}, H., {Chon}, G., \& {Collins}, C.~A. 2014, \aap, 570, A31

\bibitem[{{Bower} {et~al.}(2006){Bower}, {Benson}, {Malbon}, {Helly}, {Frenk},
  {Baugh}, {Cole}, \& {Lacey}}]{Bower+06}
{Bower}, R.~G., {Benson}, A.~J., {Malbon}, R., {et~al.} 2006, \mnras, 370, 645

\bibitem[{{Brinchmann} {et~al.}(2004){Brinchmann}, {Charlot}, {White},
  {Tremonti}, {Kauffmann}, {Heckman}, \& {Brinkmann}}]{Brinchmann+04}
{Brinchmann}, J., {Charlot}, S., {White}, S.~D.~M., {et~al.} 2004, \mnras, 351,
  1151

\bibitem[{{Calistro Rivera} {et~al.}(2016){Calistro Rivera}, {Lusso},
  {Hennawi}, \& {Hogg}}]{Calistro-Rivera+16}
{Calistro Rivera}, G., {Lusso}, E., {Hennawi}, J.~F., \& {Hogg}, D.~W. 2016,
  \apj, 833, 98

\bibitem[{{Calistro Rivera} {et~al.}(2017){Calistro Rivera}, {Williams},
  {Hardcastle}, {Duncan}, {R{\"o}ttgering}, {Best}, {Br{\"u}ggen}, {Chy{\.z}y},
  {Conselice}, {de Gasperin}, {Engels}, {G{\"u}rkan}, {Intema}, {Jarvis},
  {Mahony}, {Miley}, {Morabito}, {Prandoni}, {Sabater}, {Smith}, {Tasse}, {van
  der Werf}, \& {White}}]{Calistro-Rivera+17}
{Calistro Rivera}, G., {Williams}, W.~L., {Hardcastle}, M.~J., {et~al.} 2017,
  \mnras, 469, 3468

\bibitem[{{Callingham} {et~al.}(2017){Callingham}, {Ekers}, {Gaensler}, {Line},
  {Hurley-Walker}, {Sadler}, {Tingay}, {Hancock}, {Bell}, {Dwarakanath}, {For},
  {Franzen}, {Hindson}, {Johnston-Hollitt}, {Kapi{\'n}ska}, {Lenc}, {McKinley},
  {Morgan}, {Offringa}, {Procopio}, {Staveley-Smith}, {Wayth}, {Wu}, \&
  {Zheng}}]{Callingham+17}
{Callingham}, J.~R., {Ekers}, R.~D., {Gaensler}, B.~M., {et~al.} 2017, \apj,
  836, 174

\bibitem[{{Cavagnolo} {et~al.}(2010){Cavagnolo}, {McNamara}, {Nulsen},
  {Carilli}, {Jones}, \& {B{\^i}rzan}}]{Cavagnolo+10}
{Cavagnolo}, K.~W., {McNamara}, B.~R., {Nulsen}, P.~E.~J., {et~al.} 2010, \apj,
  720, 1066

\bibitem[{{Chambers} {et~al.}(2016){Chambers}, {Magnier}, {Metcalfe},
  {Flewelling}, {Huber}, {Waters}, {Denneau}, {Draper}, {Farrow}, {Finkbeiner},
  {Holmberg}, {Koppenhoefer}, {Price}, {Saglia}, {Schlafly}, {Smartt},
  {Sweeney}, {Wainscoat}, {Burgett}, {Grav}, {Heasley}, {Hodapp}, {Jedicke},
  {Kaiser}, {Kudritzki}, {Luppino}, {Lupton}, {Monet}, {Morgan}, {Onaka},
  {Stubbs}, {Tonry}, {Banados}, {Bell}, {Bender}, {Bernard}, {Botticella},
  {Casertano}, {Chastel}, {Chen}, {Chen}, {Cole}, {Deacon}, {Frenk},
  {Fitzsimmons}, {Gezari}, {Goessl}, {Goggia}, {Goldman}, {Grebel}, {Hambly},
  {Hasinger}, {Heavens}, {Heckman}, {Henderson}, {Henning}, {Holman}, {Hopp},
  {Ip}, {Isani}, {Keyes}, {Koekemoer}, {Kotak}, {Long}, {Lucey}, {Liu},
  {Martin}, {McLean}, {Morganson}, {Murphy}, {Nieto-Santisteban}, {Norberg},
  {Peacock}, {Pier}, {Postman}, {Primak}, {Rae}, {Rest}, {Riess}, {Riffeser},
  {Rix}, {Roser}, {Schilbach}, {Schultz}, {Scolnic}, {Szalay}, {Seitz},
  {Shiao}, {Small}, {Smith}, {Soderblom}, {Taylor}, {Thakar}, {Thiel},
  {Thilker}, {Urata}, {Valenti}, {Walter}, {Watters}, {Werner}, {White},
  {Wood-Vasey}, \& {Wyse}}]{Chambers+16}
{Chambers}, K.~C., {Magnier}, E.~A., {Metcalfe}, N., {et~al.} 2016, ArXiv
  e-prints [\eprint[arXiv]{1612.05560}]

\bibitem[{{Ching} {et~al.}(2017){Ching}, {Croom}, {Sadler}, {Robotham},
  {Brough}, {Baldry}, {Bland-Hawthorn}, {Colless}, {Driver}, {Holwerda},
  {Hopkins}, {Jarvis}, {Johnston}, {Kelvin}, {Liske}, {Loveday}, {Norberg},
  {Pracy}, {Steele}, {Thomas}, \& {Wang}}]{Ching+17}
{Ching}, J.~H.~Y., {Croom}, S.~M., {Sadler}, E.~M., {et~al.} 2017, \mnras, 469,
  4584

\bibitem[{{Condon}(1989)}]{Condon89}
{Condon}, J.~J. 1989, \apj, 338, 13

\bibitem[{{Condon} {et~al.}(1998){Condon}, {Cotton}, {Greisen}, {Yin},
  {Perley}, {Taylor}, \& {Broderick}}]{Condon+98}
{Condon}, J.~J., {Cotton}, W.~D., {Greisen}, E.~W., {et~al.} 1998, \aj, 115,
  1693

\bibitem[{{Croston} {et~al.}(2005){Croston}, {Hardcastle}, \&
  {Birkinshaw}}]{Croston+05}
{Croston}, J.~H., {Hardcastle}, M.~J., \& {Birkinshaw}, M. 2005, \mnras, 357,
  279

\bibitem[{{Croston} {et~al.}(2018{\natexlab{a}}){Croston}, {Ineson}, \&
  {Hardcastle}}]{Croston+18}
{Croston}, J.~H., {Ineson}, J., \& {Hardcastle}, M.~J. 2018{\natexlab{a}},
  \mnras, 476, 1614

\bibitem[{{Croston} {et~al.}(2018{\natexlab{b}})}]{Croston+18b}
{Croston}, J.~H. {et~al.} 2018{\natexlab{b}}, \aap\ in prep

\bibitem[{{Croton} {et~al.}(2006){Croton}, {Springel}, {White}, {De~Lucia},
  {Frenk}, {Gao}, {Jenkins}, {Kauffmann}, {Navarro}, \& {Yoshida}}]{Croton+06}
{Croton}, D., {Springel}, V., {White}, S.~D.~M., {et~al.} 2006, \mnras, 365,
  111

\bibitem[{{Croton} {et~al.}(2016){Croton}, {Stevens}, {Tonini}, {Garel},
  {Bernyk}, {Bibiano}, {Hodkinson}, {Mutch}, {Poole}, \& {Shattow}}]{Croton+16}
{Croton}, D.~J., {Stevens}, A.~R.~H., {Tonini}, C., {et~al.} 2016, \apjs, 222,
  22

\bibitem[{{da Cunha} {et~al.}(2008){da Cunha}, {Charlot}, \&
  {Elbaz}}]{daCunha+08}
{da Cunha}, E., {Charlot}, S., \& {Elbaz}, D. 2008, \mnras, 388, 1595

\bibitem[{{Daly} {et~al.}(2012){Daly}, {Sprinkle}, {O'Dea}, {Kharb}, \&
  {Baum}}]{Daly+12}
{Daly}, R.~A., {Sprinkle}, T.~B., {O'Dea}, C.~P., {Kharb}, P., \& {Baum}, S.~A.
  2012, \mnras, 423, 2498

\bibitem[{{Duncan} {et~al.}(2018)}]{Duncan+18}
{Duncan}, K.~J. {et~al.} 2018, \aap\ submitted

\bibitem[{{Eilek} \& {Owen}(2006)}]{Eilek+Owen06}
{Eilek}, J.~A. \& {Owen}, F.~N. 2006, in {Heating vs. cooling in galaxies and
  clusters of galaxies}, ed. {B\"ohringer H., Pratt G.W., Finoguenov A. \&
  Schuecker P.} ({Springer-Verlag, Heidelberg}), 0612111

\bibitem[{{Eisenstein} {et~al.}(2011){Eisenstein}, {Weinberg}, {Agol},
  {Aihara}, {Allende Prieto}, {Anderson}, {Arns}, {Aubourg}, {Bailey},
  {Balbinot}, \& et~al.}]{Eisenstein+11}
{Eisenstein}, D.~J., {Weinberg}, D.~H., {Agol}, E., {et~al.} 2011, \aj, 142, 72

\bibitem[{{English} {et~al.}(2016){English}, {Hardcastle}, \&
  {Krause}}]{English+16}
{English}, W., {Hardcastle}, M.~J., \& {Krause}, M.~G.~H. 2016, \mnras, 461,
  2025

\bibitem[{{Fabian} {et~al.}(1984){Fabian}, {Nulsen}, \&
  {Canizares}}]{Fabian+84}
{Fabian}, A.~C., {Nulsen}, P.~E.~J., \& {Canizares}, C.~R. 1984, \nat, 310, 30

\bibitem[{{Fabian} {et~al.}(2000){Fabian}, {Sanders}, {Ettori}, {Taylor},
  {Allen}, {Crawford}, {Iwasawa}, {Johnstone}, \& {Ogle}}]{Fabian+00}
{Fabian}, A.~C., {Sanders}, J.~S., {Ettori}, S., {et~al.} 2000, \mnras, 318,
  L65

\bibitem[{{Gaspari} {et~al.}(2013){Gaspari}, {Ruszkowski}, \&
  {Oh}}]{Gaspari+13}
{Gaspari}, M., {Ruszkowski}, M., \& {Oh}, S.~P. 2013, \mnras, 432, 3401

\bibitem[{{Girardi} \& {Giuricin}(2000)}]{Girardi+Giuricin00}
{Girardi}, M. \& {Giuricin}, G. 2000, \apj, 540, 45

\bibitem[{{Godfrey} {et~al.}(2017){Godfrey}, {Morganti}, \&
  {Brienza}}]{Godfrey+17}
{Godfrey}, L.~E.~H., {Morganti}, R., \& {Brienza}, M. 2017, \mnras, 471, 891

\bibitem[{{Godfrey} \& {Shabala}(2016)}]{Godfrey+Shabala16}
{Godfrey}, L.~E.~H. \& {Shabala}, S.~S. 2016, \mnras, 456, 1172

\bibitem[{{G{\"u}rkan} {et~al.}(2014){G{\"u}rkan}, {Hardcastle}, \&
  {Jarvis}}]{Gurkan+14}
{G{\"u}rkan}, G., {Hardcastle}, M.~J., \& {Jarvis}, M.~J. 2014, \mnras, 438,
  1149

\bibitem[{{G{\"u}rkan} {et~al.}(2018{\natexlab{a}}){G{\"u}rkan}, {Hardcastle},
  {Smith}, {Best}, {Bourne}, {Calistro-Rivera}, {Heald}, {Jarvis}, {Prandoni},
  {R{\"o}ttgering}, {Sabater}, {Shimwell}, {Tasse}, \& {Williams}}]{Gurkan+18}
{G{\"u}rkan}, G., {Hardcastle}, M.~J., {Smith}, D.~J.~B., {et~al.}
  2018{\natexlab{a}}, \mnras, 475, 3010

\bibitem[{{G{\"u}rkan} {et~al.}(2018{\natexlab{b}})}]{Gurkan+18b}
{G{\"u}rkan}, G. {et~al.} 2018{\natexlab{b}}, \aap\ in prep

\bibitem[{{Hardcastle}(2018)}]{Hardcastle18}
{Hardcastle}, M.~J. 2018, \mnras, 475, 2768

\bibitem[{{Hardcastle} {et~al.}(2002){Hardcastle}, {Birkinshaw}, {Cameron},
  {Harris}, {Looney}, \& {Worrall}}]{Hardcastle+02}
{Hardcastle}, M.~J., {Birkinshaw}, M., {Cameron}, R., {et~al.} 2002, \apj, 581,
  948

\bibitem[{{Hardcastle} \& {Croston}(2010)}]{Hardcastle+Croston10}
{Hardcastle}, M.~J. \& {Croston}, J.~H. 2010, \mnras, 404, 2018

\bibitem[{{Hardcastle} {et~al.}(2007){Hardcastle}, {Evans}, \&
  {Croston}}]{Hardcastle+07-2}
{Hardcastle}, M.~J., {Evans}, D.~A., \& {Croston}, J.~H. 2007, \mnras, 376,
  1849

\bibitem[{{Hardcastle} {et~al.}(2009){Hardcastle}, {Evans}, \&
  {Croston}}]{Hardcastle+09}
{Hardcastle}, M.~J., {Evans}, D.~A., \& {Croston}, J.~H. 2009, \mnras, 396,
  1929

\bibitem[{{Hardcastle} {et~al.}(2016){Hardcastle}, {G{\"u}rkan}, {van Weeren},
  {Williams}, {Best}, {de Gasperin}, {Rafferty}, {Read}, {Sabater}, {Shimwell},
  {Smith}, {Tasse}, {Bourne}, {Brienza}, {Br{\"u}ggen}, {Brunetti},
  {Chy{\.z}y}, {Conway}, {Dunne}, {Eales}, {Maddox}, {Jarvis}, {Mahony},
  {Morganti}, {Prandoni}, {R{\"o}ttgering}, {Valiante}, \&
  {White}}]{Hardcastle+16}
{Hardcastle}, M.~J., {G{\"u}rkan}, G., {van Weeren}, R.~J., {et~al.} 2016,
  \mnras, 462, 1910

\bibitem[{{Hardcastle} \& {Krause}(2013)}]{Hardcastle+Krause13}
{Hardcastle}, M.~J. \& {Krause}, M.~G.~H. 2013, \mnras, 430, 174

\bibitem[{{Hardcastle} \& {Krause}(2014)}]{Hardcastle+Krause14}
{Hardcastle}, M.~J. \& {Krause}, M.~G.~H. 2014, \mnras, 443, 1482

\bibitem[{{Hardcastle} {et~al.}(2012){Hardcastle}, {Massaro}, {Harris}, {Baum},
  {Bianchi}, {Chiaberge}, {Morganti}, {O'Dea}, \&
  {Siemiginowska}}]{Hardcastle+12}
{Hardcastle}, M.~J., {Massaro}, F., {Harris}, D.~E., {et~al.} 2012, \mnras,
  424, 1774

\bibitem[{{Hardcastle} \& {Worrall}(1999)}]{Hardcastle+Worrall99}
{Hardcastle}, M.~J. \& {Worrall}, D.~M. 1999, \mnras, 309, 969

\bibitem[{{Harvanek} {et~al.}(2001){Harvanek}, {Ellingson}, {Stocke}, \&
  {Rhee}}]{Harvanek+01}
{Harvanek}, M., {Ellingson}, E., {Stocke}, J.~T., \& {Rhee}, G. 2001, \aj, 122,
  2874

\bibitem[{{Heckman} \& {Best}(2014)}]{Heckman+Best14}
{Heckman}, T.~M. \& {Best}, P.~N. 2014, \araa, 52, 589

\bibitem[{{Heinz} {et~al.}(2006){Heinz}, {Br{\"u}ggen}, {Young}, \&
  {Levesque}}]{Heinz+06}
{Heinz}, S., {Br{\"u}ggen}, M., {Young}, A., \& {Levesque}, E. 2006, \mnras,
  373, L65

\bibitem[{{Herpich} {et~al.}(2016){Herpich}, {Mateus}, {Stasi{\'n}ska}, {Cid
  Fernandes}, \& {Vale Asari}}]{Herpich+16}
{Herpich}, F., {Mateus}, A., {Stasi{\'n}ska}, G., {Cid Fernandes}, R., \& {Vale
  Asari}, N. 2016, \mnras, 462, 1826

\bibitem[{{Hill} {et~al.}(2008){Hill}, {Gebhardt}, {Komatsu}, {Drory},
  {MacQueen}, {Adams}, {Blanc}, {Koehler}, {Rafal}, {Roth}, {Kelz}, {Gronwall},
  {Ciardullo}, \& {Schneider}}]{Hill+08}
{Hill}, G.~J., {Gebhardt}, K., {Komatsu}, E., {et~al.} 2008, in Astronomical
  Society of the Pacific Conference Series, Vol. 399, Panoramic Views of Galaxy
  Formation and Evolution, ed. T.~{Kodama}, T.~{Yamada}, \& K.~{Aoki}, 115

\bibitem[{{Hill} \& {Lilly}(1991)}]{Hill+Lilly91}
{Hill}, G.~J. \& {Lilly}, S.~J. 1991, \apj, 367, 1

\bibitem[{Hunter(2007)}]{Hunter07}
Hunter, J.~D. 2007, Computing In Science \& Engineering, 9, 90

\bibitem[{{Ineson} {et~al.}(2015){Ineson}, {Croston}, {Hardcastle}, {Kraft},
  {Evans}, \& {Jarvis}}]{Ineson+15}
{Ineson}, J., {Croston}, J.~H., {Hardcastle}, M.~J., {et~al.} 2015, \mnras,
  453, 2682

\bibitem[{{Ineson} {et~al.}(2017){Ineson}, {Croston}, {Hardcastle}, \&
  {Mingo}}]{Ineson+17}
{Ineson}, J., {Croston}, J.~H., {Hardcastle}, M.~J., \& {Mingo}, B. 2017,
  \mnras, 467, 1586

\bibitem[{{Jarrett} {et~al.}(2011){Jarrett}, {Cohen}, {Masci}, {Wright},
  {Stern}, {Benford}, {Blain}, {Carey}, {Cutri}, {Eisenhardt}, {Lonsdale},
  {Mainzer}, {Marsh}, {Padgett}, {Petty}, {Ressler}, {Skrutskie}, {Stanford},
  {Surace}, {Tsai}, {Wheelock}, \& {Yan}}]{Jarrett+11}
{Jarrett}, T.~H., {Cohen}, M., {Masci}, F., {et~al.} 2011, \apj, 735, 112

\bibitem[{{Jarvis} {et~al.}(2016){Jarvis}, {Taylor}, {Agudo}, {Allison},
  {Deane}, {Frank}, {Gupta}, {Heywood}, {Maddox}, {McAlpine}, {Santos},
  {Scaife}, {Vaccari}, {Zwart}, {Adams}, {Bacon}, {Baker}, {Bassett}, {Best},
  {Beswick}, {Blyth}, {Brown}, {Bruggen}, {Cluver}, {Colafrancesco}, {Cotter},
  {Cress}, {Dav{\'e}}, {Ferrari}, {Hardcastle}, {Hale}, {Harrison}, {Hatfield},
  {Klockner}, {Kolwa}, {Malefahlo}, {Marubini}, {Mauch}, {Moodley}, {Morganti},
  {Norris}, {Peters}, {Prandoni}, {Prescott}, {Oliver}, {Oozeer}, {Rottgering},
  {Seymour}, {Simpson}, {Smirnov}, \& {Smith}}]{Jarvis+17}
{Jarvis}, M., {Taylor}, R., {Agudo}, I., {et~al.} 2016, in Proceedings of
  MeerKAT Science: On the Pathway to the SKA. 25-27 May, 2016 Stellenbosch,
  South Africa (MeerKAT2016). Online at
  \url{https://pos.sissa.it/cgi-bin/reader/conf.cgi?confid=277}, 6

\bibitem[{{Kaiser} \& {Alexander}(1997)}]{Kaiser+Alexander97}
{Kaiser}, C.~R. \& {Alexander}, P. 1997, \mnras, 286, 215

\bibitem[{{Kaiser} {et~al.}(1997){Kaiser}, {Dennett-Thorpe}, \&
  {Alexander}}]{Kaiser+97}
{Kaiser}, C.~R., {Dennett-Thorpe}, J., \& {Alexander}, P. 1997, \mnras, 292,
  723

\bibitem[{{Kataoka} \& {Stawarz}(2005)}]{Kataoka+Stawarz05}
{Kataoka}, J. \& {Stawarz}, {\L}. 2005, \apj, 622, 797

\bibitem[{{Kauffmann} {et~al.}(2003){Kauffmann}, {Heckman}, {Tremonti},
  {Brinchmann}, {Charlot}, {White}, {Ridgway}, {Brinkmann}, {Fukugita}, {Hall},
  {Ivezi{\'c}}, {Richards}, \& {Schneider}}]{Kauffmann+03}
{Kauffmann}, G., {Heckman}, T.~M., {Tremonti}, C., {et~al.} 2003, \mnras, 346,
  1055

\bibitem[{{Kewley} {et~al.}(2006){Kewley}, {Groves}, {Kauffmann}, \&
  {Heckman}}]{Kewley+06}
{Kewley}, L.~J., {Groves}, B., {Kauffmann}, G., \& {Heckman}, T. 2006, \mnras,
  372, 961

\bibitem[{{Krause}(2005)}]{Krause05}
{Krause}, M. 2005, \aap, 431, 45

\bibitem[{{Laing} {et~al.}(1983){Laing}, {Riley}, \& {Longair}}]{Laing+83}
{Laing}, R.~A., {Riley}, J.~M., \& {Longair}, M.~S. 1983, \mnras, 204, 151

\bibitem[{{Lilly} \& {Longair}(1984)}]{Lilly+Longair84}
{Lilly}, S.~J. \& {Longair}, M.~S. 1984, \mnras, 211, 833

\bibitem[{{Lilly} {et~al.}(1984){Lilly}, {McLean}, \& {Longair}}]{Lilly+84}
{Lilly}, S.~J., {McLean}, I.~S., \& {Longair}, M.~S. 1984, \mnras, 209, 401

\bibitem[{{Luo} \& {Sadler}(2010)}]{Luo+Sadler10}
{Luo}, Q. \& {Sadler}, E.~M. 2010, \apj, 713, 398

\bibitem[{{Mainzer} {et~al.}(2011){Mainzer}, {Bauer}, {Grav}, {Masiero},
  {Cutri}, {Dailey}, {Eisenhardt}, {McMillan}, {Wright}, {Walker}, {Jedicke},
  {Spahr}, {Tholen}, {Alles}, {Beck}, {Brandenburg}, {Conrow}, {Evans},
  {Fowler}, {Jarrett}, {Marsh}, {Masci}, {McCallon}, {Wheelock}, {Wittman},
  {Wyatt}, {DeBaun}, {Elliott}, {Elsbury}, {Gautier}, {Gomillion}, {Leisawitz},
  {Maleszewski}, {Micheli}, \& {Wilkins}}]{Mainzer+11}
{Mainzer}, A., {Bauer}, J., {Grav}, T., {et~al.} 2011, \apj, 731, 53

\bibitem[{{Mateos} {et~al.}(2012){Mateos}, {Alonso-Herrero}, {Carrera},
  {Blain}, {Watson}, {Barcons}, {Braito}, {Severgnini}, {Donley}, \&
  {Stern}}]{Mateos+12}
{Mateos}, S., {Alonso-Herrero}, A., {Carrera}, F.~J., {et~al.} 2012, \mnras,
  426, 3271

\bibitem[{{McNamara} \& {Nulsen}(2012)}]{McNamara+Nulsen12}
{McNamara}, B.~R. \& {Nulsen}, P.~E.~J. 2012, New Journal of Physics, 14,
  055023

\bibitem[{{Mendygral} {et~al.}(2012){Mendygral}, {Jones}, \&
  {Dolag}}]{Mendygral+12}
{Mendygral}, P.~J., {Jones}, T.~W., \& {Dolag}, K. 2012, \apj, 750, 166

\bibitem[{{Mingo} {et~al.}(2016){Mingo}, {Watson}, {Rosen}, {Hardcastle},
  {Ruiz}, {Blain}, {Carrera}, {Mateos}, {Pineau}, \& {Stewart}}]{Mingo+16}
{Mingo}, B., {Watson}, M.~G., {Rosen}, S.~R., {et~al.} 2016, \mnras, 462, 2631

\bibitem[{{Mocz} {et~al.}(2011){Mocz}, {Fabian}, \& {Blundell}}]{Mocz+11}
{Mocz}, P., {Fabian}, A.~C., \& {Blundell}, K.~M. 2011, \mnras, 413, 1107

\bibitem[{{Mohan} \& {Rafferty}(2015)}]{Mohan+Rafferty15}
{Mohan}, N. \& {Rafferty}, D. 2015, {PyBDSM: Python Blob Detection and Source
  Measurement}, Astrophysics Source Code Library

\bibitem[{{Morganti} {et~al.}(2005){Morganti}, {Oosterloo}, {Tadhunter}, {van
  Moorsel}, \& {Emonts}}]{Morganti+05}
{Morganti}, R., {Oosterloo}, T.~A., {Tadhunter}, C.~N., {van Moorsel}, G., \&
  {Emonts}, B. 2005, \aap, 439, 521

\bibitem[{{Nesvadba} {et~al.}(2008){Nesvadba}, {Lehnert}, {De~Breuck},
  {Gilbert}, \& {van~Breugel}}]{Nesvadba+08}
{Nesvadba}, N.~P.~H., {Lehnert}, M.~D., {De~Breuck}, C., {Gilbert}, A.~M., \&
  {van~Breugel}, W. 2008, \aap, 491, 407

\bibitem[{{Norris} {et~al.}(2011){Norris}, {Hopkins}, {Afonso}, {Brown},
  {Condon}, {Dunne}, {Feain}, {Hollow}, {Jarvis}, {Johnston-Hollitt}, {Lenc},
  {Middelberg}, {Padovani}, {Prandoni}, {Rudnick}, {Seymour}, {Umana},
  {Andernach}, {Alexander}, {Appleton}, {Bacon}, {Banfield}, {Becker}, {Brown},
  {Ciliegi}, {Jackson}, {Eales}, {Edge}, {Gaensler}, {Giovannini}, {Hales},
  {Hancock}, {Huynh}, {Ibar}, {Ivison}, {Kennicutt}, {Kimball}, {Koekemoer},
  {Koribalski}, {L{\'o}pez-S{\'a}nchez}, {Mao}, {Murphy}, {Messias},
  {Pimbblet}, {Raccanelli}, {Randall}, {Reiprich}, {Roseboom},
  {R{\"o}ttgering}, {Saikia}, {Sharp}, {Slee}, {Smail}, {Thompson}, {Urquhart},
  {Wall}, \& {Zhao}}]{Norris+11}
{Norris}, R.~P., {Hopkins}, A.~M., {Afonso}, J., {et~al.} 2011, \pasa, 28, 215

\bibitem[{{O'Dea} {et~al.}(2009){O'Dea}, {Daly}, {Kharb}, {Freeman}, \&
  {Baum}}]{O'Dea+09}
{O'Dea}, C.~P., {Daly}, R.~A., {Kharb}, P., {Freeman}, K.~A., \& {Baum}, S.~A.
  2009, \aap, 494, 471

\bibitem[{{O'Sullivan} {et~al.}(2018)}]{OSullivan+18}
{O'Sullivan}, S. {et~al.} 2018, \aap\ in prep

\bibitem[{{Pizzolato} \& {Soker}(2005)}]{Pizzolato+Soker05}
{Pizzolato}, F. \& {Soker}, N. 2005, \apj, 632, 821

\bibitem[{{Prestage} \& {Peacock}(1988)}]{Prestage+Peacock88}
{Prestage}, R.~M. \& {Peacock}, J.~A. 1988, \mnras, 230, 131

\bibitem[{{Reiprich} \& {B{\"o}hringer}(2002)}]{Reiprich+Bohringer02}
{Reiprich}, T.~H. \& {B{\"o}hringer}, H. 2002, \apj, 567, 716

\bibitem[{{Reynolds} {et~al.}(2002){Reynolds}, {Heinz}, \&
  {Begelman}}]{Reynolds+02}
{Reynolds}, C.~S., {Heinz}, S., \& {Begelman}, M.~C. 2002, \mnras, 332, 271

\bibitem[{{Rovilos} {et~al.}(2014){Rovilos}, {Georgantopoulos}, {Akylas},
  {Aird}, {Alexander}, {Comastri}, {Del Moro}, {Gandhi}, {Georgakakis},
  {Harrison}, \& {Mullaney}}]{Rovilos+14}
{Rovilos}, E., {Georgantopoulos}, I., {Akylas}, A., {et~al.} 2014, \mnras, 438,
  494

\bibitem[{{Russell} {et~al.}(2017){Russell}, {McNamara}, {Fabian}, {Nulsen},
  {Combes}, {Edge}, {Hogan}, {McDonald}, {Salom{\'e}}, {Tremblay}, \&
  {Vantyghem}}]{Russell+17}
{Russell}, H.~R., {McNamara}, B.~R., {Fabian}, A.~C., {et~al.} 2017, \mnras,
  472, 4024

\bibitem[{{Sabater} {et~al.}(2018)}]{Sabater+18}
{Sabater}, P. {et~al.} 2018, \aap\ in prep

\bibitem[{{Sadler} {et~al.}(2014){Sadler}, {Ekers}, {Mahony}, {Mauch}, \&
  {Murphy}}]{Sadler+14}
{Sadler}, E.~M., {Ekers}, R.~D., {Mahony}, E.~K., {Mauch}, T., \& {Murphy}, T.
  2014, \mnras, 438, 796

\bibitem[{{Sakelliou} {et~al.}(2002){Sakelliou}, {Peterson}, {Tamura},
  {Paerels}, {Kaastra}, {Belsole}, {B{\"o}hringer}, {Branduardi-Raymont},
  {Ferrigno}, {den Herder}, {Kennea}, {Mushotzky}, {Vestrand}, \&
  {Worrall}}]{Sakelliou+02}
{Sakelliou}, I., {Peterson}, J.~R., {Tamura}, T., {et~al.} 2002, \aap, 391, 903

\bibitem[{{Schaye} {et~al.}(2015){Schaye}, {Crain}, {Bower}, {Furlong},
  {Schaller}, {Theuns}, {Dalla Vecchia}, {Frenk}, {McCarthy}, {Helly},
  {Jenkins}, {Rosas-Guevara}, {White}, {Baes}, {Booth}, {Camps}, {Navarro},
  {Qu}, {Rahmati}, {Sawala}, {Thomas}, \& {Trayford}}]{Schaye+15}
{Schaye}, J., {Crain}, R.~A., {Bower}, R.~G., {et~al.} 2015, \mnras, 446, 521

\bibitem[{{Scheuer}(1974)}]{Scheuer74}
{Scheuer}, P.~A.~G. 1974, \mnras, 166, 513

\bibitem[{{Schlafly} \& {Finkbeiner}(2011)}]{Schlafly+Finkbeiner11}
{Schlafly}, E.~F. \& {Finkbeiner}, D.~P. 2011, \apj, 737, 103

\bibitem[{{Schmidt}(1968)}]{Schmidt68}
{Schmidt}, M. 1968, \apj, 151, 393

\bibitem[{{Secrest} {et~al.}(2015){Secrest}, {Dudik}, {Dorland}, {Zacharias},
  {Makarov}, {Fey}, {Frouard}, \& {Finch}}]{Secrest+15}
{Secrest}, N.~J., {Dudik}, R.~P., {Dorland}, B.~N., {et~al.} 2015, \apjs, 221,
  12

\bibitem[{{Shabala} {et~al.}(2017){Shabala}, {Deller}, {Kaviraj}, {Middelberg},
  {Turner}, {Ting}, {Allison}, \& {Davis}}]{Shabala+17}
{Shabala}, S.~S., {Deller}, A., {Kaviraj}, S., {et~al.} 2017, \mnras, 464, 4706

\bibitem[{{Shimwell} {et~al.}(2017){Shimwell}, {R{\"o}ttgering}, {Best},
  {Williams}, {Dijkema}, {de Gasperin}, {Hardcastle}, {Heald}, {Hoang},
  {Horneffer}, {Intema}, {Mahony}, {Mandal}, {Mechev}, {Morabito}, {Oonk},
  {Rafferty}, {Retana-Montenegro}, {Sabater}, {Tasse}, {van Weeren},
  {Br{\"u}ggen}, {Brunetti}, {Chy{\.z}y}, {Conway}, {Haverkorn}, {Jackson},
  {Jarvis}, {McKean}, {Miley}, {Morganti}, {White}, {Wise}, {van Bemmel},
  {Beck}, {Brienza}, {Bonafede}, {Calistro Rivera}, {Cassano}, {Clarke},
  {Cseh}, {Deller}, {Drabent}, {van Driel}, {Engels}, {Falcke}, {Ferrari},
  {Fr{\"o}hlich}, {Garrett}, {Harwood}, {Heesen}, {Hoeft}, {Horellou},
  {Israel}, {Kapi{\'n}ska}, {Kunert-Bajraszewska}, {McKay}, {Mohan},
  {Orr{\'u}}, {Pizzo}, {Prandoni}, {Schwarz}, {Shulevski}, {Sipior}, {Smith},
  {Sridhar}, {Steinmetz}, {Stroe}, {Varenius}, {van der Werf}, {Zensus}, \&
  {Zwart}}]{Shimwell+17}
{Shimwell}, T.~W., {R{\"o}ttgering}, H.~J.~A., {Best}, P.~N., {et~al.} 2017,
  \aap, 598, A104

\bibitem[{{Shimwell} {et~al.}(2018)}]{Shimwell+18}
{Shimwell}, T.~W. {et~al.} 2018, \aap\ submitted

\bibitem[{{Smith} {et~al.}(2016){Smith}, {Best}, {Duncan}, {Hatch}, {Jarvis},
  {R{\"o}ttgering}, {Simpson}, {Stott}, {Cochrane}, {Coppin}, {Dannerbauer},
  {Davis}, {Geach}, {Hale}, {Hardcastle}, {Hatfield}, {Houghton}, {Maddox},
  {McGee}, {Morabito}, {Nisbet}, {Pandey-Pommier}, {Prandoni}, {Saxena},
  {Shimwell}, {Tarr}, {van Bemmel}, {Verma}, {White}, \& {Williams}}]{Smith+16}
{Smith}, D.~J.~B., {Best}, P.~N., {Duncan}, K.~J., {et~al.} 2016, in SF2A-2016:
  Proceedings of the Annual meeting of the French Society of Astronomy and
  Astrophysics, ed. C.~{Reyl{\'e}}, J.~{Richard}, L.~{Cambr{\'e}sy},
  M.~{Deleuil}, E.~{P{\'e}contal}, L.~{Tresse}, \& I.~{Vauglin}, 271--280

\bibitem[{{Smith} {et~al.}(2012){Smith}, {Dunne}, {da Cunha}, {Rowlands},
  {Maddox}, {Gomez}, {Bonfield}, {Charlot}, {Driver}, {Popescu}, {Tuffs},
  {Dunlop}, {Jarvis}, {Seymour}, {Symeonidis}, {Baes}, {Bourne}, {Clements},
  {Cooray}, {De Zotti}, {Dye}, {Eales}, {Scott}, {Verma}, {van der Werf},
  {Andrae}, {Auld}, {Buttiglione}, {Cava}, {Dariush}, {Fritz}, {Hopwood},
  {Ibar}, {Ivison}, {Kelvin}, {Madore}, {Pohlen}, {Rigby}, {Robotham},
  {Seibert}, \& {Temi}}]{Smith+12}
{Smith}, D.~J.~B., {Dunne}, L., {da Cunha}, E., {et~al.} 2012, \mnras, 427, 703

\bibitem[{{Smol{\v c}i{\'c}} {et~al.}(2017){Smol{\v c}i{\'c}}, {Delvecchio},
  {Zamorani}, {Baran}, {Novak}, {Delhaize}, {Schinnerer}, {Berta}, {Bondi},
  {Ciliegi}, {Capak}, {Civano}, {Karim}, {Le Fevre}, {Ilbert}, {Laigle},
  {Marchesi}, {McCracken}, {Tasca}, {Salvato}, \& {Vardoulaki}}]{Smolcic+17}
{Smol{\v c}i{\'c}}, V., {Delvecchio}, I., {Zamorani}, G., {et~al.} 2017, \aap,
  602, A2

\bibitem[{{Stern} {et~al.}(2012){Stern}, {Assef}, {Benford}, {Blain}, {Cutri},
  {Dey}, {Eisenhardt}, {Griffith}, {Jarrett}, {Lake}, {Masci}, {Petty},
  {Stanford}, {Tsai}, {Wright}, {Yan}, {Harrison}, \& {Madsen}}]{Stern+12}
{Stern}, D., {Assef}, R.~J., {Benford}, D.~J., {et~al.} 2012, \apj, 753, 30

\bibitem[{{Taylor}(2005)}]{Taylor05}
{Taylor}, M.~B. 2005, in Astronomical Society of the Pacific Conference Series,
  Vol. 347, Astronomical Data Analysis Software and Systems XIV, ed.
  P.~{Shopbell}, M.~{Britton}, \& R.~{Ebert}, 29

\bibitem[{{Turner} {et~al.}(2018){Turner}, {Rogers}, {Shabala}, \&
  {Krause}}]{Turner+18}
{Turner}, R.~J., {Rogers}, J.~G., {Shabala}, S.~S., \& {Krause}, M.~G.~H. 2018,
  \mnras, 473, 4179

\bibitem[{{Turner} \& {Shabala}(2015)}]{Turner+Shabala15}
{Turner}, R.~J. \& {Shabala}, S.~S. 2015, \apj, 806, 59

\bibitem[{{van Haarlem} {et~al.}(2013){van Haarlem}, {Wise}, {Gunst}, {Heald},
  {McKean}, {Hessels}, {de Bruyn}, {Nijboer}, {Swinbank}, {Fallows},
  {Brentjens}, {Nelles}, {Beck}, {Falcke}, {Fender}, {H{\"o}randel},
  {Koopmans}, {Mann}, {Miley}, {R{\"o}ttgering}, {Stappers}, {Wijers},
  {Zaroubi}, {van den Akker}, {Alexov}, {Anderson}, {Anderson}, {van Ardenne},
  {Arts}, {Asgekar}, {Avruch}, {Batejat}, {B{\"a}hren}, {Bell}, {Bell}, {van
  Bemmel}, {Bennema}, {Bentum}, {Bernardi}, {Best}, {B{\^i}rzan}, {Bonafede},
  {Boonstra}, {Braun}, {Bregman}, {Breitling}, {van de Brink}, {Broderick},
  {Broekema}, {Brouw}, {Br{\"u}ggen}, {Butcher}, {van Cappellen}, {Ciardi},
  {Coenen}, {Conway}, {Coolen}, {Corstanje}, {Damstra}, {Davies}, {Deller},
  {Dettmar}, {van Diepen}, {Dijkstra}, {Donker}, {Doorduin}, {Dromer}, {Drost},
  {van Duin}, {Eisl{\"o}ffel}, {van Enst}, {Ferrari}, {Frieswijk}, {Gankema},
  {Garrett}, {de Gasperin}, {Gerbers}, {de Geus}, {Grie{\ss}meier}, {Grit},
  {Gruppen}, {Hamaker}, {Hassall}, {Hoeft}, {Holties}, {Horneffer}, {van der
  Horst}, {van Houwelingen}, {Huijgen}, {Iacobelli}, {Intema}, {Jackson},
  {Jelic}, {de Jong}, {Juette}, {Kant}, {Karastergiou}, {Koers}, {Kollen},
  {Kondratiev}, {Kooistra}, {Koopman}, {Koster}, {Kuniyoshi}, {Kramer},
  {Kuper}, {Lambropoulos}, {Law}, {van Leeuwen}, {Lemaitre}, {Loose}, {Maat},
  {Macario}, {Markoff}, {Masters}, {McFadden}, {McKay-Bukowski}, {Meijering},
  {Meulman}, {Mevius}, {Middelberg}, {Millenaar}, {Miller-Jones}, {Mohan},
  {Mol}, {Morawietz}, {Morganti}, {Mulcahy}, {Mulder}, {Munk}, {Nieuwenhuis},
  {van Nieuwpoort}, {Noordam}, {Norden}, {Noutsos}, {Offringa}, {Olofsson},
  {Omar}, {Orr{\'u}}, {Overeem}, {Paas}, {Pandey-Pommier}, {Pandey}, {Pizzo},
  {Polatidis}, {Rafferty}, {Rawlings}, {Reich}, {de Reijer}, {Reitsma},
  {Renting}, {Riemers}, {Rol}, {Romein}, {Roosjen}, {Ruiter}, {Scaife}, {van
  der Schaaf}, {Scheers}, {Schellart}, {Schoenmakers}, {Schoonderbeek},
  {Serylak}, {Shulevski}, {Sluman}, {Smirnov}, {Sobey}, {Spreeuw}, {Steinmetz},
  {Sterks}, {Stiepel}, {Stuurwold}, {Tagger}, {Tang}, {Tasse}, {Thomas},
  {Thoudam}, {Toribio}, {van der Tol}, {Usov}, {van Veelen}, {van der Veen},
  {ter Veen}, {Verbiest}, {Vermeulen}, {Vermaas}, {Vocks}, {Vogt}, {de Vos},
  {van der Wal}, {van Weeren}, {Weggemans}, {Weltevrede}, {White}, {Wijnholds},
  {Wilhelmsson}, {Wucknitz}, {Yatawatta}, {Zarka}, {Zensus}, \& {van
  Zwieten}}]{vanHaarlem+13}
{van Haarlem}, M.~P., {Wise}, M.~W., {Gunst}, A.~W., {et~al.} 2013, \aap, 556,
  A2

\bibitem[{{Vogelsberger} {et~al.}(2014){Vogelsberger}, {Genel}, {Springel},
  {Torrey}, {Sijacki}, {Xu}, {Snyder}, {Nelson}, \&
  {Hernquist}}]{Vogelsberger+14}
{Vogelsberger}, M., {Genel}, S., {Springel}, V., {et~al.} 2014, \mnras, 444,
  1518

\bibitem[{{Voit} \& {Donahue}(2015)}]{Voit+Donahue14}
{Voit}, G.~M. \& {Donahue}, M. 2015, \apjl, 799, L1

\bibitem[{{Whittam} {et~al.}(2017){Whittam}, {Jarvis}, {Green}, {Heywood}, \&
  {Riley}}]{Whittam+17}
{Whittam}, I.~H., {Jarvis}, M.~J., {Green}, D.~A., {Heywood}, I., \& {Riley},
  J.~M. 2017, \mnras, 471, 908

\bibitem[{{Williams} {et~al.}(2018{\natexlab{a}}){Williams}, {Calistro Rivera},
  {Best}, {Hardcastle}, {R{\"o}ttgering}, {Duncan}, {de Gasperin}, {Jarvis},
  {Miley}, {Mahony}, {Morabito}, {Nisbet}, {Prandoni}, {Smith}, {Tasse}, \&
  {White}}]{Williams+18}
{Williams}, W.~L., {Calistro Rivera}, G., {Best}, P.~N., {et~al.}
  2018{\natexlab{a}}, \mnras, 475, 3429

\bibitem[{{Williams} {et~al.}(2018{\natexlab{b}})}]{Williams+18b}
{Williams}, W.~L. {et~al.} 2018{\natexlab{b}}, \aap\ submitted

\bibitem[{{Willott} {et~al.}(1999){Willott}, {Rawlings}, {Blundell}, \&
  {Lacy}}]{Willott+99}
{Willott}, C.~J., {Rawlings}, S., {Blundell}, K.~M., \& {Lacy}, M. 1999,
  \mnras, 309, 1017

\bibitem[{{Wright} {et~al.}(2010){Wright}, {Eisenhardt}, {Mainzer}, {Ressler},
  {Cutri}, {Jarrett}, {Kirkpatrick}, {Padgett}, {McMillan}, {Skrutskie},
  {Stanford}, {Cohen}, {Walker}, {Mather}, {Leisawitz}, {Gautier}, {McLean},
  {Benford}, {Lonsdale}, {Blain}, {Mendez}, {Irace}, {Duval}, {Liu}, {Royer},
  {Heinrichsen}, {Howard}, {Shannon}, {Kendall}, {Walsh}, {Larsen}, {Cardon},
  {Schick}, {Schwalm}, {Abid}, {Fabinsky}, {Naes}, \& {Tsai}}]{Wright+10}
{Wright}, E.~L., {Eisenhardt}, P.~R.~M., {Mainzer}, A.~K., {et~al.} 2010, \aj,
  140, 1868

\bibitem[{{Zanni} {et~al.}(2003){Zanni}, {Bodo}, {Rossi}, {Massaglia},
  {Durbala}, \& {Ferrari}}]{Zanni+03}
{Zanni}, C., {Bodo}, G., {Rossi}, P., {et~al.} 2003, \aap, 402, 949

\end{thebibliography}

\end{document}